\documentclass[%
 aip,
 amsmath,amssymb,
 reprint,%
author-numerical,%
]{revtex4-1}

\usepackage{graphicx}
\usepackage{dcolumn}
\usepackage{bm}
\usepackage[english]{babel}
\usepackage[utf8]{inputenc}
\usepackage[T1]{fontenc}
\usepackage{mathptmx}
\usepackage{longtable}
\usepackage{braket}
\usepackage{amsmath}
\usepackage{amsfonts}
\usepackage{listings}
\usepackage{graphicx}
\usepackage{color}
\usepackage[section]{placeins}

\usepackage{xr}
\makeatletter
\newcommand*{\addFileDependency}[1]{
  \typeout{(#1)}
  \@addtofilelist{#1}
  \IfFileExists{#1}{}{\typeout{No file #1.}}
}
\makeatother

\newcommand*{\myexternaldocument}[1]{
    \externaldocument{#1}
    \addFileDependency{#1.tex}
    \addFileDependency{#1.aux}
}

\myexternaldocument{supplement}
    
\begin{document}

\preprint{AIP/123-QED}

\title{Assessing MP2 frozen natural orbitals in relativistic correlated electronic structure calculations}

\author{Xiang Yuan}
    \email{xiang.yuan@univ-lille.fr}
    \affiliation{ 
        Université de Lille, CNRS, UMR 8523 - PhLAM - Physique des Lasers, Atomes et Molécules, F-59000 Lille,  France.
    }

    \affiliation{%
        Department of Chemistry and Pharmaceutical Sciences, Faculty of Science, Vrije Universiteit Amsterdam, de Boelelaan 1083, 1081 HV Amsterdam, The Netherlands.
    }%

\author{Lucas Visscher}
    \email{l.visscher@vu.nl}
    \affiliation{%
        Department of Chemistry and Pharmaceutical Sciences, Faculty of Science, Vrije Universiteit Amsterdam, de Boelelaan 1083, 1081 HV Amsterdam, The Netherlands.
    }%

\author{Andr\'{e} Severo Pereira Gomes}%
    \email{andre.gomes@univ-lille.fr}
    \affiliation{ 
        Université de Lille, CNRS, UMR 8523 - PhLAM - Physique des Lasers, Atomes et Molécules, F-59000 Lille, France.
    }%

\begin{abstract}
The high computational scaling with basis set size and the number of correlated electrons is a bottleneck limiting applications of coupled cluster (CC) algorithms, in particular for calculations based on 2- or 4-component relativistic Hamiltonians, which often employ uncontracted basis sets. This problem may be alleviated by replacing canonical Hartree-Fock (HF) virtual orbitals by natural orbitals (NOs). In this paper we describe the implementation of a module for generating NOs for correlated wavefunctions, and in particular MP2 frozen natural orbitals (MP2FNOs), as a component of our novel implementation of relativistic coupled cluster theory for massively parallel architectures [J.\ Pototschnig et.\ al.\, \emph{J.\ Chem.\ Theory Comput.}~\textbf{17}, 5509, 2021].  Our implementation can manipulate complex or quaternion density matrices, thus allowing for the generation of both Kramers-restricted and Kramers-unrestricted MP2FNOs. Furthermore, NOs are re-expressed in the parent atomic orbital (AO) basis, allowing for generating CCSD NOs in AO basis for further analysis. By investigating the truncation errors of MP2FNOs for both the correlation energy and molecular properties-electric field gradients at the nuclei, electric dipole and quadrupole moments for hydrogen halides HX (X=F-Ts), and parity-violating energy differences for H$_2$Y$_2$ (Y=O-Se)-we find MP2FNOs accelerate the convergence of the correlation energy in a roughly uniform manner across the periodic table. It is possible to obtain reliable estimates for both energies and the molecular properties considered with VMO spaces truncated to about half the size of the full spaces.
\end{abstract}

\maketitle

\section{\label{sec:level1} Introduction}
Understanding the electronic structure of large molecules or complexes containing heavy elements such as lanthanides or actinides is a problem of relevance for many technological applications. Examples are the nuclear fuel cycle, such as in the development of new extractants for separation processes\cite{Acher2016,leoncini2017ligands,sun2012ionic,Berger2020,Gogolski2020,Berger2021}, and the use of lanthanides as exceptionally strong single molecule magnets\cite{gould_2022}. 
To model such materials, Density Functional Theory (DFT) has become the most widely used correlated electronic structure theory approach~\cite{burke_perspective_2012}, even though it is difficult to systematically approach exact results with the currently available density functional approximations~\cite{burke_perspective_2012,bartlett_ab_2005}. In the particular case of relativistic electronic structure calculations, DFT energies may even for closed-shell species strongly deviate from experimental or accurate theoretical results~\cite{kervazo_accurate_2019,aebersold_considering_2021,aquino_electric_2010}. 

This also holds for molecular properties, recently ~\citeauthor{sunaga_towards_2021}\cite{sunaga_towards_2021} reported  that the performance of DFT for parity violation energy shift (PV) calculations -- a property requiring a very accurate description of the electronic wave function near the nuclei -- is somewhat disappointing, with deviations to CCSD being as large as 10\%. 
These uncertainties in the performance of  of DFT for heavy elements, especially for cases in which experimental values are absent or difficult to generate, calls for the use of state-of-the-art wavefunction methods to either provide accurate reference data, or be applied directly if the model sizes are small enough. 

Among currently available approaches, coupled cluster (CC) theory serves as a ``gold standard''~\cite{helgaker_prediction_1997,bak_accurate_2001} for its ability to give results approaching chemical accuracy for both correlation energies and properties. However, the main difficulty in employing CC theory in large-scale applications (such as molecules containing several hundred electrons) is its high computational scaling with respect to the size of the system ($N$). For example, without approximations CCSD and CCSD(T) approaches scale with $O(N^{6})$ and $O(N^{7})$, respectively, and also with approximations such as Laplace transforms\cite{almlof1991elimination}, distance screening\cite{haser1993moller} and density fitting\cite{Baerends.Ros.1973}\cite{Feyereisen1993-RI} the scaling and prefactors are still significantly higher than in a mean-field approach like DFT.

Another key ingredient in modeling heavy element species is the treatment of relativistic effects~\cite{Dyallbook2007,Reiher2014,Saue2015}. For cases in which a given molecular property is not particularly sensitive to effects such as spin-orbit coupling (SOC) (which is often so for molecular structural parameters such bond lengths and angles), or to contributions from electrons other than those in the valence (e.g.\ dipole moments), approximate treatments of relativity can be employed, through pseudopotential approaches~\cite{Batista2015,Cao2015}, or by including only scalar relativistic effects. For more challenging applications, or higher precision, a more general framework can be based on the solution of the 4-component (4C) molecular Dirac equation~\cite{visscher1994relativistic,saue_dirac_2020,belpassi2020bertha,repisky2020respect}. Nowadays this approach is often made more tractable by solving the exact 2-component (X2C) equation that can be derived from the Dirac equation after a basis set discretization~\cite{x2c:kutzelnigg2005,jensen:rehe2005,x2c:atom2mol:liu2006,ilias_infinite-order_2007,x2c:atom2mol:peng2007,x2c:liu2009,mfso:cc:liu2018,x2camf:cc:liu2018,mfso:eomcc:cheng2019,x2camf:cc:asthana2019,Sikkema2009}. Both the original 4C approach as well as its X2C approximation can deliver accurate molecular properties across the periodic table, also for properties involving core electrons~\cite{Fgri2001,pereira_gomes_influence_2004,thierfelder_scalar_2009,Pershina2010,Autschbach2015,kervazo_accurate_2019,sunaga_towards_2021,Halbert2021}. 

The one-electron functions (molecular spinors, for simplicity also referred to in the following as molecular orbitals) obtained from solving the 4C or X2C matrix equations, serve then as a basis for a correlation treatment in the so-called no-pair approximation in which contributions to the correlation energy due to admixture of states with explicit electron-positron pairs are neglected\cite{Sucher1980}. As the computational cost of 2C approaches is lower than 4C ones, the choice between which treatment to use prior to the correlated treatment will depend on a case-by-case analysis of whether the additional cost of the latter will be offset by improvements in accuracy over the former.

Recently,~\citeauthor{pototschnig_implementation_2021}\cite{pototschnig_implementation_2021} described a new, efficient relativistic coupled cluster implementation based on ExaTENSOR~\cite{lyakh_domainspecific_2019}, a distributed numerical tensor algebra library for GPU-accelerated HPC platforms. This code enabled the calculation of molecular properties with CCSD wave functions for systems such as [(UO$_2$)(NO$_3$)$_3$]$^{-}$, for which  200 electrons and around 1000 virtual molecular orbitals were included in the correlated treatment. Compared to nonrelativistic implementations, it is in particular the introduction of SOC which increases the computational cost (though not altering the overall scaling)\cite{Helmich-Paris2016ev}. Not only does this necessitate the use of complex algebra, it also makes use of contracted basis sets more difficult as one needs to be able to describe the differences in radial extent of spin-orbit split orbitals. This effectively doubles the number of functions that is needed to describe the $p$-, $d$- and $f$-type core orbitals\cite{visscher1991contraction} in the case of implementations based on expressing spinors in scalar basis sets~\cite{visscher1991contraction,saue_dirac_2020,repisky2020respect}. In this framework another complication arises in how to define contractions for the small component part of the spinors that would respect the kinetic balance condition. We note that these problems do not arise in implementations capable of directly evaluating 2-electron integrals over 2C spinor basis functions~\cite{Belpassi2005,belpassi2020bertha}. 

The lack of efficient contraction schemes for scalar basis-derived implementations, combined with the fact that in many potential applications it may be required that all, or a significant part of core electrons (e.g. in core spectroscopies~\cite{Halbert2021})  are treated explicitly, including the deep core orbitals that require many Gaussian type functions to be described correctly, leads to calculations having large virtual spaces. This directly affects the performance of CC-based approaches as these scale with the fourth power of the number of virtuals ($O(V^{4})$). In practice one therefore often alleviates the computational effort to some extent by leaving out high energy virtuals that are mostly localized in the core, but convergence with respect of such an energy cut-off threshold is slow, especially when semi-core or core correlation need to be described as well.
    
In non-relativistic electronic structure theory it has long been recognized that, in contrast to canonical Hartree-Fock molecular orbitals (CMO), natural orbitals (NOs)--the eigenvectors of the one-body reduced density matrix (1RDM)~\cite{lowdin_quantum_1955,davidson_properties_1972}--provide a more compact and quickly converging orbital representation for describing post Hartree-Fock wavefunctions. Based on this observation, the idea of replacing the CMOs by NOs to reduce the size of the virtual space and thereby computational cost was introduced\cite{barr_nature_1970}. 

Rather than using an energy threshold as is done by CMOs, one may instead omit NOs which are likely to not strongly contribute to the total correlation energy from the virtual space. This is done on by considering the magnitude of natural occupation numbers of approximate NOs, obtained from a method that quickly gives access to a reasonable approximation of the 1RDM of the correlated wave function. 
For this purpose, second order M\"oller–Plesset perturbation theory (MP2) is a particularly appealing approach to obtain the 1RDM and the  approximate NOs because of its low scaling, non-iterative $O(N^{5})$, and the ability to recover most of correlation effects. 
Within the natural orbital family of methods, the virtual frozen  natural orbitals (FNOs) approach~\cite{barr_nature_1970,sosa_selection_1989,taube_frozen_2005} has therefore become popular because of its clear concept and simple implementation. FNO theory has been thoroughly discussed for non-relativistic models such as configuration interaction (CI)\cite{barr_nature_1970}, multi-configuration self-consistent field (MCSCF)~\cite{jensen_secondorder_1988,jensen_erratum_1988} and coupled cluster~\cite{taube_frozen_2005,taube_frozen_2008}. Recently,~\citeauthor{verma_scaling_2021}\cite{verma_scaling_2021} furthermore extended the FNOs algorithm to quantum computers.
     
The ability of virtual FNOs to reduce computational cost compared to CMOs is even more appealing in relativistic electronic structure calculations, in particular in the case of contracted basis sets and/or when correlating sub-valence electrons. The main goal of this paper is therefore to describe and showcase the implementation of MP2-based FNOs (MP2FNOs) within the framework of the new relativistic coupled cluster implementation for massively parallel, GPU-accelerated platforms~\cite{pototschnig_implementation_2021}. While our code primarily aims at describing cases in which spin-orbit coupling is taken into account, we also demonstrate its use with a non-relativistic Hamiltonian.

Our second aim is to discuss the performance of MP2FNOs across the periodic table, by treating model systems containing elements ranging from first-row such as fluorine, to superheavy elements such as tennessine. Here, we shall focus in how well truncated FNO spaces describe both correlation energies as well as first-order properties such as the electric field gradient at the nuclei (EFGs), parity-violation energy shifts, and electric multipoles (dipole and quadrupole). The first two properties are chosen as representatives of properties for which a good description of wavefunctions in the region close the nuclei is important (even if only valence electrons are correlated), while the electric dipole and quadrupole moments are an example of properties for which the major contributions arise from valence electrons.

Besides being able to improve the efficiency of a calculation, natural orbitals are also interesting as tools for analysis, such as in estimating the multi-reference character of a system~\cite{gordon_natural_1999}. Their visualization in real space can furthermore provide insight into correlation effects, since by their one-particle nature they are easier to interpret than the wavefunction itself. Thus, another aim of our implementation was to provide a tool for obtaining natural orbitals for any correlated wavefunction, from a 1RDM. In this paper we shall make use of the analysis of the CCSD natural orbitals, and in subsequent work we plan to further explore the use of natural orbitals in visualization.

This article is organized as follows. In Sec. II the background of MP2FNOs theory is summarized. In Sec. III we described the details of implementation. All the sample calculations are presented and discussed in Sec.IV-V. And finally a brief summary is given in Sec. VI.

We note that upon completing this manuscript, we have become aware of another implementation of the MP2FNOs approach for relativistic correlated methods, in the BAGH code~\cite{chamoli2022lower}. While the main features of the MP2FNOs method are the same in both implementations we first note that our implementation fully exploits ExaTENSOR's single-node or distributed memory (multi-node) and GPU acceleration capabilities, and as such can be efficiently employed in systems ranging from local clusters to latest-generation supercomputers. Second, as it will be outlined below our implementation is capable of manipulating both complex and quaternion density matrices, thus allowing for the generation of both Kramers-restricted and Kramers-unrestricted MP2FNOs. Finally, it allows for re-expressing NOs in atomic orbital (AO) basis for further analysis.

\section{\label{sec:level1} Theory}
    As the MP2FNO approach is well-known in a non-relatistic context and requires essentially no modification for application in a no-pair relativistic context, we will only provide a brief description. We apply the orbital-unrelaxed MP2 approximation (for working equations for the orbital-relaxed formalism in a relativistic context see reference~\citenum{vanStralen2005}). The second-order contribution to the occupied-virtual block of the density matrix from single excitations is zero for canonical orbitals\cite{davidson_properties_1972} while the relaxation contributions to this block are ignored in orbital-unrelaxed MP2. This approximation thus decouples the occupied and virtual spaces and allows us to focus only on obtaining the virtual-virtual block of the density matrix. We want to keep occupied orbitals in their canonical form and therefore will not determine the occupied-occupied part of the density matrix, for which we simply retain the diagonal Hartree-Fock form $\bm{D^{(0)}_{oo}}=\bm{1_{oo}}$. With these approximations the second-order FNOs density matrix is given as
    \begin{equation}
        \bm{D^{FNO}}=\left[
        \begin{matrix}
            \bm{1_{oo}} & \bm{0_{ov}} \\
            \bm{0_{ov}} & \bm{D^{(2)}_{vv}}  \\
        \end{matrix}
        \right] 
     \end{equation}     
with the relevant matrix elements given as
	\begin{equation}
        D_{AB}^{(2)}= 
        \frac{1}{2}\sum_{C,I,J}\frac{\bra{AC}\ket{IJ}\bra{IJ}\ket{BC}}{\epsilon_{IJ}^{AC}\epsilon_{IJ}^{BC}}
    \label{Dvv}
    \end{equation}
    Here, following usual conventions \textit{I, J} denote occupied spinors and \textit{A, B} virtual spinors, and $\epsilon_{IJ}^{AB}=\epsilon_{I}+\epsilon_{J}-\epsilon_{A}-\epsilon_{B}$, is the energy difference between canonical Hartree-Fock spinors. In our implementation, these matrices are obtained without imposing any time-reversal symmetry on the spinor set and are thus valid for both Kramers-restricted and Kramers-unrestricted Hartree-Fock procedures. If time-reversal~\cite{Kramers1930} and spatial symmetry\cite{Visscher1996} are not present, these matrices are in general to be represented in complex algebra. 
    
    The simplest procedure to obtain natural spinors is to apply a diagonalization in complex algebra after $\bm{D^{(2)}_{vv}}$ is formed. However, for closed shell systems in which time-reversal symmetry is enforced for the orbitals at the mean-field level, as is the case in the DIRAC code, one would like to retain such Kramers pairing also for the natural spinors. This requires careful attention when degeneracies beyond the two-fold Kramers degeneracy are present, as complex diagonalization will arbitrarily mix the degenerate solutions such that Kramers pairing is not guaranteed. 
    
    In such cases, Kramers pairing of the natural spinors can be enforced by first transforming $\bm{D^{(2)}_{vv}}$ from complex to quaternion  algebra $\bm{D_{vv}^{(Q)}}$,
    \begin{equation}
        \bm{D_{vv}^{(Q)}}=\bm{U^{\dagger}}\bm{D_{vv}^{(2)}}\bm{U}
    \end{equation}
where the tranformation matrix $\bm{U}$ is given by~\cite{saue_quaternion_1999}    
    \begin{equation}
        \bm{U}=\begin{pmatrix}
            \bm{\mathbb{I}} & \check{j}\bm{\mathbb{I}} \\
            \check{j}\bm{\mathbb{I}} & \bm{\mathbb{I}} 
          \end{pmatrix}
    \end{equation}
and $\bm{\mathbb{I}}$ is a unit matrix of dimension $n\times n, n$ being the number of Kramers pairs. This transformation will block diagonalize $\bm{D^{(2)}_{vv}}$, leading to two decoupled matrix problems of half the original dimension in quaternion algebra. 
  	\begin{equation}
        \bm{D_{vv}^{(Q)}}\bm{V}=\bm{O}\bm{V}
        \label{natorb-quaternion}
    \end{equation}

Diagonalization of one of the two blocks provides a unique set (V) of quaternion FNOs in MO basis and their respective occupation numbers (O) from which the solutions for the other block can be generated via Kramers symmetry. 

At this stage, we can employ the information from the occupation numbers to reduce the original virtual space $\bm{V}$ to $\bm{\bar{V}}$ by removing from $\bm{V}$ orbitals with occupation numbers lower than a user-defined threshold. The larger such threshold is, the smaller the remaining set will be. In the numerical examples presented in the following we will try to find the optimal balance between efficiency (smaller sets) and accuracy (recovering more of the correlation energy and better reproducing first order properties).

After the basis is truncated, it is convenient to recanonize the remaining orbitals. To this end, we perform two operations: first we transform the virtual-virtual block of the Fock matrix ($\bm{F_{vv}}$) into the truncated FNOs basis ($\bm{\bar{F}_{vv}}$) and then do a (quaternion re)diagonalisation to obtain a new set of canonical orbitals $\bm{\bar{W}}$ and orbital energies $\bm{\bar{\epsilon}}$.
    \begin{align}
        \bm{\bar{F}}&=\bm{\bar{V}^{\dagger}}\bm{F}\bm{\bar{V}} \\
        \bm{\bar{F}}\bm{\bar{W}} &=\bm{\bar{\epsilon}}\bm{\bar{W}}
        \label{recanonization}
    \end{align}

The product $\bm{\bar{V}\bar{W}}$ of these two transformation matrices gives the transformation that expresses the recanonized truncated natural orbital set in the atomic orbital(AO) basis. These orbitals can then be used in any subsequent correlated calculation as replacement to the original Hartree-Fock orbitals. In summary the transformation of the original Hartree-Fock orbitals $\bm{U}$ to the new set $\bm{\bar{U}}$ is thus given by 
	\begin{equation}
	    \bm{U_{new}}=[\bm{U_{occ}},\bm{\bar{U}_{vir}}]
	\end{equation}
where
	\begin{align}
	    \bm{\bar{U}_{vir}}&=\bm{U_{vir}}(\bm{\bar{V}}\bm{\bar{W}}) \\
	    \label{transformtocanonical}
	    \bm{\bar{U}_{occ}}&=\bm{U_{occ}}
	\end{align}
and the dimension of the rectangular matrix $\bm{\bar{V}}$ and that of the square matrix $\bm{\bar{U}}$ depends on the truncation threshold.

\section{Implementation}

    The aforementioned algorithm has been implemented in the relativistic quantum chemistry package DIRAC~\cite{saue_dirac_2020}, as part of the ExaCorr code~\cite{pototschnig_implementation_2021}. Our implementation allows for calculations to be carried out either using a single-node configuration (for which ExaTENSOR provides OpenMP parallelization, on top of GPU offloading) or employing distributed memory in the case of multi-node runs.
    
    Our implementation is structured following its three main tasks: one module deals with the construction of density matrices in the MO basis, another carries out the complex or quaternion diagonalization of density matrices, and the final module takes care of the construction of the recanonized MP2FNOs. The final step should be repeated if the truncation treshold is changed, the first two are independent of this threshold.
    
    \subsection{Density matrix construction}
    
    In Exacorr all computationally expensive operations such as the tensor contractions used in the determination of amplitudes or the construction of density matrices are offloaded to ExaTENSOR library. 
    
    In the case of MP2FNOs, we have created a dedicated driver for MP2 calculations, in which we (i) calculate and store in memory two electron repulsion integrals (ERIs) in AO basis with the aid of the InteRest library\cite{repisky_respect_2020}; 
    (ii) employ the standard Yoshimine scheme\cite{yoshimine_construction_1973} (which scales as ($O(N^{5})$) to transform the AO integrals to the MO basis. Only the direct and exchange contributions to the $\bra{ij}\ket{ab}$ (OOVV-type) integrals are calculated, which makes the integral transformation step much faster than the generation of a complete set of MO integrals as is normally done in ExaCorr. After  antisymmetrization the $\bra{ij}\ket{ab}$ integrals are  stored (in memory) as ExaTENSOR tensors, which may reside on a single node or be distributed over several nodes. Finally, (iii) the MP2 amplitudes tensor and the MP2 energy are determined and the density matrix $\bm{D^{(2)}_{vv}}$ is constructed according to equation \ref{Dvv}. 

\subsection{Complex-quaternion transformation and diagonalization}
    The density matrix computed in Exacorr is generated in complex algebra. The complex-quaternion transformation is carried out following Eqn.~27 in ref.~\citenum{shee_analytic_2016} 
    \begin{equation}
        ^{Q}\gamma_{p q} = \text{Re}(\gamma_{p q}) + \check{i}\text{Im}(\gamma_{p q}) + \check{j}\text{Re}(\gamma_{p \bar{q})} + \check{k} \text{Im}(\gamma_{p \bar{q}}),
    \end{equation}
 in which lowercase symbols with (without) bars indicate the Kamers pairing of the original MO basis and the one-particle reduced density matrix (1RDM) is now indicated by $\gamma$ for consistency of notation with that of ref.~\citenum{shee_analytic_2016}. This quaternion form can be diagonalized using the quaternion diagonalization routine provided by DIRAC and be back-transformed to complex representation by the routines provided in this module. If the original basis did not possess Kramers symmetry, the diagonalization is directly carried out in complex algebra. The resulting full set of FNOs is then stored on file for analysis and processing by module (C) described below. The module responsible for the quaternion transformation and diagonalization is constructed in such a way that it can also be used with wave function models other than MP2.

\subsection{Selection and recanonization}
The third module retrieves FNOs that have occupation numbers above treshold and transforms the original Fock matrix to the truncated virtual space. After recanonization (Eqn.~\ref{recanonization}) and transformation to AO-basis (Eqn.~\ref{transformtocanonical}) the final reduced set of FNOs is stored on file. As this step takes virtually no time compared to the other steps in the procedure, this may be easily repeated to test out the effect of varying the treshold value on the generated recanonized orbitals and their energies.

\subsection{Summary of the MP2FNO Algorithm}

\begin{figure}[hbt]
 \includegraphics[width=\linewidth]{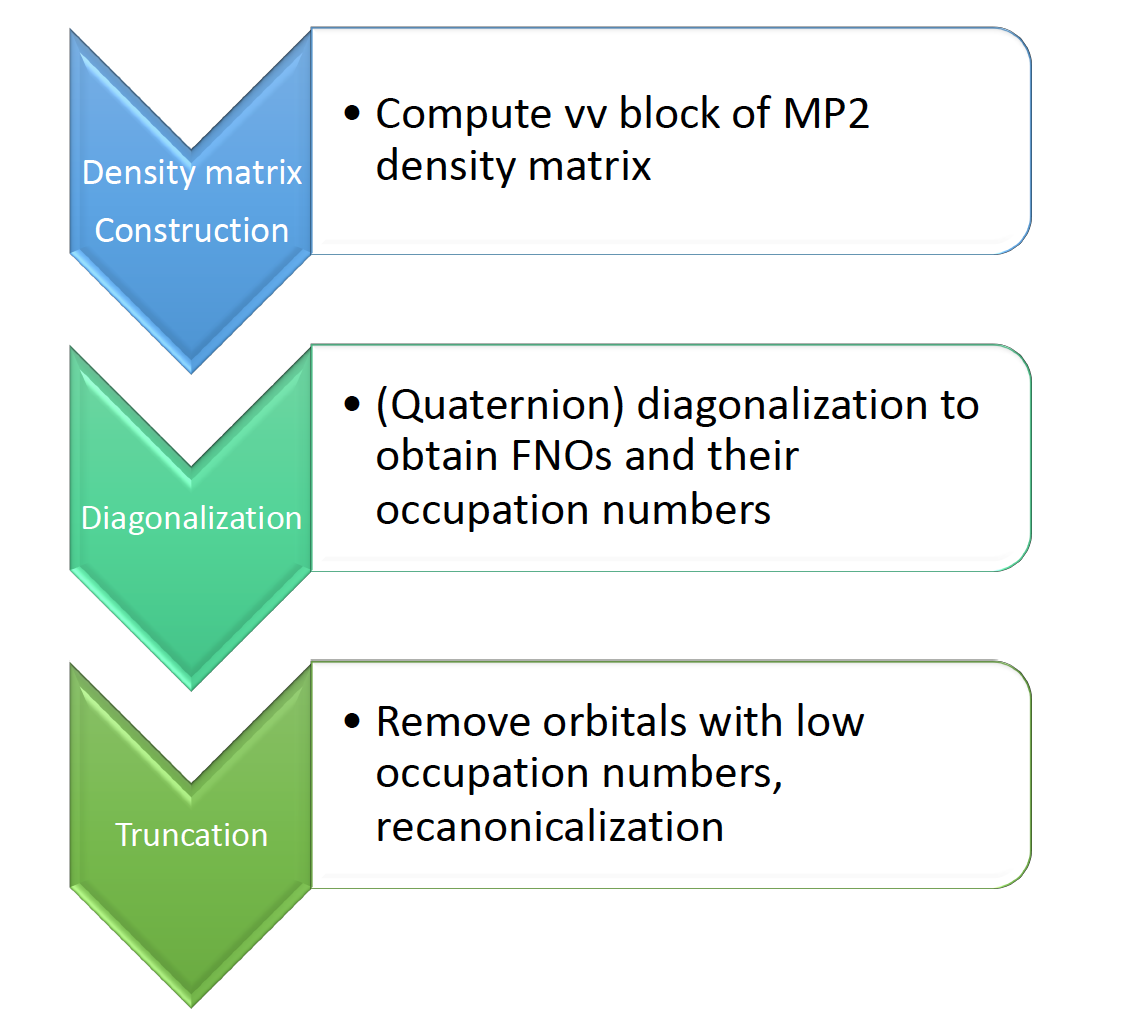}
 \caption{\label{fig:workflow-mp2no} Work flow of a MP2 Frozen natural orbitals calculation.}
\end{figure}

The MP2FNO workflow is schematically represented in Figure~\ref{fig:workflow-mp2no}, and consists of the following steps:
    \begin{enumerate}
        \item A calculation in the full basis set is performed to get a set of occupied and virtual molecular orbitals.
        \item The virtual-virtual block of MP2 density matrix is generated. We note that in this step is possible to either consider the full virtual set, or already employ a preliminary screening in which for instance very high-energy virtuals are neglected.
        \item Optional, only for restricted starting orbitals: Transform the complex $\bm{D^{(2)}_{vv}}$ into the quaternion representation, $\bm{D^{(Q)}_{vv}}$.
        \item Diagonalize $\bm{D_{vv}}$ or $\bm{D^{(Q)}_{vv}}$ to determine FNOs and occupation numbers.
        \item For a given threshold: truncate the FNO space and transform the Fock matrix to recanonize and write the orbitals out in the AO basis.
        \item Use the recanonized FNOs in higher level wave-function models such as coupled cluster. 
\end{enumerate}

\section{Computational details}

All MP2 and CC calculations are carried out with development versions  (revisions~\texttt{2e659d7, 0f8e9f2, 9e10bc667, 8b81f8a, e7d2d4d, ec415a5, 0ff0d6f, d70bbe2}) of the DIRAC code\cite{Dirac_22,Dirac_21,saue_dirac_2020}, and of the ExaTENSOR~\cite{lyakh_domainspecific_2019,exatensor-url} library (revision~\texttt{d304c03b7}), employing Dyall basis sets of triple-zeta quality (dyall.av3z)~\cite{dyall_relativistic_2002,dyall_relativistic_2006,dyall_relativistic_2012} for the heavy elements (Br, I, At, Ts, Se), and Dunning aug-cc-pVTZ sets~\cite{Dunning1989,kendall_electron_1992,woon_gaussian_1993} for (H, F, Cl, O, S), all of which are left uncontracted unless otherwise noted. The aug-cc-pCVTZ basis set was also employed in the HCl case~\cite{Woon1995,Peterson2002}, for investigating how the additional core correlating functions in aug-cc-pCVTZ affect the results. Finally, for investigating the convergence of expectation values for HI, we employed double-zeta quality Dyall~\cite{dyall_relativistic_2002,dyall_relativistic_2006} (dyall.av2z) and Dunning~\cite{Dunning1989,kendall_electron_1992,woon_gaussian_1993} (aug-cc-pVDZ) basis sets for I and H respectively.

Here we make use in all calculations of the exact two component Hamiltonian (X2C)~\cite{ilias_infinite-order_2007}, in which we include two-electron spin-orbit contributions via to the untransformed two-electron potential via atomic mean-field contributions calculated with the \textsc{AMFI} code~\cite{HE1996365,marian2001,prog:amfi}. As part of the supplementary information, we provide in table S2 results with different Hamiltonians (Dirac-Coulomb, X2C, spinfree X2C and Non-relativistic) for expectation values for the HTs molecule. We note the code is equally capable of handling orbitals obtained with the so-called molecular-mean-field approach, in which the transformation to 2C is carried out after a 4C mean-field calculation~\cite{Sikkema2009}. In the case of the HCl molecule, we have also employed the non-relativistic Hamiltonian (as specified by the \texttt{.NONREL} keyword), both with contracted and uncontracted bases sets.

The molecular structures employed in all calculations have been taken from the literature: In the case of hydrogen halides, from~\citeauthor{huber1979constants}\cite{huber1979constants} for HF to HI, from~\citeauthor{pereira_gomes_influence_2004}\cite{pereira_gomes_influence_2004} for HAt and from ~\citeauthor{thierfelder_scalar_2009}\cite{thierfelder_scalar_2009} for HTs. The internuclear distances employed are thus H-F(0.9168 \AA); H-Cl(1.27455 \AA); H-Br(1.41443 \AA); H-I(1.609 \AA); H-At(1.722 \AA); H-Ts(1.941 \AA). 
For the chiral molecules H$_{2}$Y$_{2}$, the Y-Y bond length, H-Y bond length and H-Y-Y bond angle are taken from Table I of~\citeauthor{laerdahl_fully_1999}\cite{laerdahl_fully_1999} and the dihedral angle is fixed at 45 degrees throughout the computation. All calculations have been carried out with a point charge nucleus model. 

Besides calculating MP2, CCSD and CCSD(T) energies, we have obtained electric dipole moment (EDM), electric quadrupole moment (EQM) and electric field gradient (EFG) for the HX systems, and parity violation energy differences (PV) for H$_{2}$Y$_{2}$ systems. These properties have been obtained analytically for CCSD wavefunctions using the implementation described in ref.~\citenum{shee_analytic_2016}.

In the calculations, the size of the complete virtual spinor spaces is 156 (HF), 182 (HCl), 310 (HBr), 382 (HI), 544 (HAt) and 588 (HTs), 314 (H$_{2}$O$_{2}$), 366 (H$_{2}$S$_{2}$) and 622 (H$_{2}$Se$_{2}$). Unless otherwise noted, all electrons were correlated in the calculations. 
For the smaller systems HF, HCl, HBr, H$_{2}$O$_{2}$ and H$_{2}$S$_{2}$ we were able to perform calculations on a single node of the laboratory compute cluster in Lille. For HI, HAt, H$_{2}$Se$_{2}$ and HTs we employed, respectively, 32, 32, 49 and 64 nodes of the Summit supercomputer. For HI calculations with double-zeta quality basis sets, we were able to perform single-node calculations on the Jean Zay supercomputer.

The data presented in this manuscript is available at the Zenodo repository of reference ~\citenum{dataset-mp2fnos}.

\section{Results and discussion}

    In this section we discuss the performance of the MP2FNOs approach for the CCSD correlation energy and molecular properties, and the impact of employing different occupation number thresholds on the results. In order to minimize computational cost on our systematic studies, we carried out CCSD(T) calculations only when discussing the effect of FNO and CMO truncation on HCl bond lengths.
    
    For minimizing the bias in the CMO and FNO comparison, we report results in which we employ the closest possible number of CMO to that of FNO, such that for CMO we avoid truncations that would remove close-lying orbitals such as those belonging to a same atomic shell. In practice, this means CMO calculations generally contain a few more orbitals than FNO ones.
    
    In figure S1 in the supplemental information, we present for the EFG of HI the difference between the approach taken here for CMO, and that of considering striclty the same number of CMO and FNO. We can see that arbitrarily truncating the virtual space for CMO can lead to large oscillations in the expectation values, a point that will be addressed in detail in the following. 
    Additionally, from the comparison of Hamiltonians for HTs in table S2 in the supplemental information, we see that our choice of the X2C Hamiltonian is a suitable one for our purposes; while total energies obviously are very different between Hamiltonians, we see that for those taking into account SOC, relative energies (HOMO-LUMO gap) and expectation values are very close to each other. This is in stark contrast with scalar relativistic and non-relativistic calculations, which still show fairly good agreement with 2C/4C approaches for the quadrupole moment and even the HOMO-LUMO gap, but are completely off the mark for the EFG and dipole moment. This is due to the very strong spin-orbit splitting in the valence p-shell of the superheavy tenessine. 

    \subsection{Correlation Energy}
\begin{figure*}[htb]
            \includegraphics[width=\linewidth]{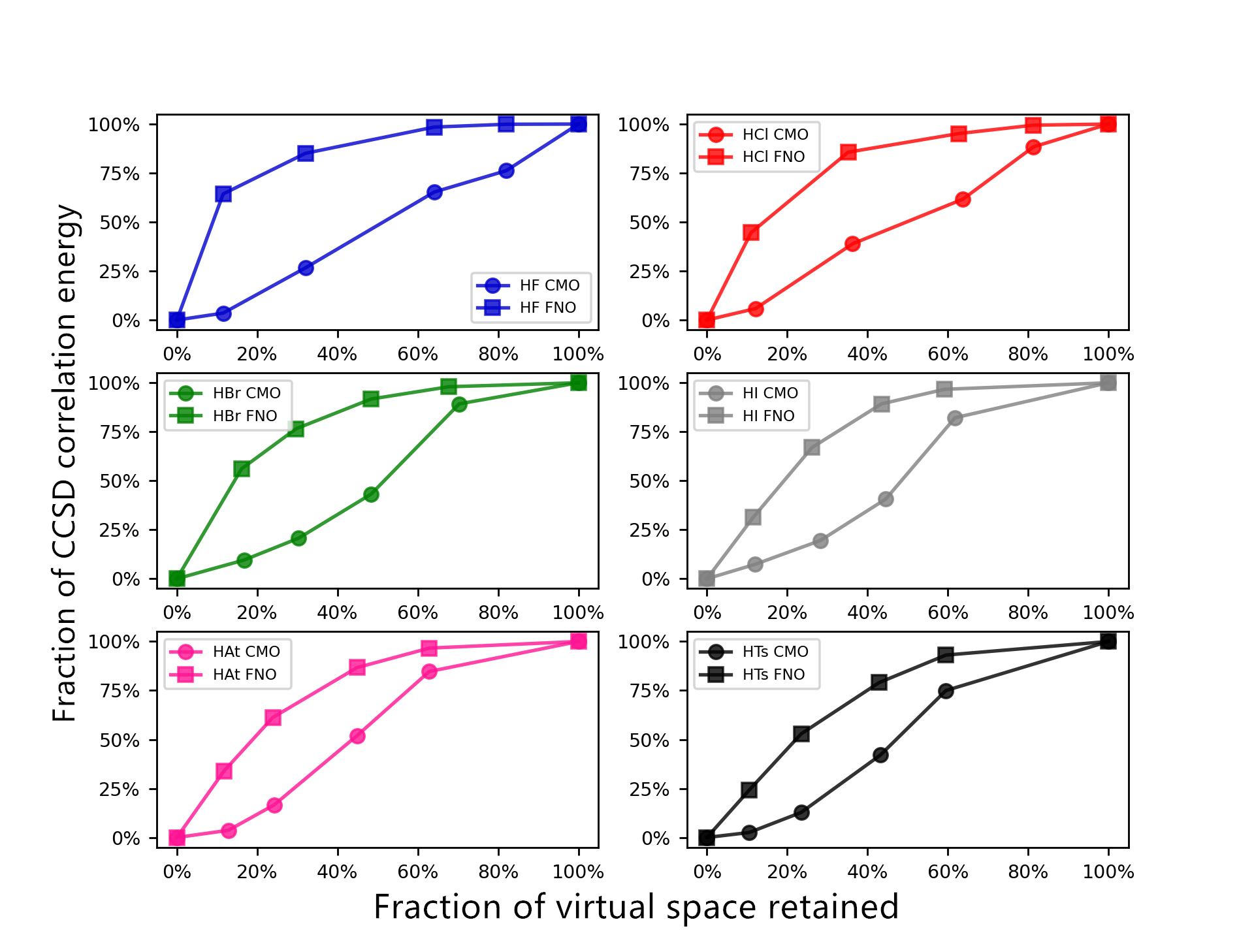}
            \caption{\label{fig:ccsd-convergence} Convergence of the CCSD correlation energy with respect to the size of the virtual orbital space, for the X2C Hamiltonian. The X axis indicates the fraction of the virtual space retained, while the Y axis gives the fraction of the correlation energy recovered with respect to the value obtained with the untruncated virtual space.}
        \end{figure*}

        CCSD correlation energies obtained with CMO and FNOs are displayed in Figure~\ref{fig:ccsd-convergence} for the hydrogen halides. In these, we show values computed for four values of natural occupation number threshold ($1.0 \times 10^{-3}, 1.0 \times 10^{-4}, 1.0 \times 10^{-5}, 1.0 \times 10^{-6}$), as well as the two extrema: a point without any virtual orbitals (corresponding to the Hartree-Fock solution), and another without any truncation of the virtual space. Since the number of virtuals varies for each system, we provide relative measures in terms of the percentage of the virtual space included in the calculations for each point, and likewise for the amount of correlation energy recovered at each point. 

        From the plots, it is clearly the case that FNOs show a more rapid convergence than CMO across the periodic table, as could be expected from the non-relativistic literature~\cite{taube_frozen_2005,taube_frozen_2008,DePrince2013}. For example, to recover only 50\% correlation energy already 40\% to 50\% of the virtual CMO space is required, while only  20\% of the FNO space suffices. Also for the more realistic goal of attaining at least 90\% of the correlation energy use of FNOs can reduce the required space by 10s of \% thereby introducing significant savings in (memory) storage of tensors with virtual indexes, as well as in operation counts for contractions involving virtual indexes in the CCSD equations. 

        An interesting finding is that the area enclosed by two lines slightly decreases as we go down the periodic table from F to Ts. This means that the FNOs approach recovers less correlation, for a given truncation, as the number of electrons increases. For example, a threshold of $1.0 \times 10^{-5}$ (the 4th point in Figure~\ref{fig:ccsd-convergence}) means, for HF and HCl, that more than  60\% of the virtual space is included in the calculation, whereas for HTs the same threshold only includes 43\%; this means that for HTs  virtual orbitals with low occupation numbers are still important for the correlation treatment.

\begin{figure}[htb]
            \includegraphics[width=8.5cm,height=8.0cm]{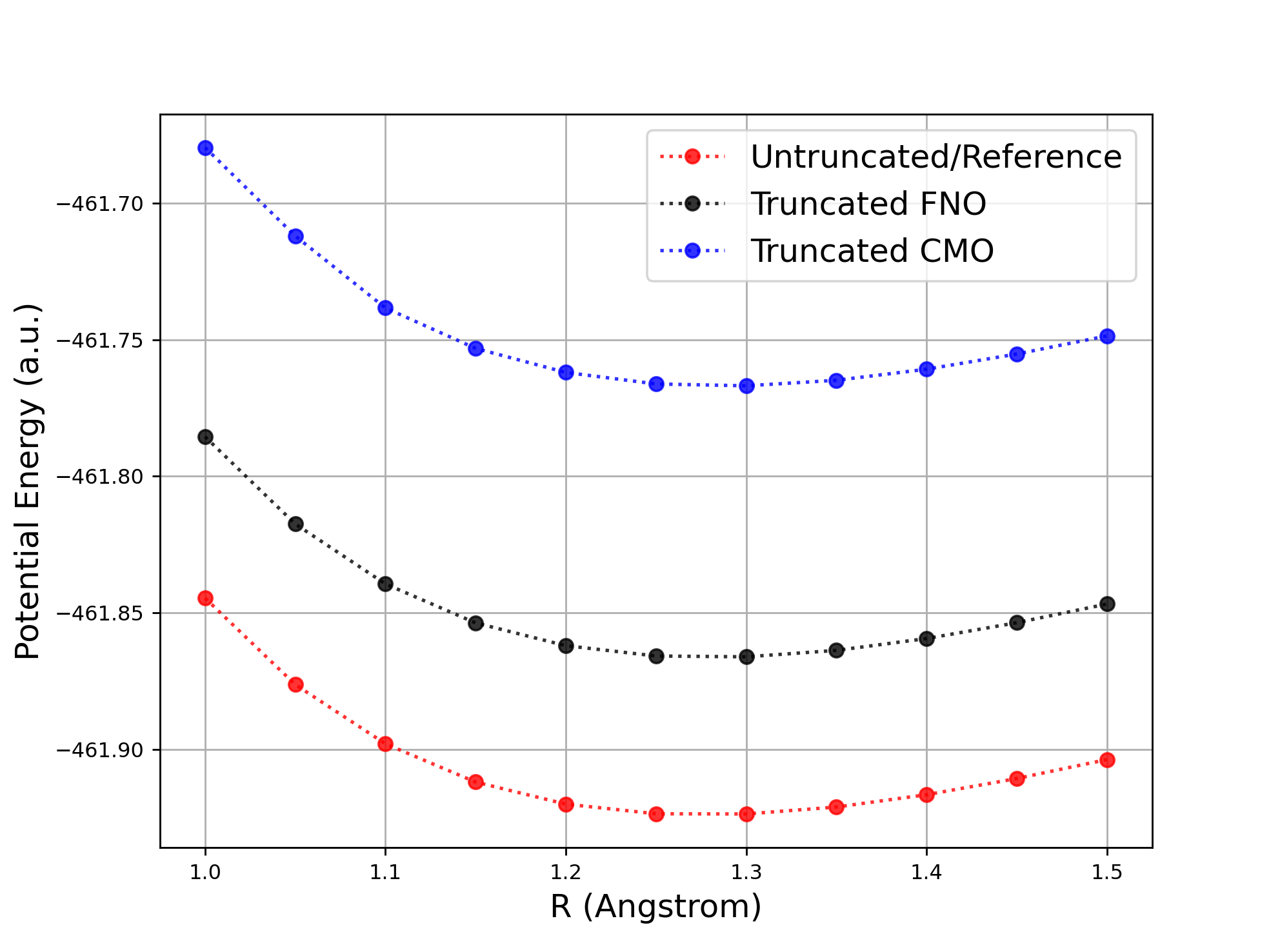}
            \caption{\label{fig:pes-scan} Potential energy curves of HCl at CCSD(T) level using untruncated orbitals (red) as reference, truncated FNO (black) and truncated CMO (blue), respectively. The X axis is the internuclear distance and the Y axis is the total energy in Hartrees.}
        \end{figure}

        Apart from testing the performance for single point at the potential energy surface, it is also important to verify the performance of FNOs at different geometric structures. To this end, in Figure~\ref{fig:pes-scan} we present the potential energy curves around equilibrium for HCl at CCSD(T) level, comparing
        sets of FNOs and CMO virtuals that are truncated to the same number of virtuals (corresponding to about 50\% of the complete virtual space).  We see that, except for a global energy shift, the FNOs follow the curvature of full virtual space potential energy curve more closely than the CMO curve does.
        
        The better agreement of FNOs relative to CMOs can be quantified by a comparison of spectroscopic constants for the three curves, shown in Table~\ref{tab:spectroscopic-hcl}. There, we report the equilibrium distance ($R_{e}$) and vibrational constant ($\omega_{e}$), calculated with the LEVEL program~\cite{le_roy_level_2017}. Taking the full valence space result as a reference, the truncated orbitals overestimate the $R_{e}$ but underestimate the $\omega_{e}$. However, the error of truncated FNOs is 0.0037\AA  ($R_{e}$) and 17 cm$^{-1}$ ($\omega_{e}$), which is only half that of truncated CMOs. 
        
        \begin{table}[htb]
            \center
            \caption{Spectroscopic constants of ground state of HCl}\label{tab:spectroscopic-hcl}

        \setlength{\tabcolsep}{6.5mm}{
            \begin{tabular}{ccc}
             & $R_{e}$(\AA)    & $\omega_{e}$ (cm$^{-1}$)    \\
            \hline
            Exp                  & 1.2746 & 2991 \\
            Untruncated Orbitals & 1.2756 & 2986 \\
            Truncated CMO       & 1.2859 & 2942 \\
            Truncated FNO        & 1.2793 & 2969 \\
            \hline
            \end{tabular}}
        \end{table}

        As we correlated all electrons, but employed basis sets without specific core correlating functions, one may ask what would be the effect of including such functions. In table S1 of the supplemental information, we provide a comparison of correlation energies and expectation vales between the uncontracted aug-pVTZ and aug-pCVTZ  basis sets for the HCl molecule without using virtual space truncation. These results clearly indicate that, although the additional high angular momentum core correlating functions are quite important for increasing the amount of correlation energy for the core and core-valence interactions, the two types of uncontracted basis sets agree well for the computed property values.
         Another interesting point to note is that, as can be seen from table S3 of the supplemental information, the percentage of correlation energy recovered at each virtual space truncation point is nearly the same for MP2 and CCSD. This indicates that the information obtained from the full virtual space MP2 calculation preceding the truncation can potentially be used to correct the CCSD correlation energy for the effect of truncation. While going beyond the scope of this paper, this point merits further investigation, in which also correction for truncation errors in higher order methods such as CCSD(T) could be investigated. 

    \subsection{Molecular properties}

\subsubsection{Electric Dipole and Quadrupole Moments}
        
To asses performance of FNOs for molecular properties we first considered the molecular EDM and EQM because these do well characterize the overall electronic charge distribution within molecules. These properties sample the regions away from the nuclei, as is clear from their operator forms in Eqs.~\ref{dipole-operator} and~\ref{quadrupole-operator}, respectively:
            \begin{align}
                D_\mu &= e\sum_{i}(\vec{r}_{i})_\mu \label{dipole-operator}\\
                Q_{\mu \nu} &= e\sum_{i}((\vec{r}_{i})_{\mu}(\vec{r}_{i})_{\nu}-r^{2}\delta_{\mu \nu}) \label{quadrupole-operator}
            \end{align}

\begin{figure*}[htb]
                \includegraphics[width=\linewidth]{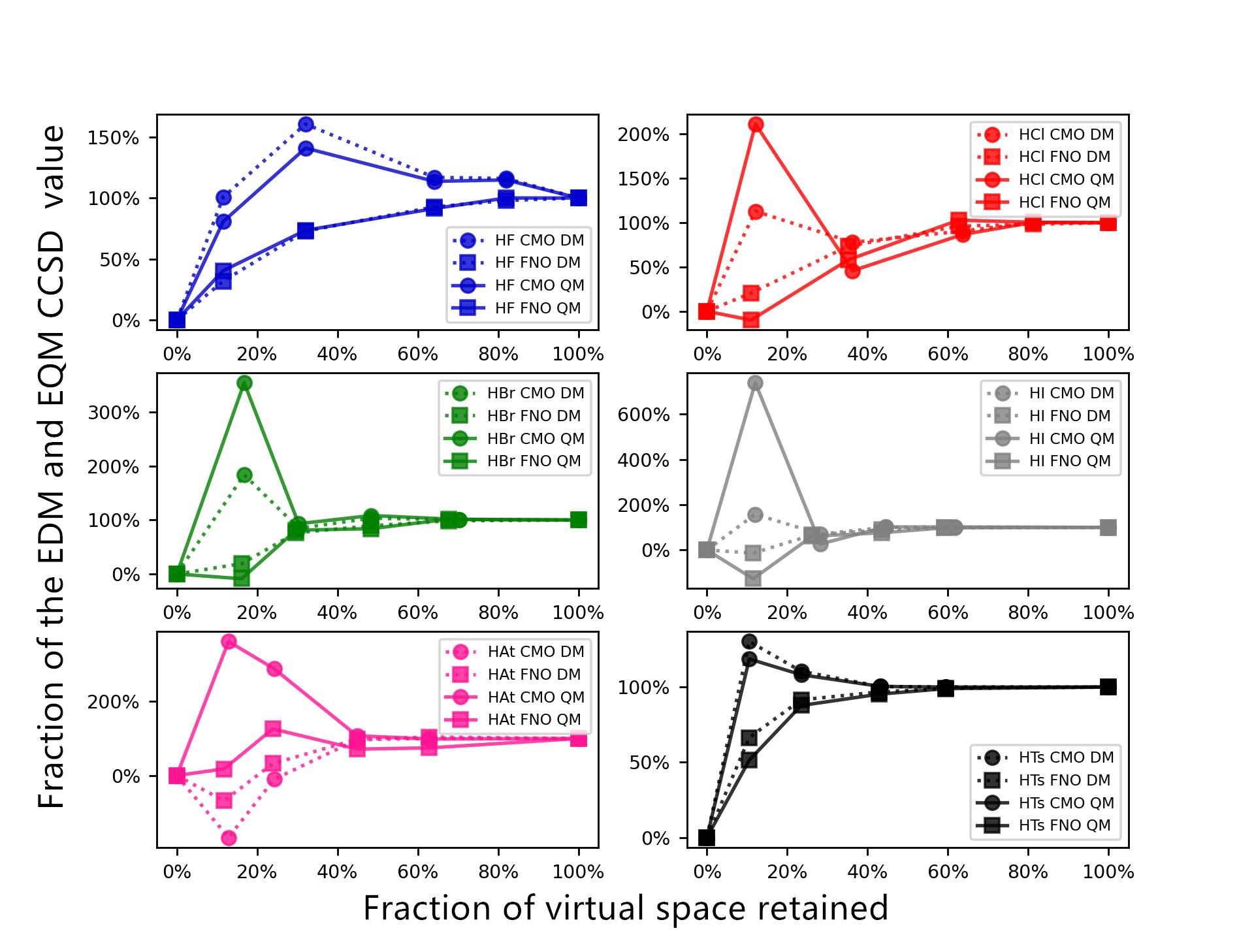}
                \caption{\label{fig:dipole-quadrupole-convergence} Convergence of the CCSD electric dipole and quadrupole moments with respect to the size of the virtual orbital space, for the X2C Hamiltonian. The X axis indicates the fraction of the virtual space retained, while the Y axis gives the fraction of the expectation values recovered with respect to the value obtained with the untruncated virtual space.}
            \end{figure*}

Figure~\ref{fig:dipole-quadrupole-convergence} shows how the EDM and EQM correlation correction varies with the truncation of the virtual orbital spaces, with  dotted and solid lines representing EDM and EQM, respectively. As before FNOs results are plotted with square markers. For these properties, the convergence is non-monotonic, unlike energy that we considered before. For both CMO and FNO truncation, the use of a high truncation value and consequently small virtual space (less than 30\% virtual orbitals) may lead to a strong overestimation or even a wrong sign of the correlation correction to the molecular property. Especially for CMO truncation, it is almost impossible to estimate the truncation error from a sequence of results for small virtual spaces. For FNO truncation oscillations are much less pronounced than for CMO, which indicates that also for these properties it is advantageous to use FNO truncation rather than CMO truncation. The similar performance for the correlation energy could thereby be used as a guideline. Taking the HF molecule as example, we note that with a threshold of $1.0 \times 10^{-4}$, the FNOs recover 72\% of the correlation contribution to EDM while recovering 85\% of the correlation energy. This indicates that the simplification of using a common cutoff for both properties could be a suitable strategy. In addition, we note that the performance of FNOs in EDM also applies to the heavier elements in the studied series. Even for HTs, setting the truncation threshold to $1.0 \times 10^{-4}$ (using only 23\% of the orbital space) already recovers 91\% of the correlation contribution to the EDM.
            
The EQM shows a similar behaviour as the EDM, albeit with stronger oscillations. The largest oscillations for the EQM occur at the same position as for the  EDM, but the amplitude thereof is much larger. This is again most pronounced in the CMO case. Taking HI as an example, retaining 12\% of the CMO virtual space overestimates the correlation contribution to EDM by 150\%, while the EQM contribution is even seven times too large. As for the EDMs, the oscillations resulting from FNO truncation are smaller and convergence is more smooth, which should make it possible to use a more aggressive cut-off strategy than is possible with CMOs.

\subsubsection{Electric field gradient} 

To also consider the effect of truncation on properties that probe the regions close to atomic nuclei we now turn to the electric field gradients (EFGs) at the halogen nuclei. EFGs couple with nuclear quadrupole moments~\cite{kello_determination_nodate} for nuclei with spin greater or equal to one and are important in the analysis of nuclear magnetic resonance (NMR) experiments~\cite{Autschbach2010}. For EFGs both (semicore) correlation and SOC effects are known to be of importance\cite{larsson1986relativistically} and they form therefore a good second test for the applicability of FNO truncation in relativistic calculations. The EFG is defined as the second derivative of the electric potential V(\textbf{R}) with respect to the nuclear position vector, taken at the nuclear position $R_{N}$
                \begin{equation}
                    q_{\mu\nu}(\textbf{R}_{N}) = -\frac{\partial V(\textbf{R})}{\partial R_{\mu} \partial R_{\nu}}|_{\textbf{R}=\textbf{R}_{N}}
                \end{equation}
By introducing the EFG tensor operator\cite{visscher_molecular_1998} 
\begin{equation}
\hat{q}_{\mu\nu}^{e}= \frac{3(\vec{r}-\vec{R}_{N})_{\mu}(\vec{r}-\vec{R}_{N})_{\nu}-|\vec{r}-\vec{R}_{N}|^{2}\delta_{\mu\nu}}{|\vec{r}-\vec{R}_{N}|^{5}}
\end{equation}
the EFG can be expressed as the expectation value of a one-body operator which makes evaluating its value similar to computing the EDM and EQM. 

\begin{figure*}[htb]
                \includegraphics[width=\linewidth]{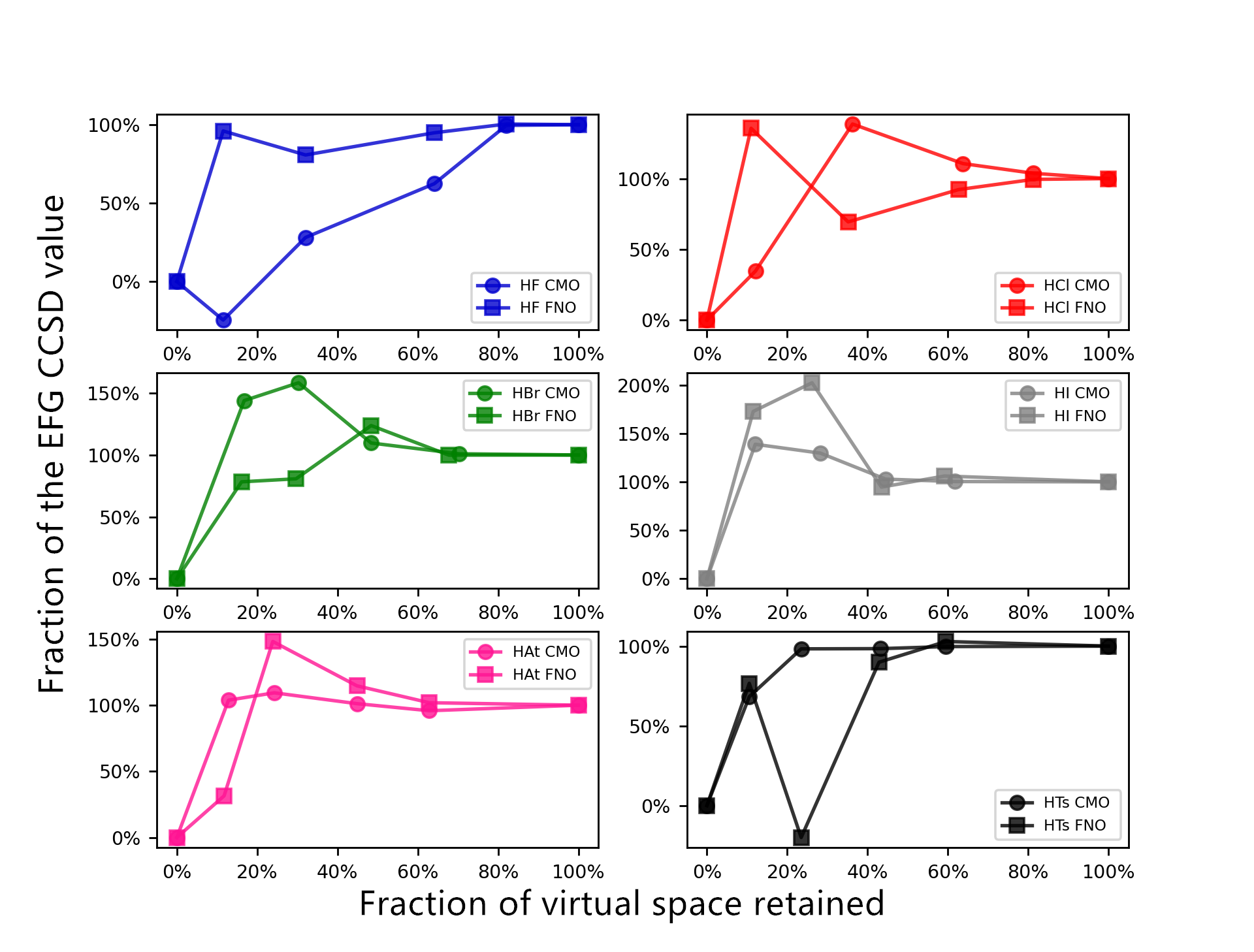}
                \caption{\label{fig:epsart5} Convergence of the CCSD electric field gradient at the halogen nucleus with respect to the size of the virtual orbital space, for the X2C Hamiltonian. The X axis indicates the fraction of the virtual space retained, while the Y axis gives the fraction of the expectation value recovered with respect to the value obtained with the untruncated virtual space.}
            \end{figure*}

For linear molecules it suffices to compute only the zz-component (with the z-axis chosen along the molecular bond) as the other non-zero parts of the tensor can then be determined by symmetry.  Figure~\ref{fig:epsart5} plots this zz component of the EFG and shows for both FNO and CMO oscillations upon truncating the orbital spaces. In this case FNO truncation can lead to similar and even more rapid oscillations than with CMO truncation. Recalling that the only difference between computing the EFG and EDM properties is in the operators used, it is of interest to consider their differences. The EFG tensor operator scales as $r^{-3}$ in contrast to the $r^1$ and $r^2$ scaling of the EDM and EQM operators. Density changes in the core region due to correlation are thus magnified by the EFG operator while they are hardly of influence for the EDM and EQM. Such changes may both come from core correlation as well as from correlating the valence electrons (through the tails of the valence orbitals in the core region). As core correlation may be harder to converge than valence correlation, we will separate the two effects by comparing all-electron and valence-only electrons calculations in the next subsection.

    \subsubsection{Convergence analysis for HCl and HI}
    
\begin{figure*}
        \centering
        
        \begin{minipage}[b]{.485\textwidth}
            \includegraphics[width=\textwidth,height=6.0cm]{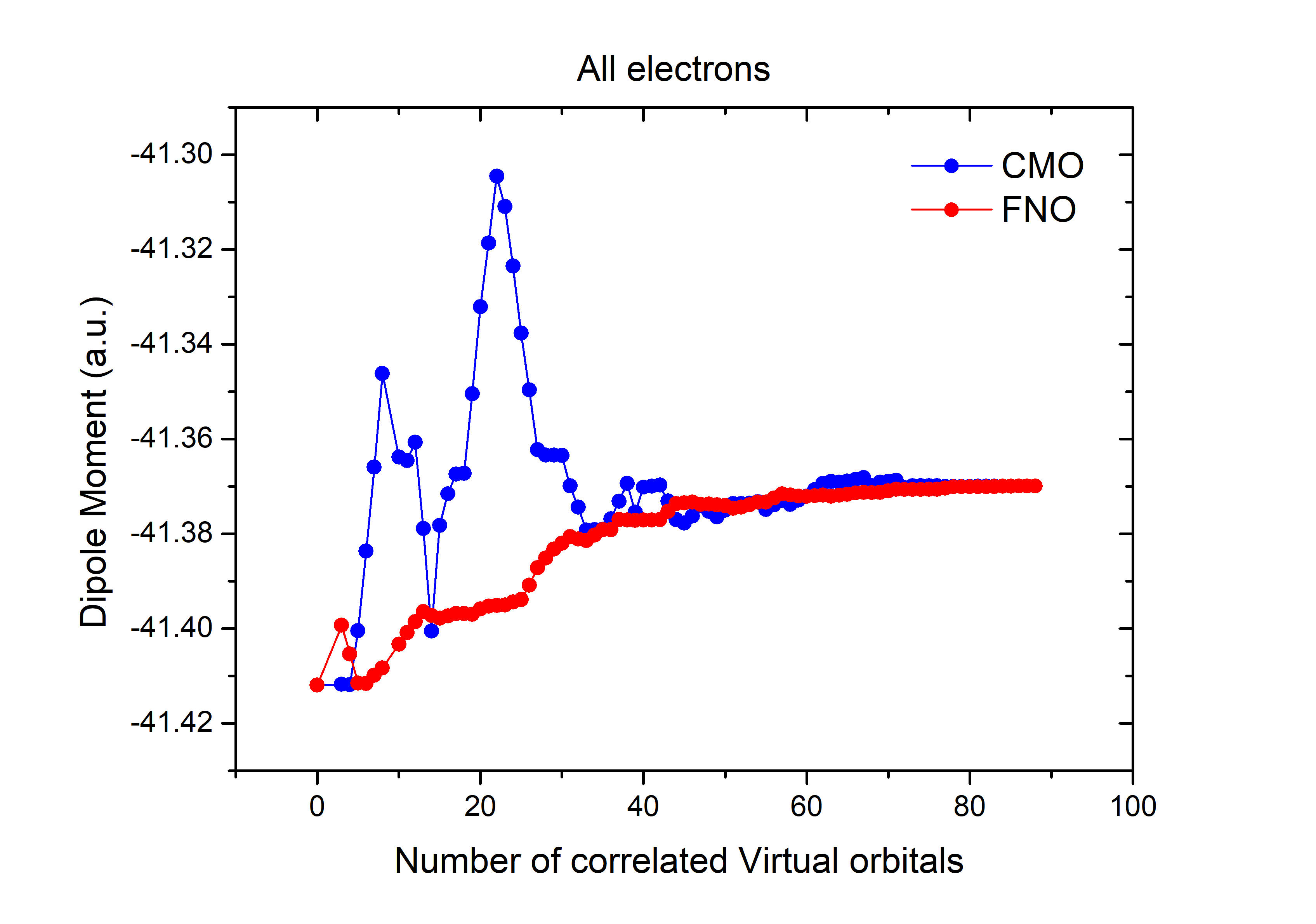}
        \end{minipage}
        \hfill
        \begin{minipage}[b]{.485\textwidth}
            \includegraphics[width=\textwidth,height=6.0cm]{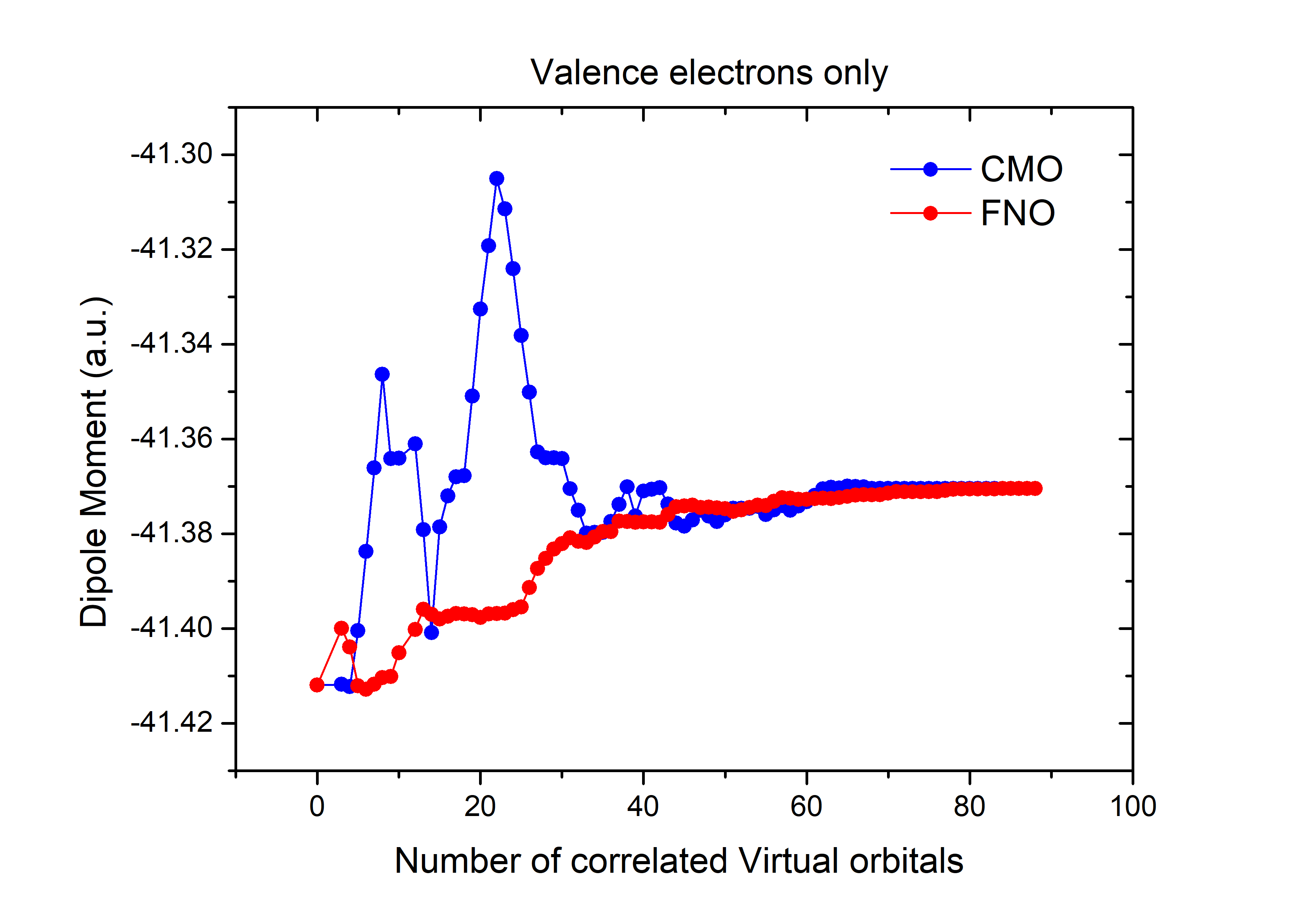}
        \end{minipage}
        \hfill
        \begin{minipage}[b]{.485\textwidth}
            \includegraphics[width=\textwidth,height=6.0cm]{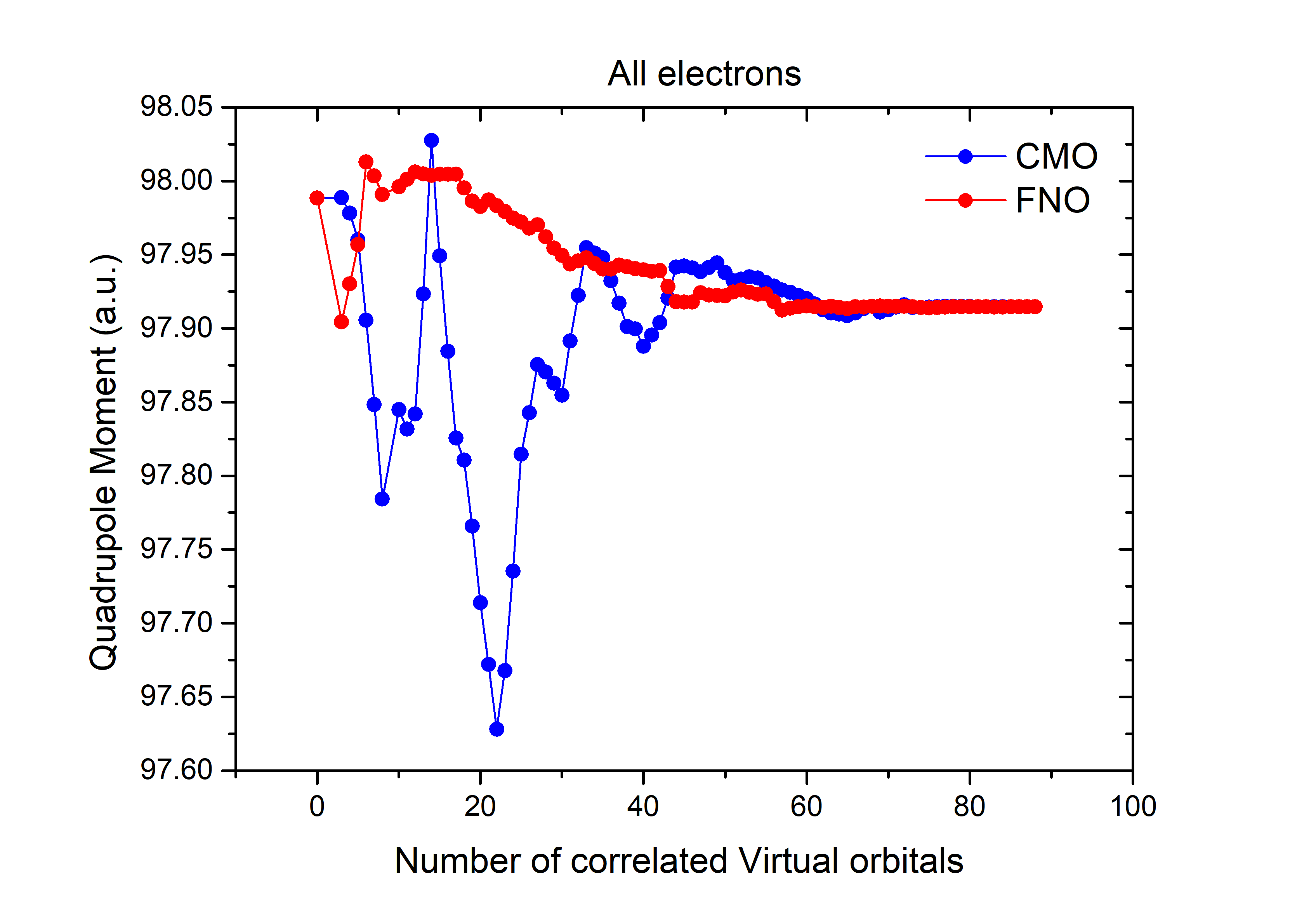}
        \end{minipage}
        \hfill
        \begin{minipage}[b]{.485\textwidth}
            \includegraphics[width=\textwidth,height=6.0cm]{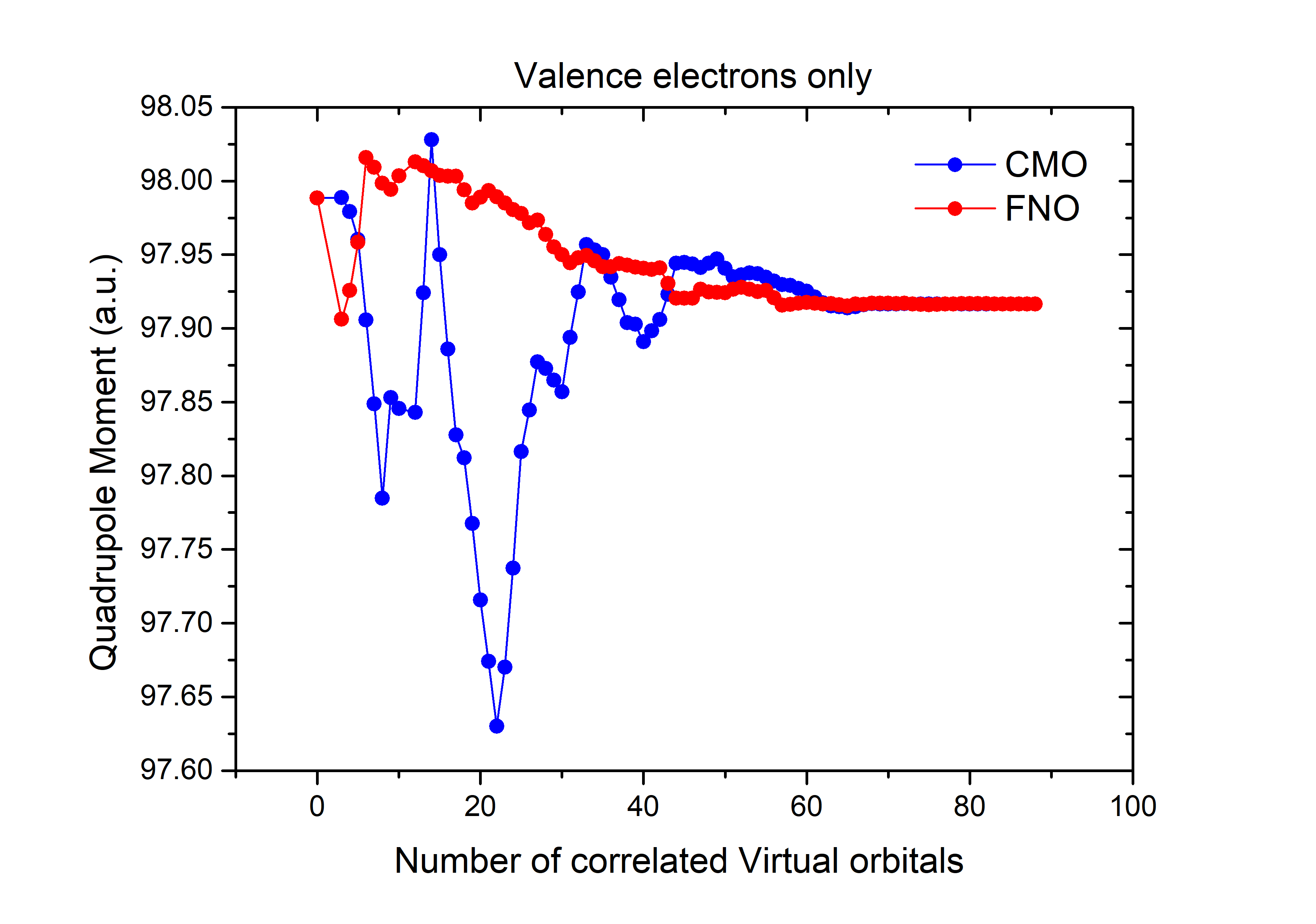}
        \end{minipage}
        \hfill
        \begin{minipage}[b]{.485\textwidth}
            \includegraphics[width=\textwidth,height=6.0cm]{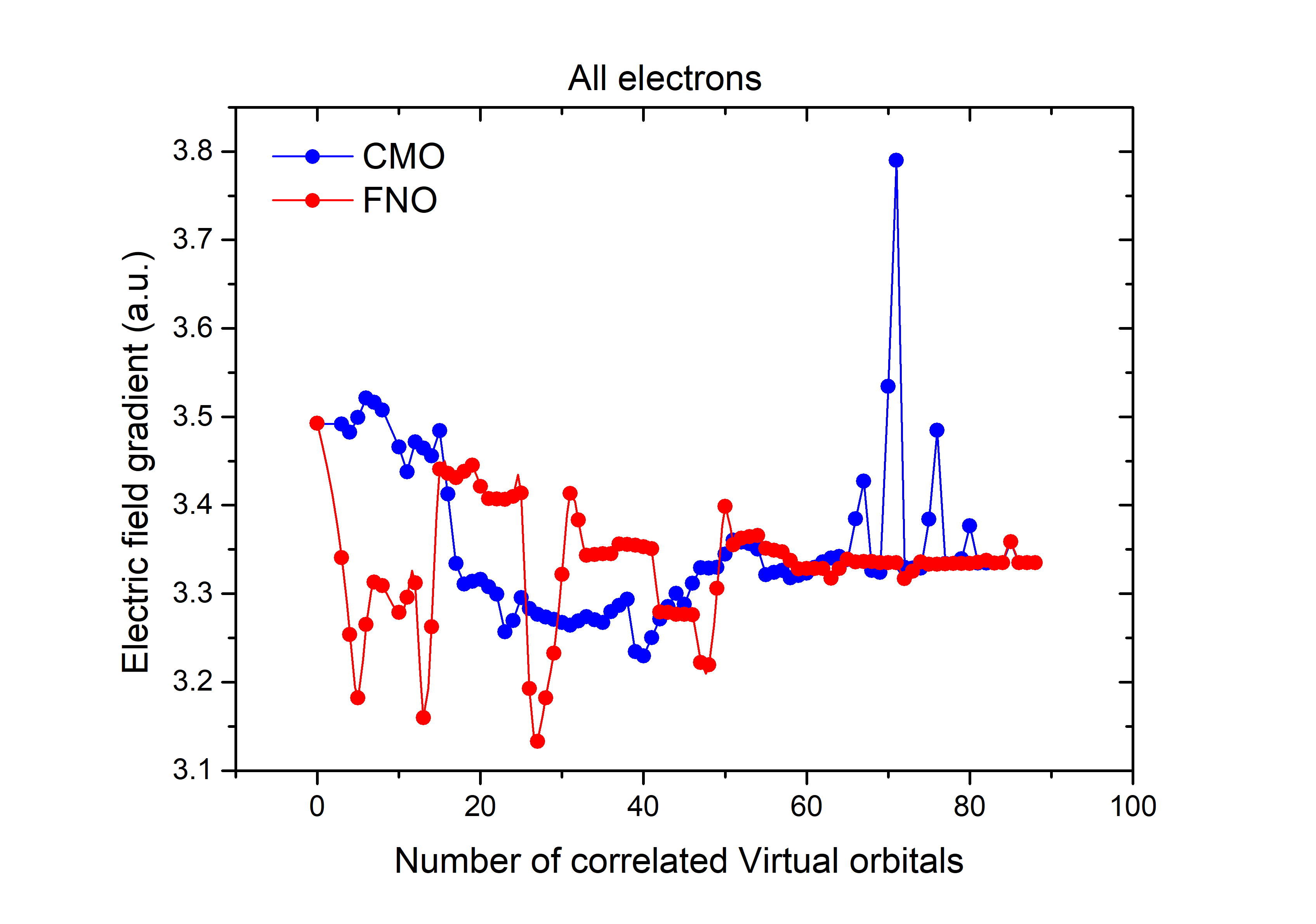}
        \end{minipage}
        \hfill
        \begin{minipage}[b]{.485\textwidth}
            \includegraphics[width=\textwidth,height=6.0cm]{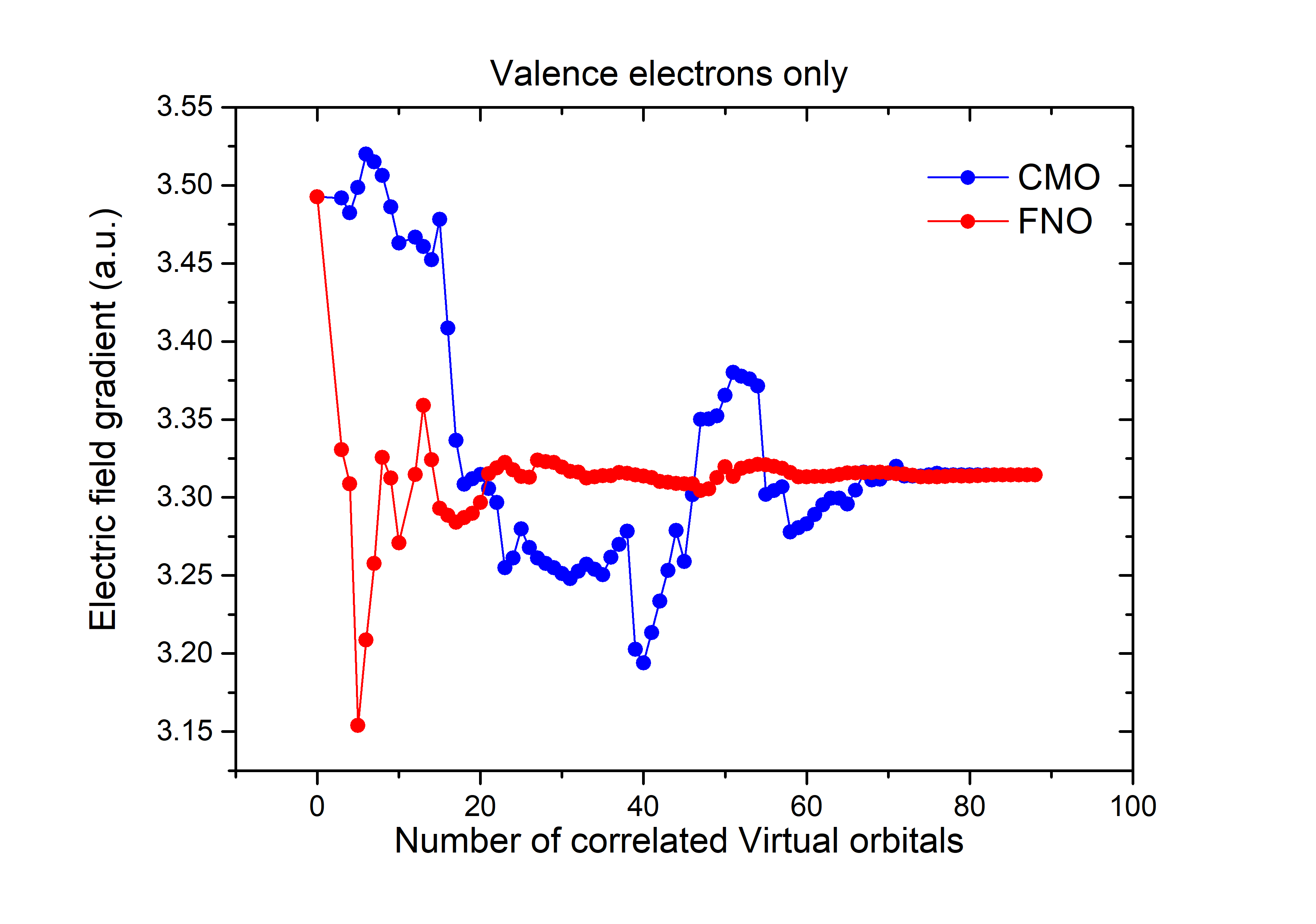}
        \end{minipage}
        \caption{\label{fig:epsart6} Effect of virtual space truncation on the expectation values (top: EDM; middle: EQM; bottom: EFG) of HCl, for the X2C Hamiltonian. The X axis gives the number of correlated virtual orbitals, up to and including the full virtual space. Figures to the left correspond to calculations correlating all electrons, and to the right correlating valence electrons only.}
    \end{figure*}

    The issue of oscillating convergence can be conveniently analyzed for the HCl molecule as chlorine is large enough to investigate the effect of core correlation, yet small enough to allow for quick calculations. Rather than taking only a few truncation values, we now systematically extend the size of the virtual space by adding individual FNOs and show the effect thereof on the EDM, EQM and EFG of HCl in Figure~\ref{fig:epsart6} for both all-electron and valence-only correlation calculations. 
    
    For the all-electron case, we see in the EDM and EQM plots for CMO truncation peaks when 13 and 23 orbitals are used. This is probably due to quasi-degeneracies, as the orbital energies of the 12th, 13th, 14th orbital are 0.42418, 0.43657 and 0.43673 Hartree, respectively. This set of virtual orbitals is primarily a diffuse chlorine p-type orbital shell that is split due to the formation of the hydrogen-chlorine bond as well as due to spin-orbit coupling. Taking only one of the two almost degenerate $\pi$-type orbitals in the correlation space, gives an unbalanced description and impacts these valence properties significantly. For the EFG this particular set of three CMOs is less important because diffuse orbitals only contribute indirectly to this property. For the EFG one may notice (Figure~\ref{fig:epsart6}, lower left panel, blue line) a large oscillation around virtual orbital number 70 in the all-electron calculation. This is due to the more core-like p-orbitals 70, 71, and 72 with energies of respectively 35.63, 35.73 and 36.10 Hartree, that are impacting the correlation of core electrons. Restricting the correlation treatment to valence electrons only strongly reduces the effect of these virtuals (Figure~\ref{fig:epsart6}, lower right panel).
    
    The FNO curves for the EDM and EQM show only an initial oscillation that is followed by rather smooth convergence. This initial oscillation can also be traced to quasi-degenerate orbitals (but now in terms of occupation numbers). For the FNOs, the occupation number of the 2nd, 3rd, 4th, 5th orbital are 0.008692, 0.008676, 0.008508, 0.008491, respectively, and like with near-degenerate CMOs it appears recommendable to either include all or none of a degenerate set. For the EFG, the situtation is unfortunately more complicated, also with FNOs. 
    
    In the all-electron calculation values only stabilize after inclusion of about 50 orbitals, while in the valence-only calculation stable convergence is reached after addition of about 20 FNOs. Compared to the CMO truncation scheme, the advantage of FNO truncation appears to be absence of "late" oscillations (at low threshold values) that could cause artefacts in the CMO truncation schemes. Such oscillations are typical for EFGs, for which an indivdual orbital may provide a significant contribution but where the contribution of full, spherically symmetric shells of orbitals add up to zero due to symmetry. This is well-known for CMOs, but also holds for FNOs. An example are the EFG integrals of the 13th, 14th, 15th FNOs which evaluate to -78, 44 and 26 au. These large values get multiplied by very similar occupation numbers (resp. 7.9 10$^{-4}$, 7.0 10$^{-4}$, 7.0\ 10$^{-4}$) so that their contributions to the total EFG nearly cancel.
    
    A similar argument can be used to rationalize the difference between the valence and all-electron calculations. In the correlation of core orbitals one effectively changes the relative occupation of the three $2p$ orbitals in the chlorine core by making their occupations slightly smaller than 2. Imbalances in the correlation get thereby for the EFG multiplied by large integral values (about 80 au for the $2p_{3/2}$ orbitals) which creates the large oscillation seen in these calculations. For EDMs and EQMs the integral values are much more alike and smaller in magnitude, leading to the observed smoother convergence.

\begin{figure*}
        \centering
        
        \begin{minipage}[b]{.485\textwidth}
            \includegraphics[width=\textwidth,height=6.0cm]{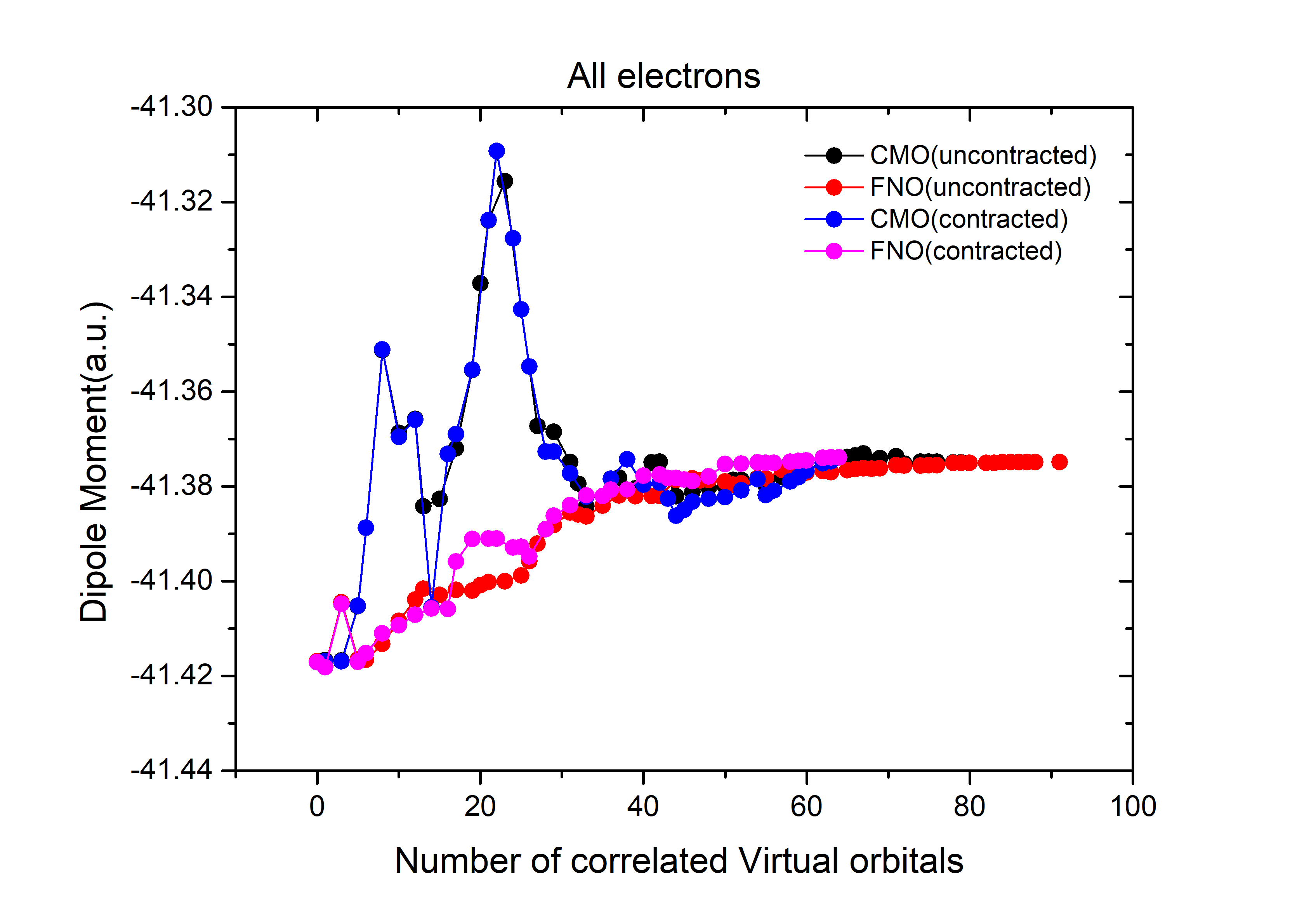}
        \end{minipage}
        \hfill
        \begin{minipage}[b]{.485\textwidth}
            \includegraphics[width=\textwidth,height=6.0cm]{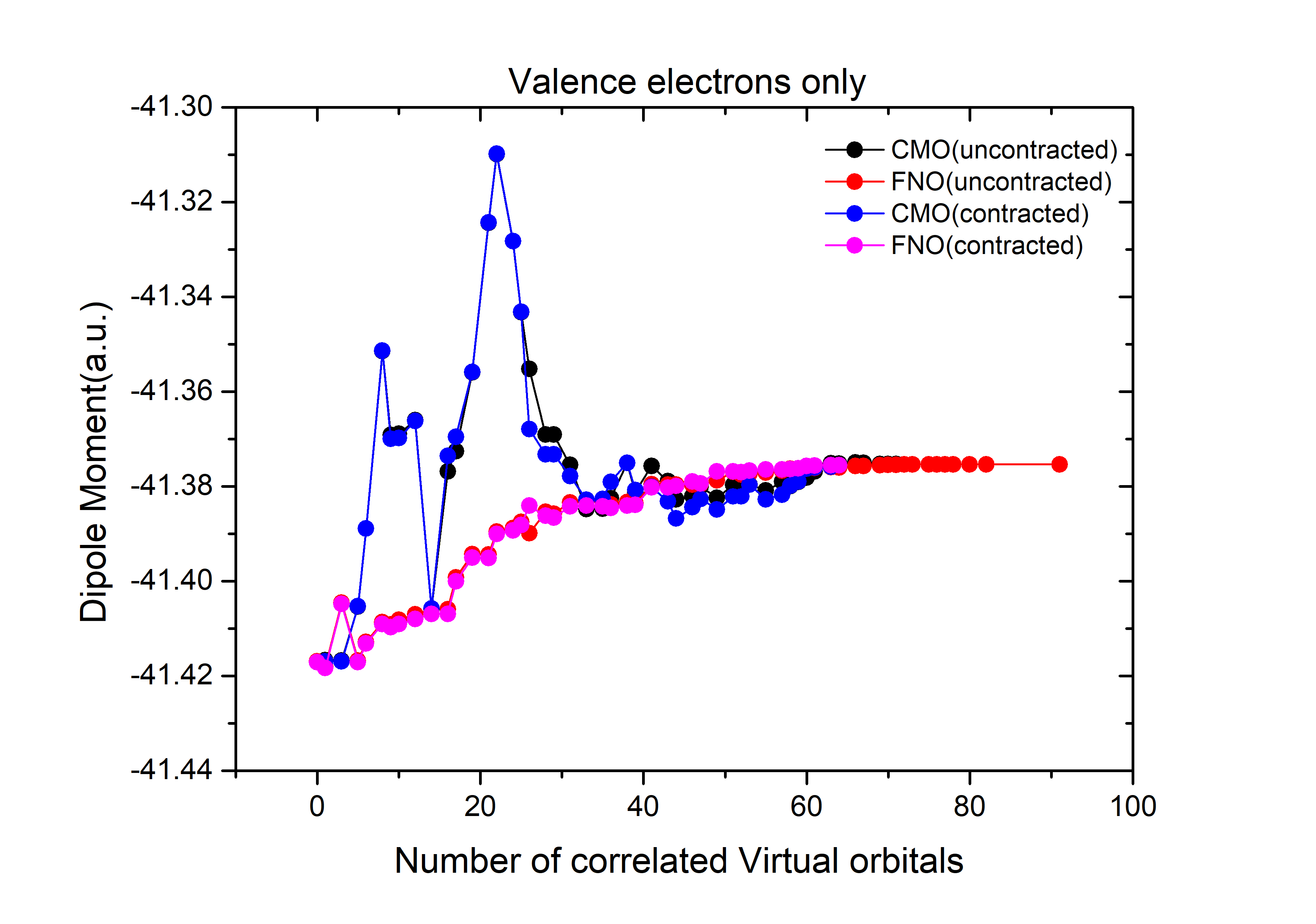}
        \end{minipage}
        \hfill
        \begin{minipage}[b]{.485\textwidth}
            \includegraphics[width=\textwidth,height=6.0cm]{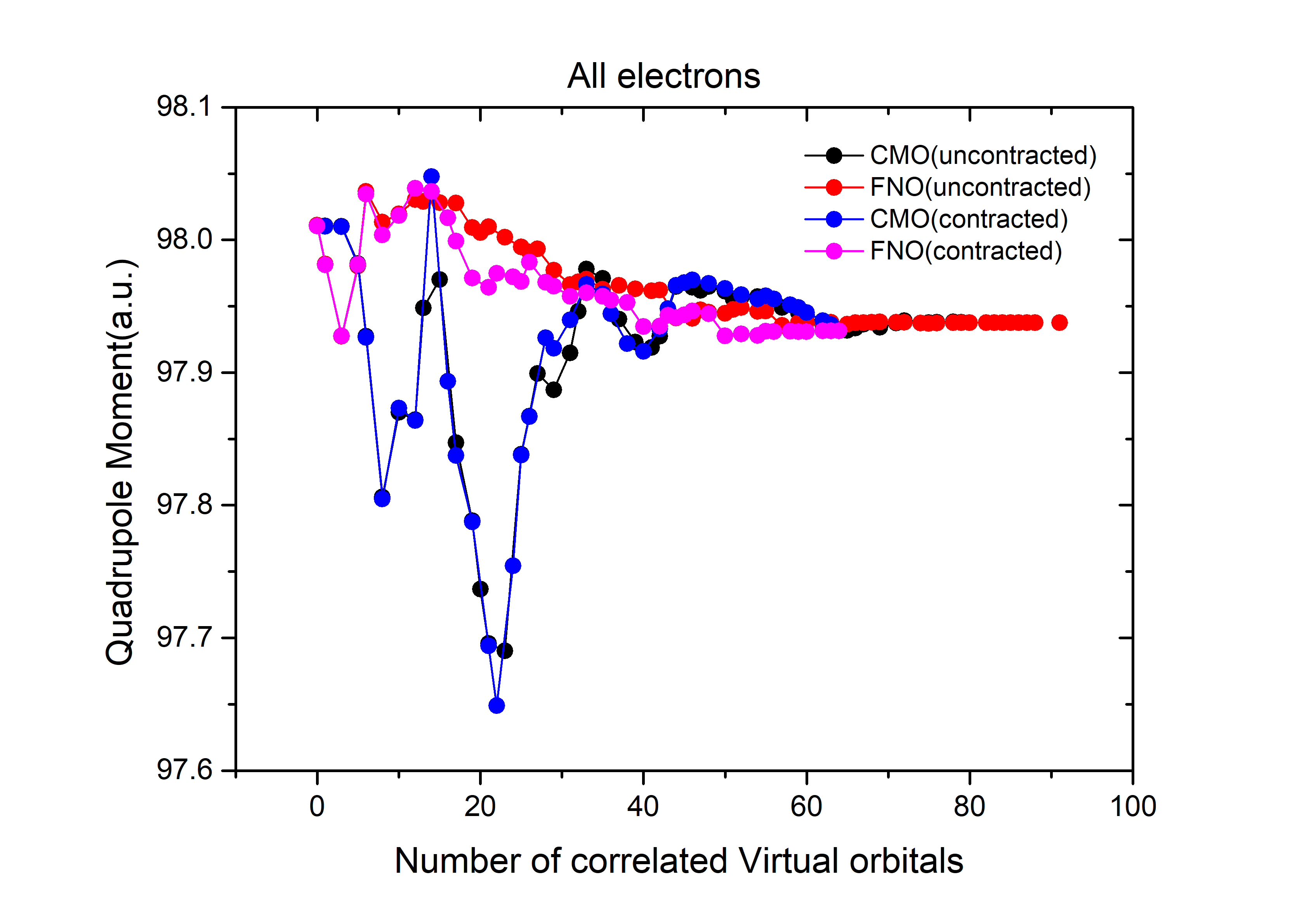}
        \end{minipage}
        \hfill
        \begin{minipage}[b]{.485\textwidth}
            \includegraphics[width=\textwidth,height=6.0cm]{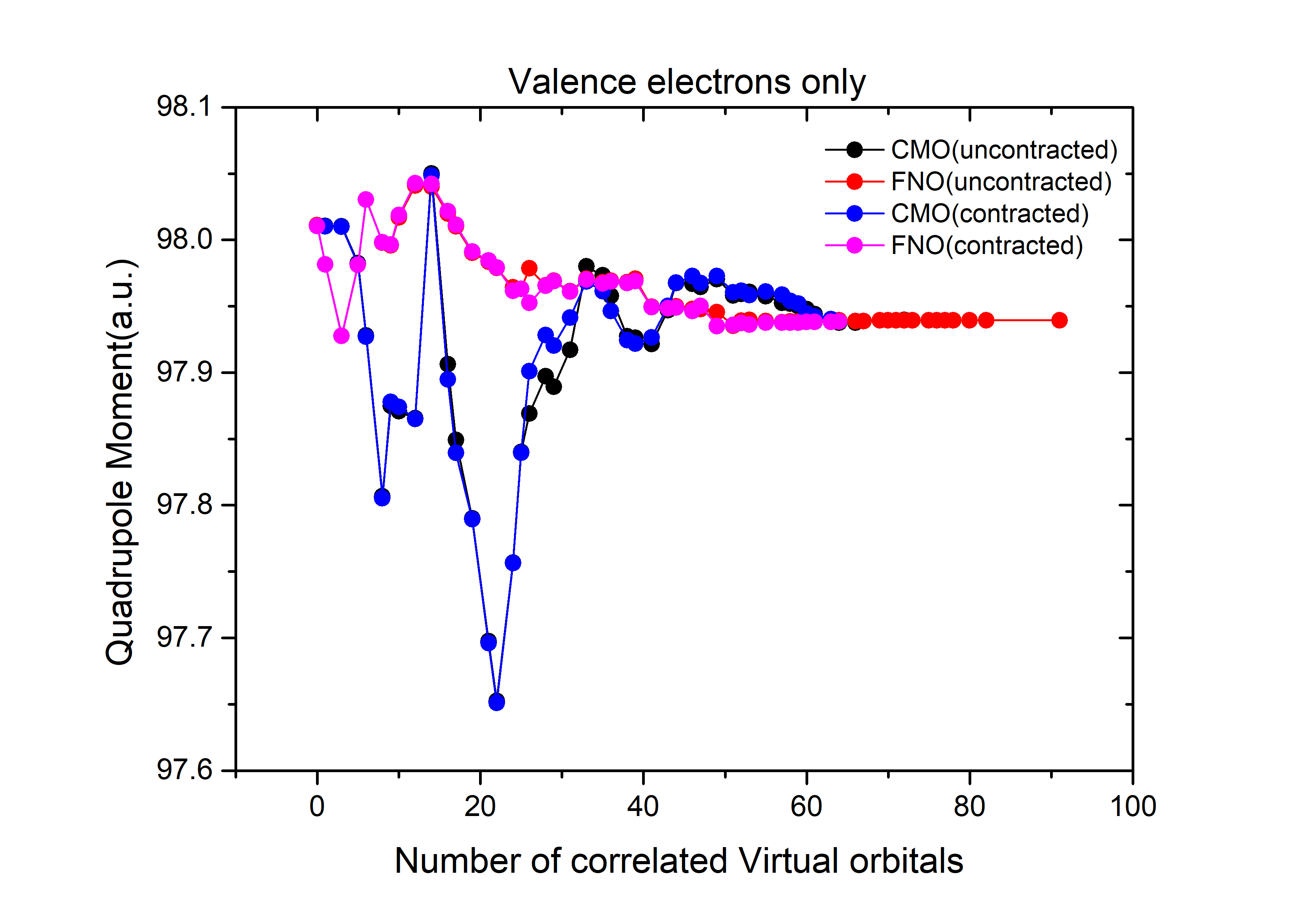}
        \end{minipage}
        \hfill
        \begin{minipage}[b]{.485\textwidth}
            \includegraphics[width=\textwidth,height=6.0cm]{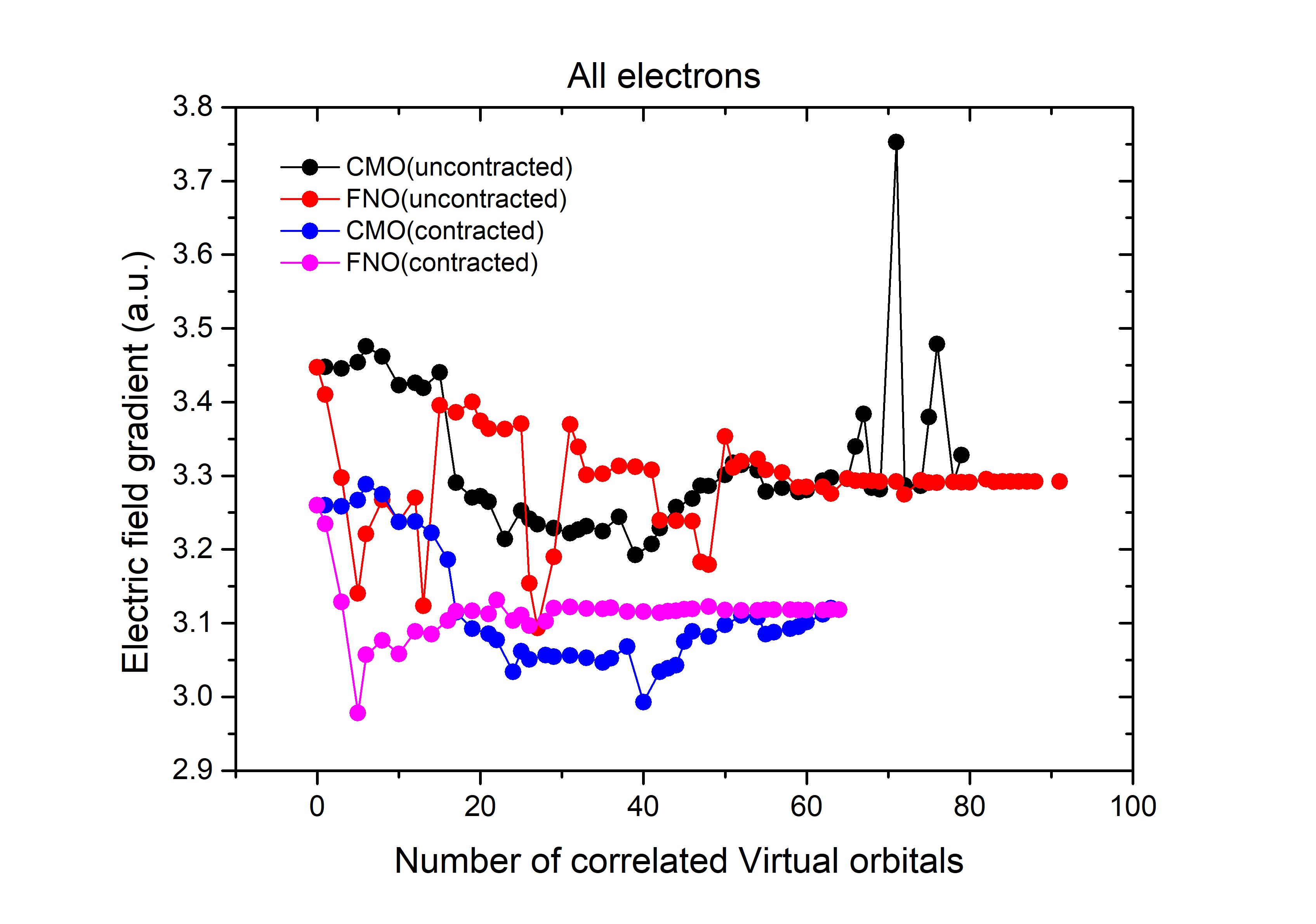}
        \end{minipage}
        \hfill
        \begin{minipage}[b]{.485\textwidth}
            \includegraphics[width=\textwidth,height=6.0cm]{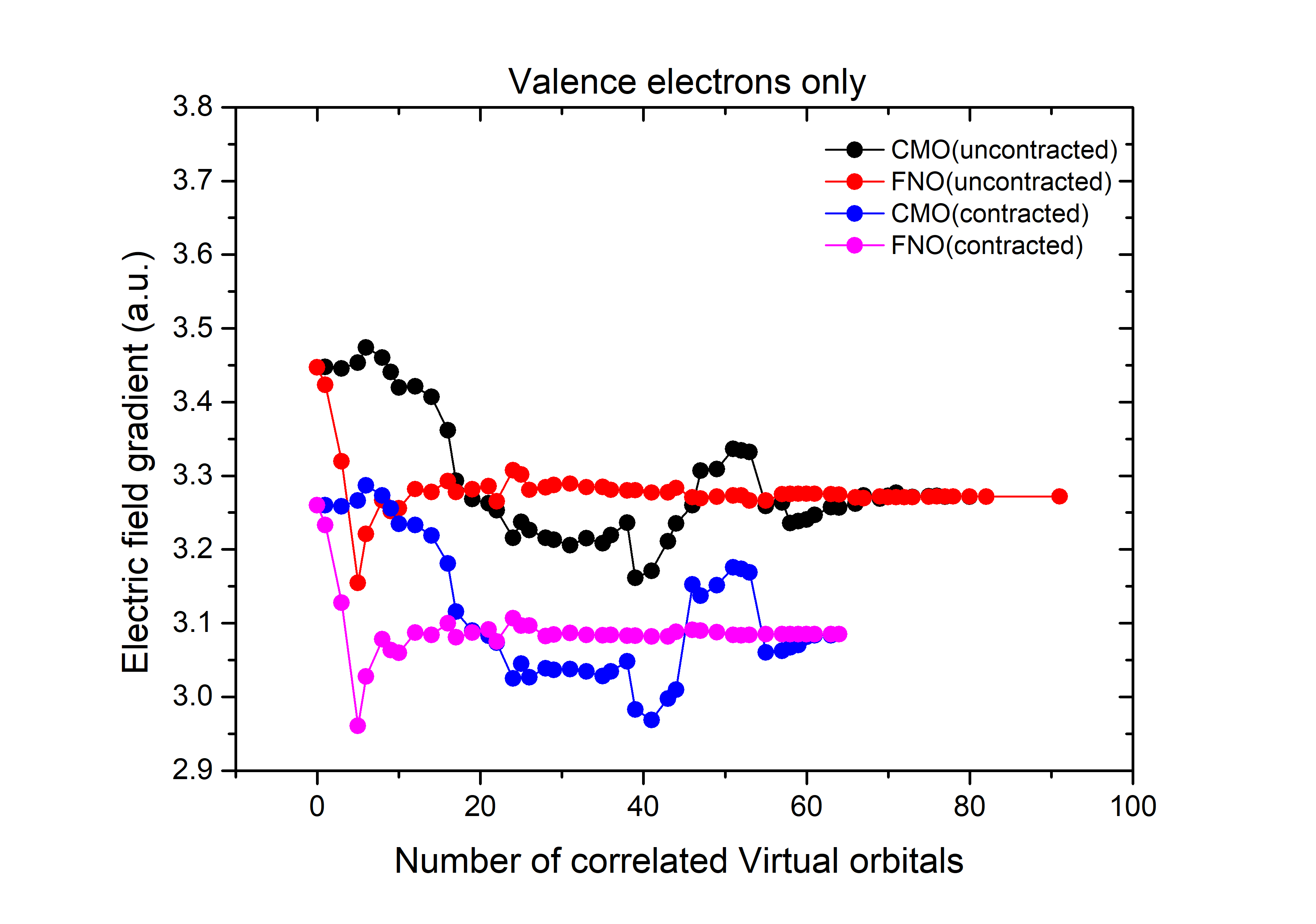}
        \end{minipage}
        \caption{\label{fig:epsart7} Effect of virtual space truncation on the expectation values (top: EDM; middle: EQM; bottom: EFG) of HCl, for the non-relativistic Hamiltonian and employing contracted and uncontracted Dunning basis sets. The X axis gives the number of correlated virtual orbitals, up to and including the full virtual space. Figures to the left correspond to calculations correlating all electrons, and to the right correlating valence electrons only.}
    \end{figure*}

    To check if such oscillation problems are particular for the relativistic domain, or can also be found in the non-relativistic case, we have carried out the same analysis as above by employing a non-relativistic Hamiltonian, employing both contracted and uncontracted (valence) basis sets and correlating both valence and all electrons. The results are shown in Fig.~\ref{fig:epsart7}.
    
    From these, it can be seen that for the electric dipole and quadrupole moments, the non-relativistic calculations basically show the same oscillations as for X2C, and results for contracted and uncontracted basis sets are nearly indistinguishable from each other. For the EFG there is also no noticeable difference for the convergence patterns between X2C and non-relativistic results for uncontracted basis sets.
    
    On the other hand, we observe significant differences between EFG calculations employing contracted and uncontracted basis sets. The first one is that, already at the Hartree-Fock level, there is a significant difference (nearly 10\%) on the absolute value of the EFG, with contracted values underestimating uncontracted ones. Second, we note that with contracted basis sets, there are very few oscillations in FNO EFG values (even when correlating all electrons), and FNO results are already quite stable for a much lower number of virtuals than CMO results (CMO EFG values are still not completely converged at nearly full virtual spaces for the contracted sets). 
    
    This can be understood as a manifestation of the degree of atomic symmetry that is imposed by the contraction for the different orbital shells, particularly for the occupied core orbitals. In the contracted sets these are forced to maintain their atomic-like nature during the molecular calculation because electron correlation cannot as easily deform the orbitals as is possible when the basis set is uncontracted. This is underscored by the nearly identical behavior of the contracted all electron and valence calculations, as in the latter core orbitals are kept frozen.

\begin{figure*}
        \centering
        
        \begin{minipage}[b]{.485\textwidth}
            \includegraphics[width=\textwidth,height=6.0cm]{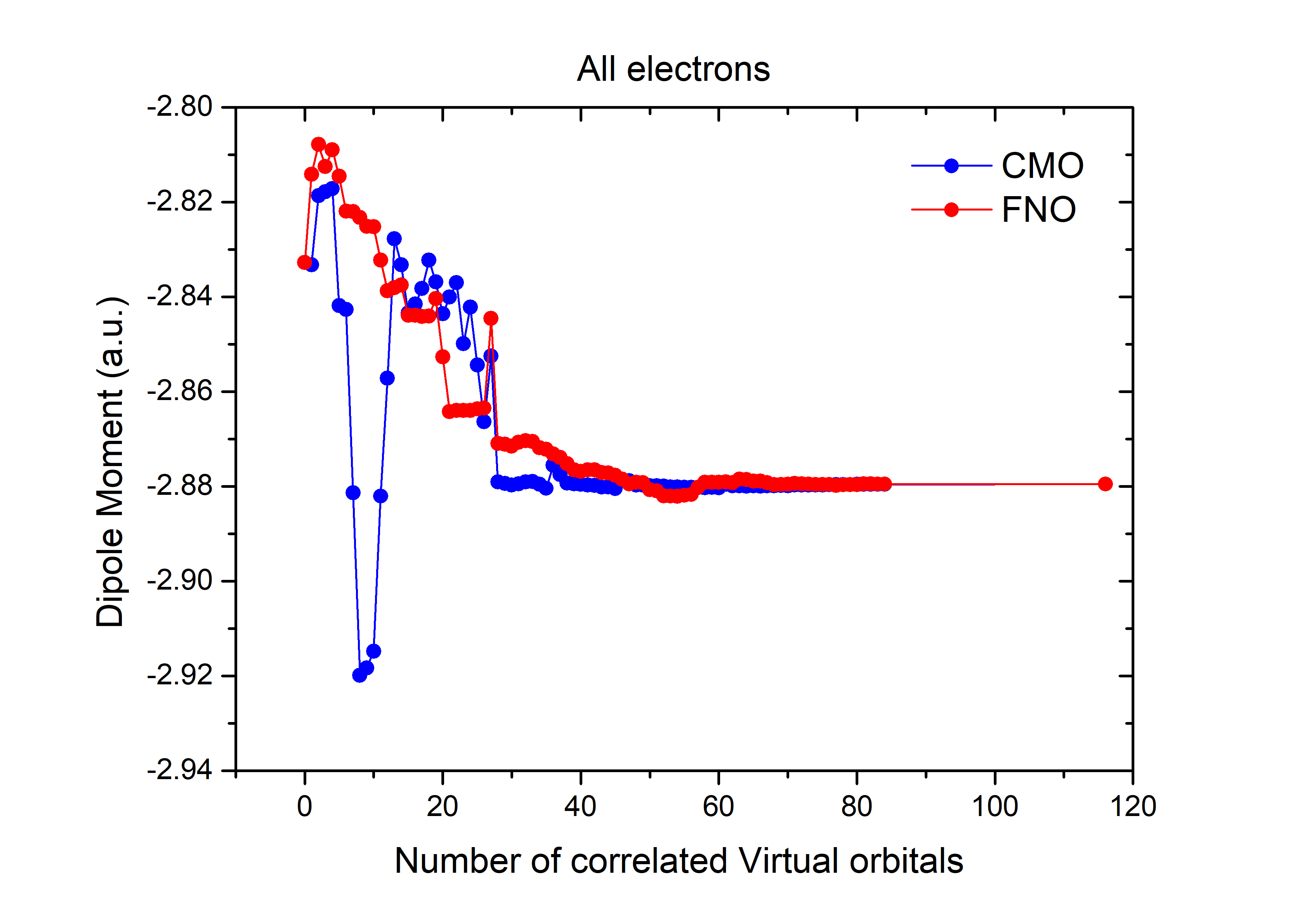}
        \end{minipage}
        \hfill
        \begin{minipage}[b]{.485\textwidth}
            \includegraphics[width=\textwidth,height=6.0cm]{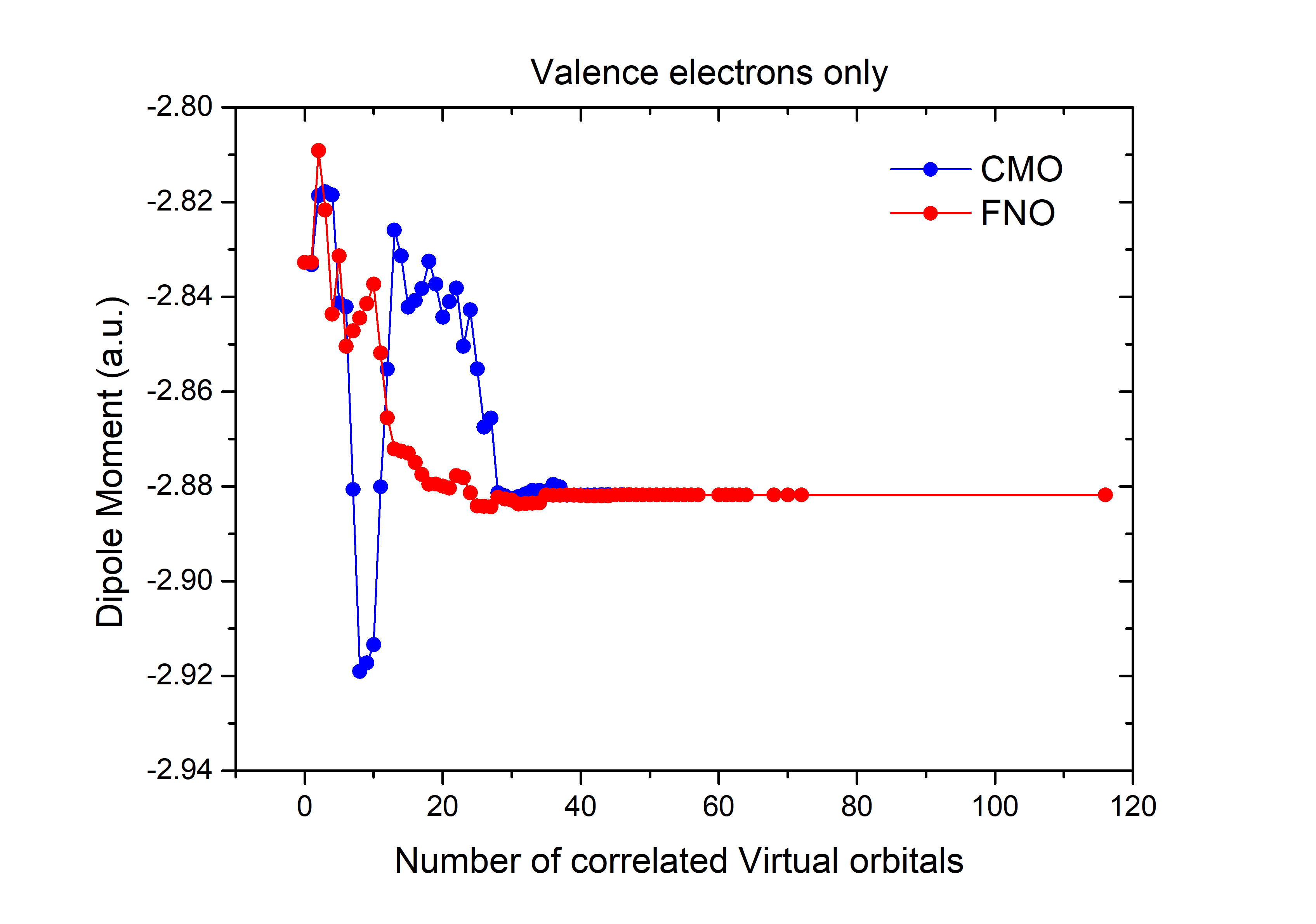}
        \end{minipage}
        \hfill
        \begin{minipage}[b]{.485\textwidth}
            \includegraphics[width=\textwidth,height=6.0cm]{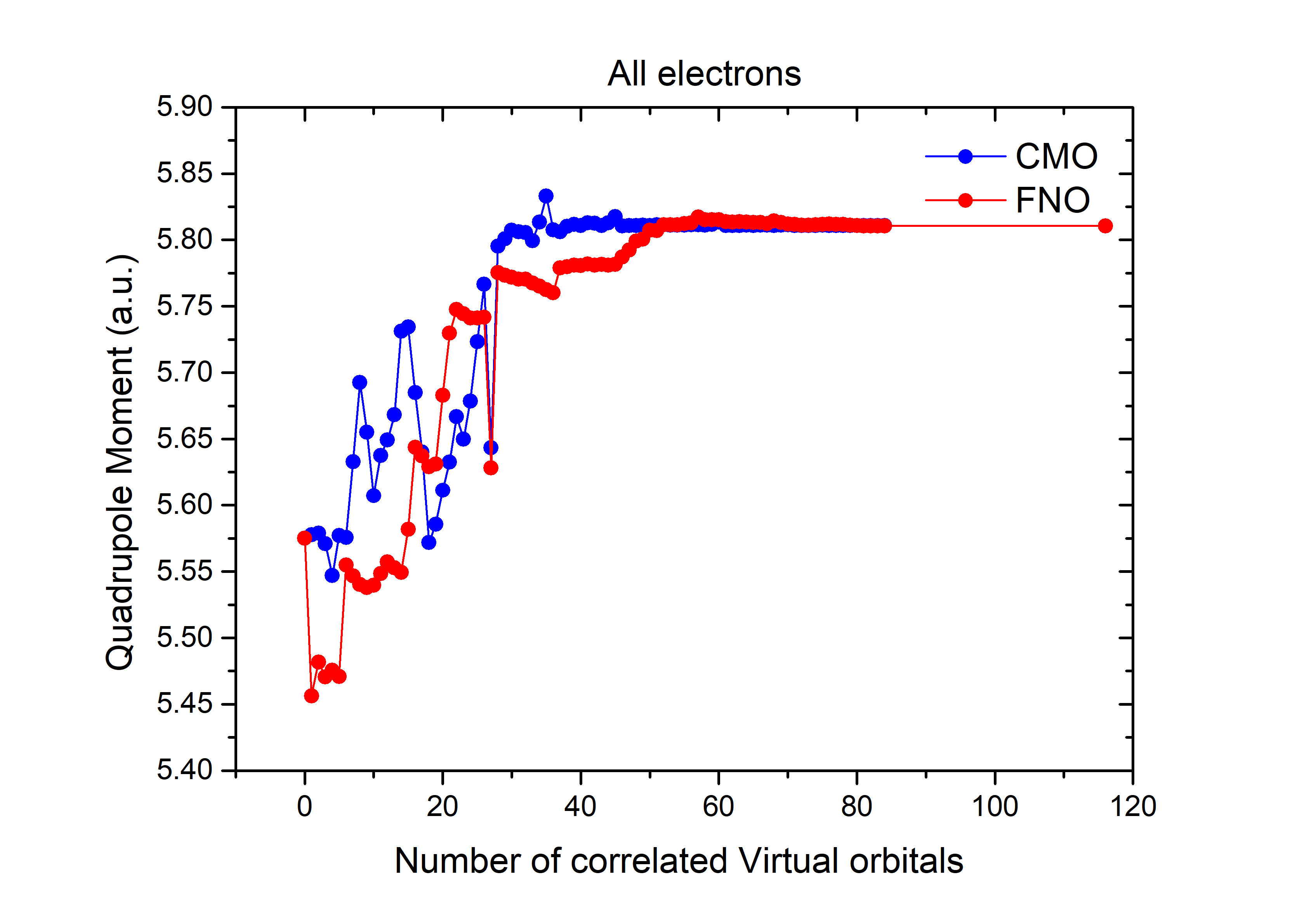}
        \end{minipage}
        \hfill
        \begin{minipage}[b]{.485\textwidth}
            \includegraphics[width=\textwidth,height=6.0cm]{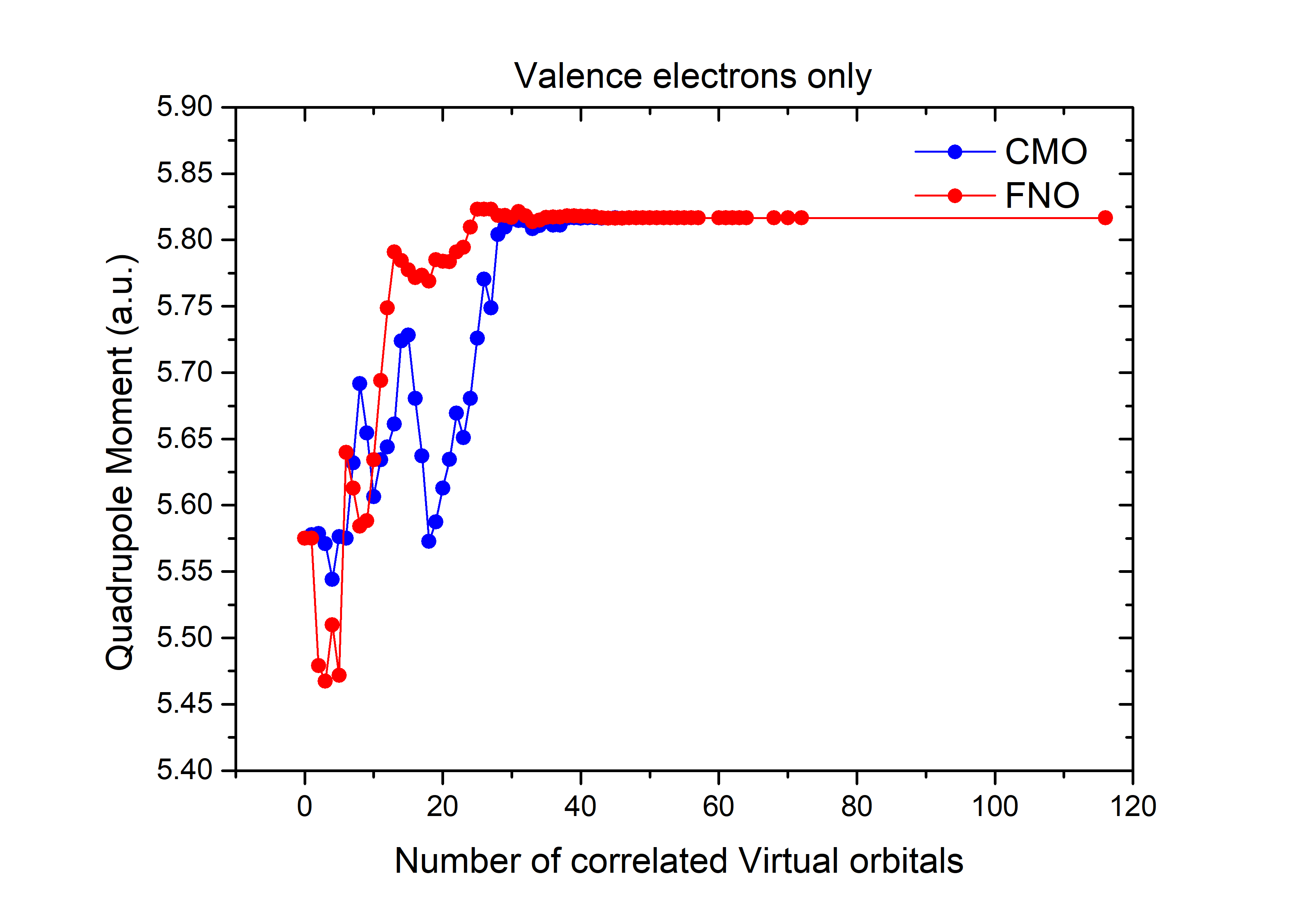}
        \end{minipage}
        \hfill
        \begin{minipage}[b]{.485\textwidth}
            \includegraphics[width=\textwidth,height=6.0cm]{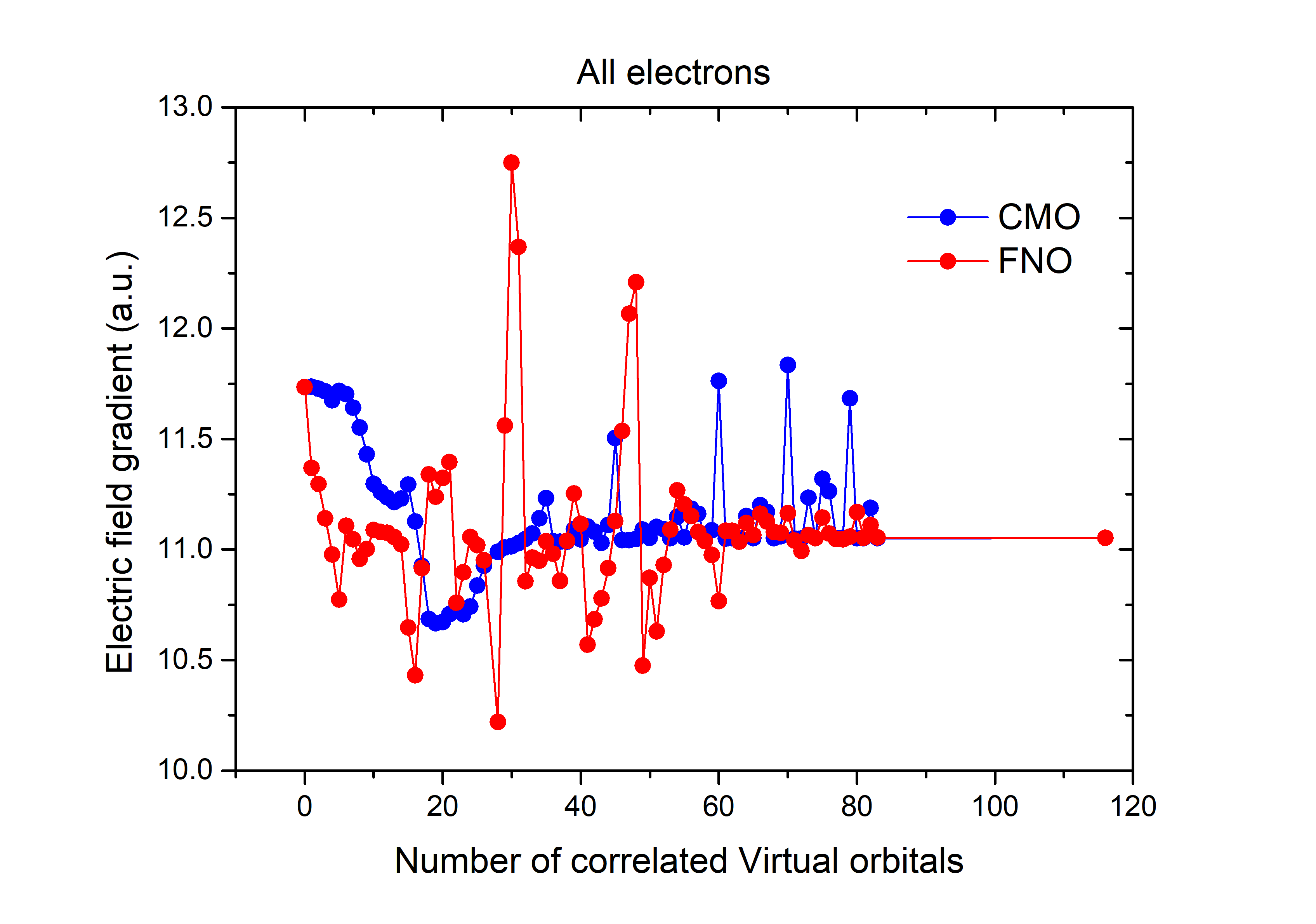}
        \end{minipage}
        \hfill
        \begin{minipage}[b]{.485\textwidth}
            \includegraphics[width=\textwidth,height=6.0cm]{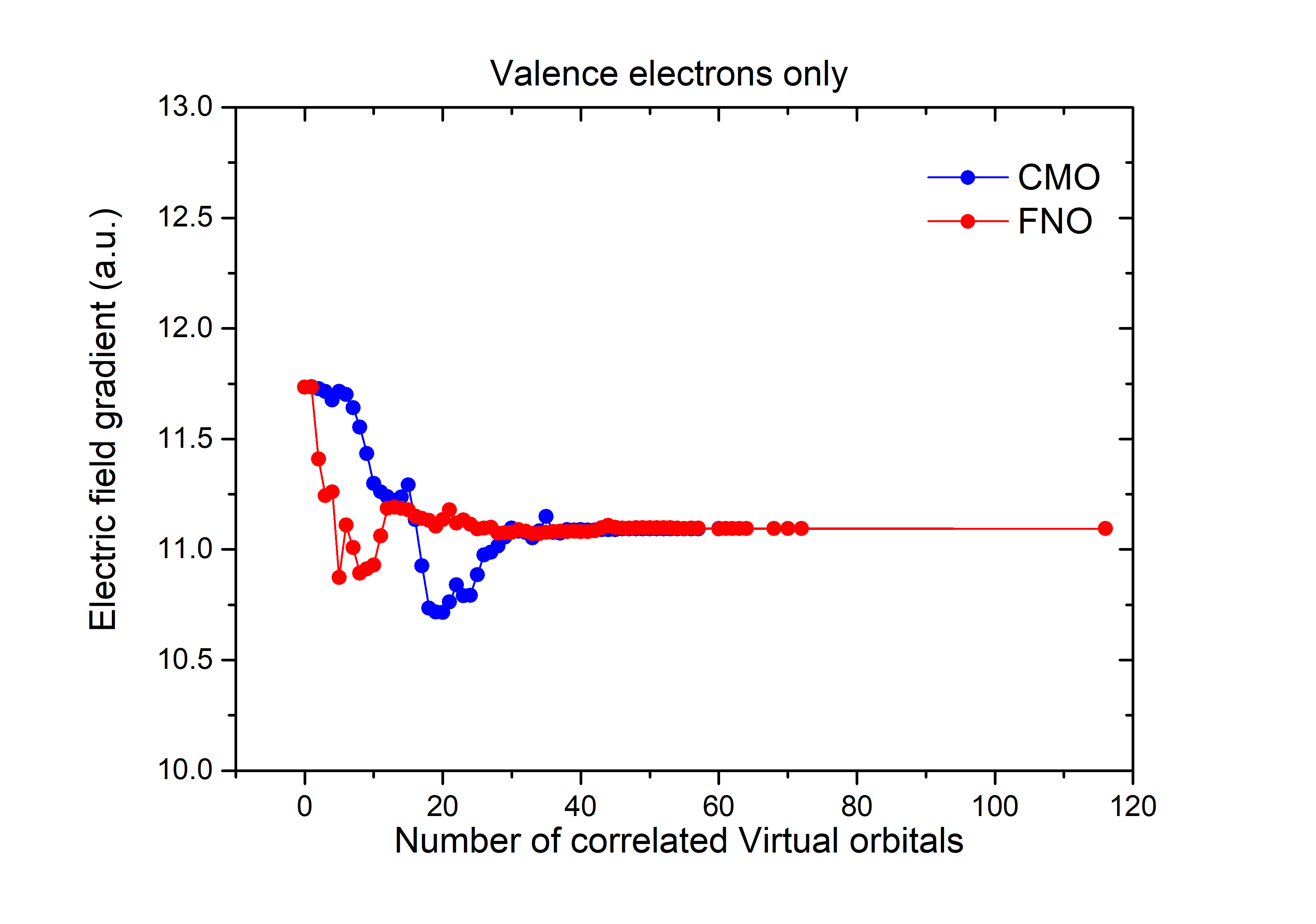}
        \end{minipage}
        \caption{\label{fig:epsart8} Effect of virtual space truncation on the expectation values (top: EDM; middle: EQM; bottom: EFG) of HI, for the X2C Hamiltonian. The X axis gives the number of correlated virtual orbitals, up to and including the full virtual space. Figures to the left correspond to calculations correlating all electrons, and to the right correlating valence electrons only.}
    \end{figure*}

    To check if the convergence behavior mentioned above is also found for a heavier system, we carried out similar calculations for HI, but using an uncontracted double zeta basis set. It can be seen from Fig \ref{fig:epsart8} that although HI has many more electrons, it still shows a very similar convergence as observed for HCl. For instance, to get a converged EFG value of HI, in the valence only computation, one just needs to correlate 20 FNOs. 
    
    Looking at these tests, we may conclude that for properties like EDM and EQM, which are dominated by valence electron contributions, one can safely use FNOs and their occupation numbers to cut off the orbital space, both in valence-only and all-electron correlation calculations. For sensitive properties for which core electrons may give large and nearly cancelling contributions, it is also with FNOs numerically more stable to correlate only the valence electrons.
    
\subsection{Parity violation}
Detection of parity violation (PV) effects, associated with weak force in atoms and molecules is an active field of research\cite{bast_analysis_2011,rauhut_parity-violation_2021,sunaga_towards_2021}. While this property can also be computed with perturbation theory starting from nonrelativistic theory\cite{BergerPV2000}, it is advantageous to use a  relativistic quantum chemistry framework because the PV energy can then be formulated as an expectation value of an effective one-body operator\cite{laerdahl_fully_1999}:
             \begin{equation}
                 E_{PV}=\sum_{A}\bra{\Psi}\hat{H}^{A}_{PV}\ket{\Psi}
             \end{equation}
 with           
             \begin{equation}
                 \hat{H}^{A}_{PV}=\frac{G_{F}}{2 \sqrt{2}}Q_{W}^{A}\sum_{i} \gamma_{i}^{5}\rho^{A}(r_{i})
             \end{equation}
and $G_{F}$=1.16637$\times$10$^{-11}$MeV$^{-2}$ being the Fermi coupling constant. The A and i label nuclei and electrons, respectively. The weak charge $Q_{W}^{A}=-N_{A}+Z_{A}(1-4sin^{2}\theta_{\omega})$, where $N_{A}$ and Z$_{A}$ is the number of neutrons and protons in each nucleus. $\theta_{\omega}$ is Weinberg mixing angle, that is set to 0.2319 for $sin^{2}\theta_{\omega}$. $\rho^{A}$ and $\gamma_{i}^{5}$ are normalized nucleon density and 4-dimensional chirality operator, respectively. 
             \begin{equation}
                  \gamma^{5}=\begin{pmatrix}
             \mathbb{O} &\mathbb{I} \\
             \mathbb{I} & \mathbb{O} 
           \end{pmatrix}
             \end{equation}
        
Sunaga et al.\cite{sunaga_towards_2021} already investigated use of CMO truncation at CCSD level with two different threshold values and showed that truncation is well possible for this property. In current work, we also employ FNO truncation and test more thresholds for CMO truncation and its effect on the PV value.

 \begin{figure*}
 \includegraphics[width=\linewidth]{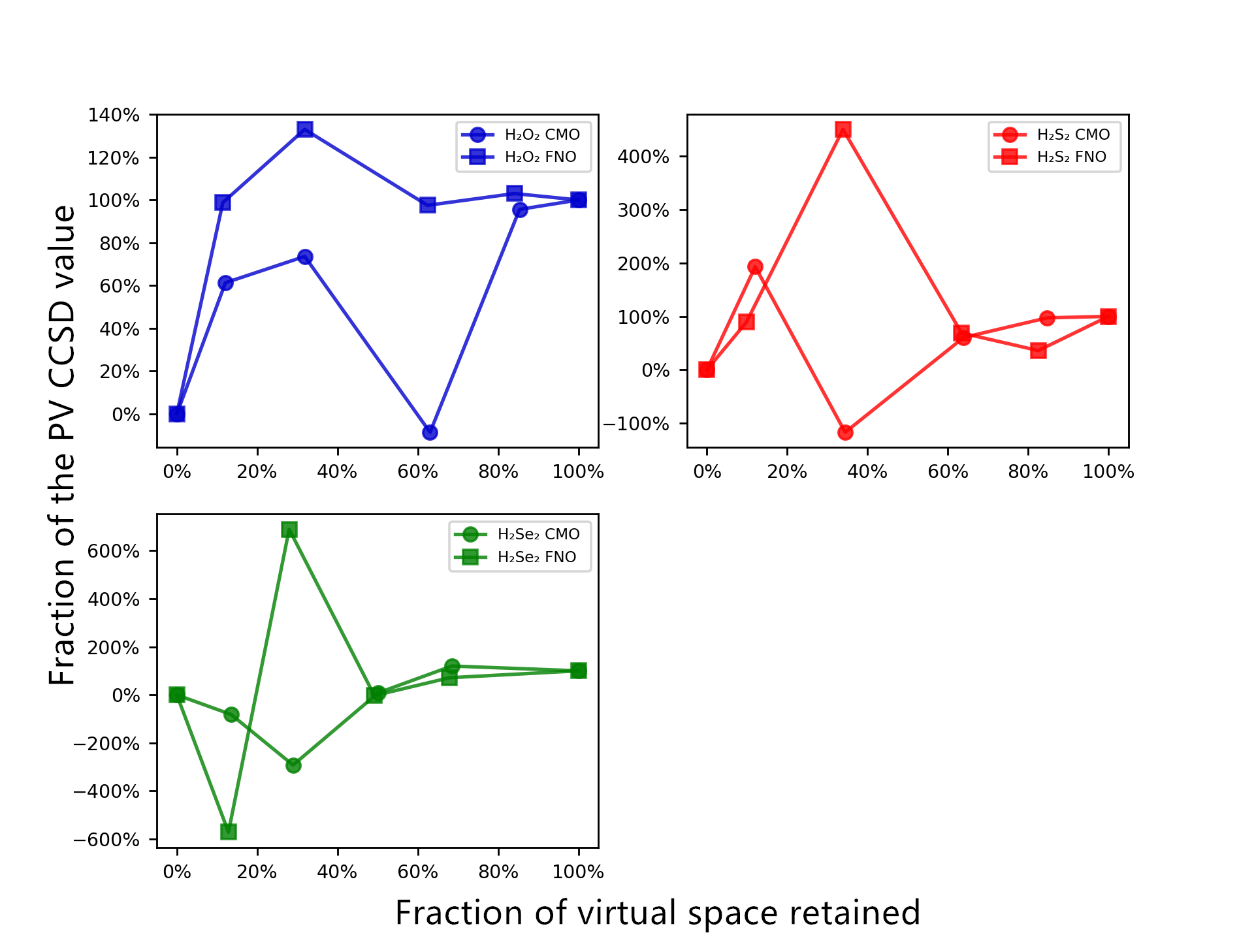}
\caption{\label{fig:epsarts1} Convergence of the CCSD PV energy with respect to the size of the virtual orbital space, for the X2C Hamiltonian. The X axis indicates the fraction of the virtual space retained, while the Y axis gives the fraction of the expectation value recovered with respect to the value obtained with the untruncated virtual space.}
\end{figure*}
            
The results of the two truncation schemes for the H$_{2}$Y$_{2}$ (Y=O, S, Se) molecules are displayed in Figure~\ref{fig:epsarts1}. We find that both FNO and CMO truncation leads to strong oscillations like observed in the EFG case. Again, these are most pronounced for aggressive truncation in which more than 50\% of the virtual orbital space is removed. Looking more closely at these systems we find that orbital energies (for CMO) or occupation numbers (for FNO) show a number of near-degeneracies. These orbitals give individually large but partially cancelling contributions. We have not investigated valence-only correlation explicitly for this property as core contributions are probably more important than for the EFG and it is more difficult to define a representative small test case. For our current tests we note that the convergence behaviour is consistent with results of reference~\citenum{sunaga_towards_2021}, in which rather conservative CMO truncation tresholds of 100 and 500 Hartree were used for H$_{2}$Se$_{2}$.  With both CMO and FNO truncation a 50 \% reduction appears realistic, leading to a significant reduction of the computational effort. For instance, in the current test the H$_{2}$Se$_{2}$ calculation on 49 nodes of Summit with the full orbital space took 100 minutes while the 50 \% reduced calculation too only 20 minutes.
        
Nevertheless, for the particular case of PV energy evaluation, FNO truncation does not appear to give a significant advantage over the simpler CMO truncation scheme. Apart from the sensitivity of PV operator on core correlation already observed in the EFG evaluation, another possible reason is our use of the unrelaxed MP2 density matrix. \citeauthor{shee_analytic_2016}~\cite{shee_analytic_2016} found for H$_{2}$S$_{2}$ that the contribution of orbital relaxation to the density matrix is significant when used to evaluate this property. It would be interesting to investigate how these contributions impact the generated FNOs are thereby the convergence of the FNO truncation scheme.

\section{Conclusions}

In this work we describe the formulation and implementation of the MP2 frozen natural orbitals (MP2FNOs) method for relativistic electronic structure calculations, which is an appealing approach for truncating the large virtual orbital spaces typically associated with relativistic calculations, while retaining the high accuracy expected of wavefunction-based approaches such as coupled cluster. 

This implementation was carried out in the massively parallel coupled cluster module of the DIRAC program, with the help of a framework to manipulate 1RDMs obtained from correlated wavefunction calculations. A particularity of our code is its ability to generate a set of canonical occupied and truncated virtual orbitals as well as natural orbitals in the atomic basis, so that these can be conveniently employed in post-Hartree-Fock calculations, and also used for analysis.

We employed our code to investigate the performance of MP2FNOs for the calculation of correlation energies and ground-state first-order properties such as dipole and quadrupole moments, electric field gradient at the nuclei, and parity violation energy shifts. As model systems, we considered species containing elements from the first row up to and including the superheavy element tennessine.

We have found that though MP2FNOs are always capable of recovering more correlation energy than their canonical counterparts for a given truncation, this advantage is slightly diminished for the heaviest systems considered, containing Astatine and Tennessine. In spite of that, it is generally found that  MP2FNOs can reduce virtual spaces to about 50\% of their original size without significant errors in the energy.

Among the properties considered, a truncation of about 50\% of the original virtual space has also been shown to provide values that are nearly converged to the results without truncation. That said, different properties exhibit very different convergence behavior; for valence properties such as the electric dipole moment MP2FNOs show a fairly smooth convergence to the reference values, whereas for the electric field gradient and in particular the parity violation energy shifts, for which regions close to the nuclei are important (and thus higher-lying virtuals are more important in the correlation treatment), significant variations on the calculated values are found for virtual spaces smaller than 50\% of the original virtual space.

We have found that for EFGs, correlating only valence electrons provides a pragmatic solution that recovers the smooth convergence seen for valence properties by removing difficult to converge, individually large but in total nearly cancelling, contributions from the core electrons. Whether such a strategy can also work for the PV energy shifts remains to be investigated.

As a final point and perspective we note that improvement of the efficiency of the current scheme is well possible by implementing efficient approximate schemes like Cholesky decomposition and Laplace transforms to generate the MP2 1RDM. Work along these lines is in progress and should enable treatment of larger systems in the future. 

\begin{acknowledgments}
We acknowledge funding from projects Labex CaPPA (ANR-11-LABX-0005-01) and CompRIXS (ANR-19-CE29-0019, DFG JA 2329/6-1), the I-SITE ULNE project OVERSEE and MESONM International Associated Laboratory (LAI) (ANR-16-IDEX-0004), and support from the French national supercomputing facilities (grant DARI A0090801859, A0110801859).

This research used resources of the Oak Ridge Leadership Computing Facility, which is a DOE Office of Science User Facility supported under Contract DE-AC05-00OR22725. 

\end{acknowledgments}





\bibliographystyle{aipnum4-1}
\bibliography{MP2FNO.bib}

\begin{thebibliography}{95}%
\makeatletter
\providecommand \@ifxundefined [1]{%
 \@ifx{#1\undefined}
}%
\providecommand \@ifnum [1]{%
 \ifnum #1\expandafter \@firstoftwo
 \else \expandafter \@secondoftwo
 \fi
}%
\providecommand \@ifx [1]{%
 \ifx #1\expandafter \@firstoftwo
 \else \expandafter \@secondoftwo
 \fi
}%
\providecommand \natexlab [1]{#1}%
\providecommand \enquote  [1]{``#1''}%
\providecommand \bibnamefont  [1]{#1}%
\providecommand \bibfnamefont [1]{#1}%
\providecommand \citenamefont [1]{#1}%
\providecommand \href@noop [0]{\@secondoftwo}%
\providecommand \href [0]{\begingroup \@sanitize@url \@href}%
\providecommand \@href[1]{\@@startlink{#1}\@@href}%
\providecommand \@@href[1]{\endgroup#1\@@endlink}%
\providecommand \@sanitize@url [0]{\catcode `\\12\catcode `\$12\catcode
  `\&12\catcode `\#12\catcode `\^12\catcode `\_12\catcode `\%12\relax}%
\providecommand \@@startlink[1]{}%
\providecommand \@@endlink[0]{}%
\providecommand \url  [0]{\begingroup\@sanitize@url \@url }%
\providecommand \@url [1]{\endgroup\@href {#1}{\urlprefix }}%
\providecommand \urlprefix  [0]{URL }%
\providecommand \Eprint [0]{\href }%
\providecommand \doibase [0]{http://dx.doi.org/}%
\providecommand \selectlanguage [0]{\@gobble}%
\providecommand \bibinfo  [0]{\@secondoftwo}%
\providecommand \bibfield  [0]{\@secondoftwo}%
\providecommand \translation [1]{[#1]}%
\providecommand \BibitemOpen [0]{}%
\providecommand \bibitemStop [0]{}%
\providecommand \bibitemNoStop [0]{.\EOS\space}%
\providecommand \EOS [0]{\spacefactor3000\relax}%
\providecommand \BibitemShut  [1]{\csname bibitem#1\endcsname}%
\let\auto@bib@innerbib\@empty
\bibitem [{\citenamefont {Acher}\ \emph {et~al.}(2016)\citenamefont {Acher},
  \citenamefont {Cherkaski}, \citenamefont {Dumas}, \citenamefont {Tamain},
  \citenamefont {Guillaumont}, \citenamefont {Boubals}, \citenamefont
  {Javierre}, \citenamefont {Hennig}, \citenamefont {Solari},\ and\
  \citenamefont {Charbonnel}}]{Acher2016}%
  \BibitemOpen
  \bibfield  {author} {\bibinfo {author} {\bibfnamefont {E.}~\bibnamefont
  {Acher}}, \bibinfo {author} {\bibfnamefont {Y.~H.}\ \bibnamefont
  {Cherkaski}}, \bibinfo {author} {\bibfnamefont {T.}~\bibnamefont {Dumas}},
  \bibinfo {author} {\bibfnamefont {C.}~\bibnamefont {Tamain}}, \bibinfo
  {author} {\bibfnamefont {D.}~\bibnamefont {Guillaumont}}, \bibinfo {author}
  {\bibfnamefont {N.}~\bibnamefont {Boubals}}, \bibinfo {author} {\bibfnamefont
  {G.}~\bibnamefont {Javierre}}, \bibinfo {author} {\bibfnamefont
  {C.}~\bibnamefont {Hennig}}, \bibinfo {author} {\bibfnamefont {P.~L.}\
  \bibnamefont {Solari}}, \ and\ \bibinfo {author} {\bibfnamefont {M.-C.}\
  \bibnamefont {Charbonnel}},\ }\href {\doibase 10.1021/acs.inorgchem.6b00592}
  {\bibfield  {journal} {\bibinfo  {journal} {Inorg. Chem.}\ }\textbf {\bibinfo
  {volume} {55}},\ \bibinfo {pages} {5558} (\bibinfo {year}
  {2016})}\BibitemShut {NoStop}%
\bibitem [{\citenamefont {Leoncini}, \citenamefont {Huskens},\ and\
  \citenamefont {Verboom}(2017)}]{leoncini2017ligands}%
  \BibitemOpen
  \bibfield  {author} {\bibinfo {author} {\bibfnamefont {A.}~\bibnamefont
  {Leoncini}}, \bibinfo {author} {\bibfnamefont {J.}~\bibnamefont {Huskens}}, \
  and\ \bibinfo {author} {\bibfnamefont {W.}~\bibnamefont {Verboom}},\
  }\href@noop {} {\bibfield  {journal} {\bibinfo  {journal} {Chem. Soc. Rev.}\
  }\textbf {\bibinfo {volume} {46}},\ \bibinfo {pages} {7229} (\bibinfo {year}
  {2017})}\BibitemShut {NoStop}%
\bibitem [{\citenamefont {Sun}, \citenamefont {Luo},\ and\ \citenamefont
  {Dai}(2012)}]{sun2012ionic}%
  \BibitemOpen
  \bibfield  {author} {\bibinfo {author} {\bibfnamefont {X.}~\bibnamefont
  {Sun}}, \bibinfo {author} {\bibfnamefont {H.}~\bibnamefont {Luo}}, \ and\
  \bibinfo {author} {\bibfnamefont {S.}~\bibnamefont {Dai}},\ }\href@noop {}
  {\bibfield  {journal} {\bibinfo  {journal} {Chem. Rev.}\ }\textbf {\bibinfo
  {volume} {112}},\ \bibinfo {pages} {2100} (\bibinfo {year}
  {2012})}\BibitemShut {NoStop}%
\bibitem [{\citenamefont {Berger}\ \emph {et~al.}(2020)\citenamefont {Berger},
  \citenamefont {Marie}, \citenamefont {Guillaumont}, \citenamefont {Tamain},
  \citenamefont {Dumas}, \citenamefont {Dirks}, \citenamefont {Boubals},
  \citenamefont {Acher}, \citenamefont {Laszczyk},\ and\ \citenamefont
  {Berthon}}]{Berger2020}%
  \BibitemOpen
  \bibfield  {author} {\bibinfo {author} {\bibfnamefont {C.}~\bibnamefont
  {Berger}}, \bibinfo {author} {\bibfnamefont {C.}~\bibnamefont {Marie}},
  \bibinfo {author} {\bibfnamefont {D.}~\bibnamefont {Guillaumont}}, \bibinfo
  {author} {\bibfnamefont {C.}~\bibnamefont {Tamain}}, \bibinfo {author}
  {\bibfnamefont {T.}~\bibnamefont {Dumas}}, \bibinfo {author} {\bibfnamefont
  {T.}~\bibnamefont {Dirks}}, \bibinfo {author} {\bibfnamefont
  {N.}~\bibnamefont {Boubals}}, \bibinfo {author} {\bibfnamefont
  {E.}~\bibnamefont {Acher}}, \bibinfo {author} {\bibfnamefont
  {M.}~\bibnamefont {Laszczyk}}, \ and\ \bibinfo {author} {\bibfnamefont
  {L.}~\bibnamefont {Berthon}},\ }\href {\doibase
  10.1021/acs.inorgchem.9b03024} {\bibfield  {journal} {\bibinfo  {journal}
  {Inorg. Chem.}\ }\textbf {\bibinfo {volume} {59}},\ \bibinfo {pages} {1823}
  (\bibinfo {year} {2020})}\BibitemShut {NoStop}%
\bibitem [{\citenamefont {Gogolski}\ and\ \citenamefont
  {Jensen}(2020)}]{Gogolski2020}%
  \BibitemOpen
  \bibfield  {author} {\bibinfo {author} {\bibfnamefont {J.~M.}\ \bibnamefont
  {Gogolski}}\ and\ \bibinfo {author} {\bibfnamefont {M.~P.}\ \bibnamefont
  {Jensen}},\ }\href {\doibase 10.1080/01496395.2020.1845209} {\bibfield
  {journal} {\bibinfo  {journal} {Sep. Sci. Technol}\ }\textbf {\bibinfo
  {volume} {56}},\ \bibinfo {pages} {2775} (\bibinfo {year}
  {2020})}\BibitemShut {NoStop}%
\bibitem [{\citenamefont {Berger}\ \emph {et~al.}(2021)\citenamefont {Berger},
  \citenamefont {Moreau}, \citenamefont {Marie}, \citenamefont {Guillaumont},
  \citenamefont {Beillard},\ and\ \citenamefont {Berthon}}]{Berger2021}%
  \BibitemOpen
  \bibfield  {author} {\bibinfo {author} {\bibfnamefont {C.}~\bibnamefont
  {Berger}}, \bibinfo {author} {\bibfnamefont {E.}~\bibnamefont {Moreau}},
  \bibinfo {author} {\bibfnamefont {C.}~\bibnamefont {Marie}}, \bibinfo
  {author} {\bibfnamefont {D.}~\bibnamefont {Guillaumont}}, \bibinfo {author}
  {\bibfnamefont {A.}~\bibnamefont {Beillard}}, \ and\ \bibinfo {author}
  {\bibfnamefont {L.}~\bibnamefont {Berthon}},\ }\href {\doibase
  10.1080/07366299.2021.1920167} {\bibfield  {journal} {\bibinfo  {journal}
  {Solvent. Extr. Ion. Exc}\ ,\ \bibinfo {pages} {1}} (\bibinfo {year}
  {2021})}\BibitemShut {NoStop}%
\bibitem [{\citenamefont {Gould}\ \emph {et~al.}(2022)\citenamefont {Gould},
  \citenamefont {McClain}, \citenamefont {Reta}, \citenamefont {Kragskow},
  \citenamefont {Marchiori}, \citenamefont {Lachman}, \citenamefont {Choi},
  \citenamefont {Analytis}, \citenamefont {Britt}, \citenamefont {Chilton},
  \citenamefont {Harvey},\ and\ \citenamefont {Long}}]{gould_2022}%
  \BibitemOpen
  \bibfield  {author} {\bibinfo {author} {\bibfnamefont {C.~A.}\ \bibnamefont
  {Gould}}, \bibinfo {author} {\bibfnamefont {K.~R.}\ \bibnamefont {McClain}},
  \bibinfo {author} {\bibfnamefont {D.}~\bibnamefont {Reta}}, \bibinfo {author}
  {\bibfnamefont {J.~G.~C.}\ \bibnamefont {Kragskow}}, \bibinfo {author}
  {\bibfnamefont {D.~A.}\ \bibnamefont {Marchiori}}, \bibinfo {author}
  {\bibfnamefont {E.}~\bibnamefont {Lachman}}, \bibinfo {author} {\bibfnamefont
  {E.-S.}\ \bibnamefont {Choi}}, \bibinfo {author} {\bibfnamefont {J.~G.}\
  \bibnamefont {Analytis}}, \bibinfo {author} {\bibfnamefont {R.~D.}\
  \bibnamefont {Britt}}, \bibinfo {author} {\bibfnamefont {N.~F.}\ \bibnamefont
  {Chilton}}, \bibinfo {author} {\bibfnamefont {B.~G.}\ \bibnamefont {Harvey}},
  \ and\ \bibinfo {author} {\bibfnamefont {J.~R.}\ \bibnamefont {Long}},\
  }\href {\doibase 10.1126/science.abl5470} {\bibfield  {journal} {\bibinfo
  {journal} {Science}\ }\textbf {\bibinfo {volume} {375}},\ \bibinfo {pages}
  {198} (\bibinfo {year} {2022})}\BibitemShut {NoStop}%
\bibitem [{\citenamefont {Burke}(2012)}]{burke_perspective_2012}%
  \BibitemOpen
  \bibfield  {author} {\bibinfo {author} {\bibfnamefont {K.}~\bibnamefont
  {Burke}},\ }\href {\doibase 10.1063/1.4704546} {\bibfield  {journal}
  {\bibinfo  {journal} {J. Chem. Phys.}\ }\textbf {\bibinfo {volume} {136}},\
  \bibinfo {pages} {150901} (\bibinfo {year} {2012})}\BibitemShut {NoStop}%
\bibitem [{\citenamefont {Bartlett}, \citenamefont {Lotrich},\ and\
  \citenamefont {Schweigert}(2005)}]{bartlett_ab_2005}%
  \BibitemOpen
  \bibfield  {author} {\bibinfo {author} {\bibfnamefont {R.~J.}\ \bibnamefont
  {Bartlett}}, \bibinfo {author} {\bibfnamefont {V.~F.}\ \bibnamefont
  {Lotrich}}, \ and\ \bibinfo {author} {\bibfnamefont {I.~V.}\ \bibnamefont
  {Schweigert}},\ }\href {\doibase 10.1063/1.1904585} {\bibfield  {journal}
  {\bibinfo  {journal} {J. Chem. Phys.}\ }\textbf {\bibinfo {volume} {123}},\
  \bibinfo {pages} {062205} (\bibinfo {year} {2005})}\BibitemShut {NoStop}%
\bibitem [{\citenamefont {Kervazo}\ \emph {et~al.}(2019)\citenamefont
  {Kervazo}, \citenamefont {Réal}, \citenamefont {Virot}, \citenamefont
  {Gomes},\ and\ \citenamefont {Vallet}}]{kervazo_accurate_2019}%
  \BibitemOpen
  \bibfield  {author} {\bibinfo {author} {\bibfnamefont {S.}~\bibnamefont
  {Kervazo}}, \bibinfo {author} {\bibfnamefont {F.}~\bibnamefont {Réal}},
  \bibinfo {author} {\bibfnamefont {F.}~\bibnamefont {Virot}}, \bibinfo
  {author} {\bibfnamefont {A.~S.~P.}\ \bibnamefont {Gomes}}, \ and\ \bibinfo
  {author} {\bibfnamefont {V.}~\bibnamefont {Vallet}},\ }\href {\doibase
  10.1021/acs.inorgchem.9b02096} {\bibfield  {journal} {\bibinfo  {journal}
  {Inorg. Chem.}\ }\textbf {\bibinfo {volume} {58}},\ \bibinfo {pages} {14507}
  (\bibinfo {year} {2019})}\BibitemShut {NoStop}%
\bibitem [{\citenamefont {Aebersold}\ and\ \citenamefont
  {Wilson}(2021)}]{aebersold_considering_2021}%
  \BibitemOpen
  \bibfield  {author} {\bibinfo {author} {\bibfnamefont {L.~E.}\ \bibnamefont
  {Aebersold}}\ and\ \bibinfo {author} {\bibfnamefont {A.~K.}\ \bibnamefont
  {Wilson}},\ }\href {\doibase 10.1021/acs.jpca.1c06155} {\bibfield  {journal}
  {\bibinfo  {journal} {J. Phys. Chem. A}\ }\textbf {\bibinfo {volume} {125}},\
  \bibinfo {pages} {7029} (\bibinfo {year} {2021})}\BibitemShut {NoStop}%
\bibitem [{\citenamefont {Aquino}, \citenamefont {Govind},\ and\ \citenamefont
  {Autschbach}(2010)}]{aquino_electric_2010}%
  \BibitemOpen
  \bibfield  {author} {\bibinfo {author} {\bibfnamefont {F.}~\bibnamefont
  {Aquino}}, \bibinfo {author} {\bibfnamefont {N.}~\bibnamefont {Govind}}, \
  and\ \bibinfo {author} {\bibfnamefont {J.}~\bibnamefont {Autschbach}},\
  }\href {\doibase 10.1021/ct1002847} {\bibfield  {journal} {\bibinfo
  {journal} {J. Chem. Theory. Comput.}\ }\textbf {\bibinfo {volume} {6}},\
  \bibinfo {pages} {2669} (\bibinfo {year} {2010})}\BibitemShut {NoStop}%
\bibitem [{\citenamefont {Sunaga}\ and\ \citenamefont
  {Saue}(2021)}]{sunaga_towards_2021}%
  \BibitemOpen
  \bibfield  {author} {\bibinfo {author} {\bibfnamefont {A.}~\bibnamefont
  {Sunaga}}\ and\ \bibinfo {author} {\bibfnamefont {T.}~\bibnamefont {Saue}},\
  }\href {\doibase 10.1080/00268976.2021.1974592} {\bibfield  {journal}
  {\bibinfo  {journal} {Mol. Phys}\ }\textbf {\bibinfo {volume} {119}},\
  \bibinfo {pages} {e1974592} (\bibinfo {year} {2021})}\BibitemShut {NoStop}%
\bibitem [{\citenamefont {Helgaker}\ \emph {et~al.}(1997)\citenamefont
  {Helgaker}, \citenamefont {Gauss}, \citenamefont {Jørgensen},\ and\
  \citenamefont {Olsen}}]{helgaker_prediction_1997}%
  \BibitemOpen
  \bibfield  {author} {\bibinfo {author} {\bibfnamefont {T.}~\bibnamefont
  {Helgaker}}, \bibinfo {author} {\bibfnamefont {J.}~\bibnamefont {Gauss}},
  \bibinfo {author} {\bibfnamefont {P.}~\bibnamefont {Jørgensen}}, \ and\
  \bibinfo {author} {\bibfnamefont {J.}~\bibnamefont {Olsen}},\ }\href
  {\doibase 10.1063/1.473634} {\bibfield  {journal} {\bibinfo  {journal} {J.
  Chem. Phys.}\ }\textbf {\bibinfo {volume} {106}},\ \bibinfo {pages} {6430}
  (\bibinfo {year} {1997})}\BibitemShut {NoStop}%
\bibitem [{\citenamefont {Bak}\ \emph {et~al.}(2001)\citenamefont {Bak},
  \citenamefont {Gauss}, \citenamefont {Jørgensen}, \citenamefont {Olsen},
  \citenamefont {Helgaker},\ and\ \citenamefont {Stanton}}]{bak_accurate_2001}%
  \BibitemOpen
  \bibfield  {author} {\bibinfo {author} {\bibfnamefont {K.~L.}\ \bibnamefont
  {Bak}}, \bibinfo {author} {\bibfnamefont {J.}~\bibnamefont {Gauss}}, \bibinfo
  {author} {\bibfnamefont {P.}~\bibnamefont {Jørgensen}}, \bibinfo {author}
  {\bibfnamefont {J.}~\bibnamefont {Olsen}}, \bibinfo {author} {\bibfnamefont
  {T.}~\bibnamefont {Helgaker}}, \ and\ \bibinfo {author} {\bibfnamefont
  {J.~F.}\ \bibnamefont {Stanton}},\ }\href {\doibase 10.1063/1.1357225}
  {\bibfield  {journal} {\bibinfo  {journal} {J. Chem. Phys.}\ }\textbf
  {\bibinfo {volume} {114}},\ \bibinfo {pages} {6548} (\bibinfo {year}
  {2001})}\BibitemShut {NoStop}%
\bibitem [{\citenamefont {Alml{\"o}f}(1991)}]{almlof1991elimination}%
  \BibitemOpen
  \bibfield  {author} {\bibinfo {author} {\bibfnamefont {J.}~\bibnamefont
  {Alml{\"o}f}},\ }\href@noop {} {\bibfield  {journal} {\bibinfo  {journal}
  {Chem. Phys. letters}\ }\textbf {\bibinfo {volume} {181}},\ \bibinfo {pages}
  {319} (\bibinfo {year} {1991})}\BibitemShut {NoStop}%
\bibitem [{\citenamefont {H{\"a}ser}(1993)}]{haser1993moller}%
  \BibitemOpen
  \bibfield  {author} {\bibinfo {author} {\bibfnamefont {M.}~\bibnamefont
  {H{\"a}ser}},\ }\href@noop {} {\bibfield  {journal} {\bibinfo  {journal}
  {Theor Chem Acc}\ }\textbf {\bibinfo {volume} {87}},\ \bibinfo {pages} {147}
  (\bibinfo {year} {1993})}\BibitemShut {NoStop}%
\bibitem [{\citenamefont {Baerends}, \citenamefont {Ellis},\ and\ \citenamefont
  {Ros}(1973)}]{Baerends.Ros.1973}%
  \BibitemOpen
  \bibfield  {author} {\bibinfo {author} {\bibfnamefont {E.~J.}\ \bibnamefont
  {Baerends}}, \bibinfo {author} {\bibfnamefont {D.~E.}\ \bibnamefont {Ellis}},
  \ and\ \bibinfo {author} {\bibfnamefont {P.}~\bibnamefont {Ros}},\ }\href
  {\doibase 10.1016/0301-0104(73)80059-x} {\bibfield  {journal} {\bibinfo
  {journal} {Chem. Phys.}\ }\textbf {\bibinfo {volume} {2}},\ \bibinfo {pages}
  {41 } (\bibinfo {year} {1973})}\BibitemShut {NoStop}%
\bibitem [{\citenamefont {Feyereisen}, \citenamefont {Fitzgerald},\ and\
  \citenamefont {Komornicki}(1993)}]{Feyereisen1993-RI}%
  \BibitemOpen
  \bibfield  {author} {\bibinfo {author} {\bibfnamefont {M.}~\bibnamefont
  {Feyereisen}}, \bibinfo {author} {\bibfnamefont {G.}~\bibnamefont
  {Fitzgerald}}, \ and\ \bibinfo {author} {\bibfnamefont {A.}~\bibnamefont
  {Komornicki}},\ }\href {\doibase 10.1016/0009-2614(93)87156-w} {\bibfield
  {journal} {\bibinfo  {journal} {Chem. Phys. Lett.}\ }\textbf {\bibinfo
  {volume} {208}},\ \bibinfo {pages} {359} (\bibinfo {year}
  {1993})}\BibitemShut {NoStop}%
\bibitem [{\citenamefont {Dyall}\ and\ \citenamefont
  {{Faegri}}(2007)}]{Dyallbook2007}%
  \BibitemOpen
  \bibfield  {author} {\bibinfo {author} {\bibfnamefont {K.~G.}\ \bibnamefont
  {Dyall}}\ and\ \bibinfo {author} {\bibfnamefont {K.}~\bibnamefont {{Faegri}},
  \bibfnamefont {Jr.}},\ }\href
  {https://oxford.universitypressscholarship.com/view/10.1093/oso/9780195140866.001.0001/isbn-9780195140866}
  {\emph {\bibinfo {title} {Introduction to Relativistic Quantum Chemistry}}}\
  (\bibinfo  {publisher} {Oxford University Press},\ \bibinfo {year}
  {2007})\BibitemShut {NoStop}%
\bibitem [{\citenamefont {Reiher}\ and\ \citenamefont
  {Wolf}(2014)}]{Reiher2014}%
  \BibitemOpen
  \bibfield  {author} {\bibinfo {author} {\bibfnamefont {M.}~\bibnamefont
  {Reiher}}\ and\ \bibinfo {author} {\bibfnamefont {A.}~\bibnamefont {Wolf}},\
  }\href {\doibase 10.1002/9783527667550} {\emph {\bibinfo {title}
  {Relativistic Quantum Chemistry}}}\ (\bibinfo  {publisher} {Wiley},\ \bibinfo
  {year} {2014})\BibitemShut {NoStop}%
\bibitem [{\citenamefont {Saue}\ and\ \citenamefont
  {Visscher}(2015)}]{Saue2015}%
  \BibitemOpen
  \bibfield  {author} {\bibinfo {author} {\bibfnamefont {T.}~\bibnamefont
  {Saue}}\ and\ \bibinfo {author} {\bibfnamefont {L.}~\bibnamefont
  {Visscher}},\ }in\ \href {\doibase 10.1002/9781118688304.ch3} {\emph
  {\bibinfo {booktitle} {Computational Methods in Lanthanide and Actinide
  Chemistry}}}\ (\bibinfo  {publisher} {John Wiley {\&} Sons Ltd},\ \bibinfo
  {year} {2015})\ pp.\ \bibinfo {pages} {55--87}\BibitemShut {NoStop}%
\bibitem [{\citenamefont {Batista}, \citenamefont {Martin},\ and\ \citenamefont
  {Yang}(2015)}]{Batista2015}%
  \BibitemOpen
  \bibfield  {author} {\bibinfo {author} {\bibfnamefont {E.~R.}\ \bibnamefont
  {Batista}}, \bibinfo {author} {\bibfnamefont {R.~L.}\ \bibnamefont {Martin}},
  \ and\ \bibinfo {author} {\bibfnamefont {P.}~\bibnamefont {Yang}},\ }in\
  \href {\doibase 10.1002/9781118688304.ch14} {\emph {\bibinfo {booktitle}
  {Computational Methods in Lanthanide and Actinide Chemistry}}}\ (\bibinfo
  {publisher} {John Wiley {\&} Sons Ltd},\ \bibinfo {year} {2015})\ pp.\
  \bibinfo {pages} {375--400}\BibitemShut {NoStop}%
\bibitem [{\citenamefont {Cao}\ and\ \citenamefont {Weigand}(2015)}]{Cao2015}%
  \BibitemOpen
  \bibfield  {author} {\bibinfo {author} {\bibfnamefont {X.}~\bibnamefont
  {Cao}}\ and\ \bibinfo {author} {\bibfnamefont {A.}~\bibnamefont {Weigand}},\
  }in\ \href {\doibase 10.1002/9781118688304.ch6} {\emph {\bibinfo {booktitle}
  {Computational Methods in Lanthanide and Actinide Chemistry}}}\ (\bibinfo
  {publisher} {John Wiley {\&} Sons Ltd},\ \bibinfo {year} {2015})\ pp.\
  \bibinfo {pages} {147--179}\BibitemShut {NoStop}%
\bibitem [{\citenamefont {Visscher}\ \emph {et~al.}(1994)\citenamefont
  {Visscher}, \citenamefont {Visser}, \citenamefont {Aerts}, \citenamefont
  {Merenga},\ and\ \citenamefont {Nieuwpoort}}]{visscher1994relativistic}%
  \BibitemOpen
  \bibfield  {author} {\bibinfo {author} {\bibfnamefont {L.}~\bibnamefont
  {Visscher}}, \bibinfo {author} {\bibfnamefont {O.}~\bibnamefont {Visser}},
  \bibinfo {author} {\bibfnamefont {P.~J.}\ \bibnamefont {Aerts}}, \bibinfo
  {author} {\bibfnamefont {H.}~\bibnamefont {Merenga}}, \ and\ \bibinfo
  {author} {\bibfnamefont {W.}~\bibnamefont {Nieuwpoort}},\ }\href@noop {}
  {\bibfield  {journal} {\bibinfo  {journal} {Comput. Phys. Commun}\ }\textbf
  {\bibinfo {volume} {81}},\ \bibinfo {pages} {120} (\bibinfo {year}
  {1994})}\BibitemShut {NoStop}%
\bibitem [{\citenamefont {Saue}\ \emph {et~al.}(2020)\citenamefont {Saue},
  \citenamefont {Bast}, \citenamefont {Gomes}, \citenamefont {Jensen},
  \citenamefont {Aucar}, \citenamefont {Remigio}, \citenamefont {Dyall},
  \citenamefont {Eliav}, \citenamefont {Fasshauer}, \citenamefont {Fleig},
  \citenamefont {Halbert}, \citenamefont {Hedegård}, \citenamefont {Helmich},
  \citenamefont {Jacob}, \citenamefont {Knecht}, \citenamefont {Laerdahl},
  \citenamefont {Nayak}, \citenamefont {Olsen}, \citenamefont {Senjean},
  \citenamefont {Shee},\ and\ \citenamefont {Sunaga}}]{saue_dirac_2020}%
  \BibitemOpen
  \bibfield  {author} {\bibinfo {author} {\bibfnamefont {T.}~\bibnamefont
  {Saue}}, \bibinfo {author} {\bibfnamefont {R.}~\bibnamefont {Bast}}, \bibinfo
  {author} {\bibfnamefont {A.~S.~P.}\ \bibnamefont {Gomes}}, \bibinfo {author}
  {\bibfnamefont {H.~J.~A.}\ \bibnamefont {Jensen}}, \bibinfo {author}
  {\bibfnamefont {I.~A.}\ \bibnamefont {Aucar}}, \bibinfo {author}
  {\bibfnamefont {R.~D.}\ \bibnamefont {Remigio}}, \bibinfo {author}
  {\bibfnamefont {K.~G.}\ \bibnamefont {Dyall}}, \bibinfo {author}
  {\bibfnamefont {E.}~\bibnamefont {Eliav}}, \bibinfo {author} {\bibfnamefont
  {E.}~\bibnamefont {Fasshauer}}, \bibinfo {author} {\bibfnamefont
  {T.}~\bibnamefont {Fleig}}, \bibinfo {author} {\bibfnamefont
  {L.}~\bibnamefont {Halbert}}, \bibinfo {author} {\bibfnamefont {E.~D.}\
  \bibnamefont {Hedegård}}, \bibinfo {author} {\bibfnamefont {B.}~\bibnamefont
  {Helmich}}, \bibinfo {author} {\bibfnamefont {C.~R.}\ \bibnamefont {Jacob}},
  \bibinfo {author} {\bibfnamefont {S.}~\bibnamefont {Knecht}}, \bibinfo
  {author} {\bibfnamefont {J.~K.}\ \bibnamefont {Laerdahl}}, \bibinfo {author}
  {\bibfnamefont {M.~K.}\ \bibnamefont {Nayak}}, \bibinfo {author}
  {\bibfnamefont {J.~M.~H.}\ \bibnamefont {Olsen}}, \bibinfo {author}
  {\bibfnamefont {B.}~\bibnamefont {Senjean}}, \bibinfo {author} {\bibfnamefont
  {A.}~\bibnamefont {Shee}}, \ and\ \bibinfo {author} {\bibfnamefont
  {A.}~\bibnamefont {Sunaga}},\ }\href@noop {} {\bibfield  {journal} {\bibinfo
  {journal} {J. Chem. Phys.}\ }\textbf {\bibinfo {volume} {152}},\ \bibinfo
  {pages} {204104} (\bibinfo {year} {2020})}\BibitemShut {NoStop}%
\bibitem [{\citenamefont {Belpassi}\ \emph {et~al.}(2020)\citenamefont
  {Belpassi}, \citenamefont {De~Santis}, \citenamefont {Quiney}, \citenamefont
  {Tarantelli},\ and\ \citenamefont {Storchi}}]{belpassi2020bertha}%
  \BibitemOpen
  \bibfield  {author} {\bibinfo {author} {\bibfnamefont {L.}~\bibnamefont
  {Belpassi}}, \bibinfo {author} {\bibfnamefont {M.}~\bibnamefont {De~Santis}},
  \bibinfo {author} {\bibfnamefont {H.~M.}\ \bibnamefont {Quiney}}, \bibinfo
  {author} {\bibfnamefont {F.}~\bibnamefont {Tarantelli}}, \ and\ \bibinfo
  {author} {\bibfnamefont {L.}~\bibnamefont {Storchi}},\ }\href@noop {}
  {\bibfield  {journal} {\bibinfo  {journal} {J. Chem. Phys.}\ }\textbf
  {\bibinfo {volume} {152}},\ \bibinfo {pages} {164118} (\bibinfo {year}
  {2020})}\BibitemShut {NoStop}%
\bibitem [{\citenamefont {Repisky}\ \emph
  {et~al.}(2020{\natexlab{a}})\citenamefont {Repisky}, \citenamefont
  {Komorovsky}, \citenamefont {Kadek}, \citenamefont {Konecny}, \citenamefont
  {Ekstr{\"o}m}, \citenamefont {Malkin}, \citenamefont {Kaupp}, \citenamefont
  {Ruud}, \citenamefont {Malkina},\ and\ \citenamefont
  {Malkin}}]{repisky2020respect}%
  \BibitemOpen
  \bibfield  {author} {\bibinfo {author} {\bibfnamefont {M.}~\bibnamefont
  {Repisky}}, \bibinfo {author} {\bibfnamefont {S.}~\bibnamefont {Komorovsky}},
  \bibinfo {author} {\bibfnamefont {M.}~\bibnamefont {Kadek}}, \bibinfo
  {author} {\bibfnamefont {L.}~\bibnamefont {Konecny}}, \bibinfo {author}
  {\bibfnamefont {U.}~\bibnamefont {Ekstr{\"o}m}}, \bibinfo {author}
  {\bibfnamefont {E.}~\bibnamefont {Malkin}}, \bibinfo {author} {\bibfnamefont
  {M.}~\bibnamefont {Kaupp}}, \bibinfo {author} {\bibfnamefont
  {K.}~\bibnamefont {Ruud}}, \bibinfo {author} {\bibfnamefont {O.~L.}\
  \bibnamefont {Malkina}}, \ and\ \bibinfo {author} {\bibfnamefont {V.~G.}\
  \bibnamefont {Malkin}},\ }\href@noop {} {\bibfield  {journal} {\bibinfo
  {journal} {J. Chem. Phys.}\ }\textbf {\bibinfo {volume} {152}},\ \bibinfo
  {pages} {184101} (\bibinfo {year} {2020}{\natexlab{a}})}\BibitemShut
  {NoStop}%
\bibitem [{\citenamefont {Kutzelnigg}\ and\ \citenamefont
  {Liu}(2005)}]{x2c:kutzelnigg2005}%
  \BibitemOpen
  \bibfield  {author} {\bibinfo {author} {\bibfnamefont {W.}~\bibnamefont
  {Kutzelnigg}}\ and\ \bibinfo {author} {\bibfnamefont {W.}~\bibnamefont
  {Liu}},\ }\href {\doibase 10.1063/1.2137315} {\bibfield  {journal} {\bibinfo
  {journal} {J. Chem. Phys.}\ }\textbf {\bibinfo {volume} {123}},\ \bibinfo
  {pages} {241102} (\bibinfo {year} {2005})}\BibitemShut {NoStop}%
\bibitem [{\citenamefont {Jensen}(2005)}]{jensen:rehe2005}%
  \BibitemOpen
  \bibfield  {author} {\bibinfo {author} {\bibfnamefont {H.~J.~{\relax Aa}.}\
  \bibnamefont {Jensen}},\ }\href@noop {} {} (\bibinfo {year} {2005}),\
  \bibinfo {note} {{{\textit{Douglas--Kroll the Easy Way}}, Talk at Conference
  on Relativistic Effects in Heavy Elements - REHE, M{\"u}lheim, Germany,
  April, 2005. Available at
  \url{https://doi.org/10.6084/m9.figshare.12046158}}}\BibitemShut {NoStop}%
\bibitem [{\citenamefont {Liu}\ and\ \citenamefont
  {Peng}(2006)}]{x2c:atom2mol:liu2006}%
  \BibitemOpen
  \bibfield  {author} {\bibinfo {author} {\bibfnamefont {W.}~\bibnamefont
  {Liu}}\ and\ \bibinfo {author} {\bibfnamefont {D.}~\bibnamefont {Peng}},\
  }\href {\doibase 10.1063/1.2222365} {\bibfield  {journal} {\bibinfo
  {journal} {J. Chem. Phys.}\ }\textbf {\bibinfo {volume} {125}},\ \bibinfo
  {pages} {044102} (\bibinfo {year} {2006})}\BibitemShut {NoStop}%
\bibitem [{\citenamefont {Iliaš}\ and\ \citenamefont
  {Saue}(2007)}]{ilias_infinite-order_2007}%
  \BibitemOpen
  \bibfield  {author} {\bibinfo {author} {\bibfnamefont {M.}~\bibnamefont
  {Iliaš}}\ and\ \bibinfo {author} {\bibfnamefont {T.}~\bibnamefont {Saue}},\
  }\href {\doibase 10.1063/1.2436882} {\bibfield  {journal} {\bibinfo
  {journal} {J. Chem. Phys.}\ }\textbf {\bibinfo {volume} {126}},\ \bibinfo
  {pages} {064102} (\bibinfo {year} {2007})}\BibitemShut {NoStop}%
\bibitem [{\citenamefont {Peng}\ \emph {et~al.}(2007)\citenamefont {Peng},
  \citenamefont {Liu}, \citenamefont {Xiao},\ and\ \citenamefont
  {Cheng}}]{x2c:atom2mol:peng2007}%
  \BibitemOpen
  \bibfield  {author} {\bibinfo {author} {\bibfnamefont {D.}~\bibnamefont
  {Peng}}, \bibinfo {author} {\bibfnamefont {W.}~\bibnamefont {Liu}}, \bibinfo
  {author} {\bibfnamefont {Y.}~\bibnamefont {Xiao}}, \ and\ \bibinfo {author}
  {\bibfnamefont {L.}~\bibnamefont {Cheng}},\ }\href {\doibase
  10.1063/1.2772856} {\bibfield  {journal} {\bibinfo  {journal} {J. Chem.
  Phys.}\ }\textbf {\bibinfo {volume} {127}},\ \bibinfo {pages} {104106}
  (\bibinfo {year} {2007})}\BibitemShut {NoStop}%
\bibitem [{\citenamefont {Liu}\ and\ \citenamefont {Peng}(2009)}]{x2c:liu2009}%
  \BibitemOpen
  \bibfield  {author} {\bibinfo {author} {\bibfnamefont {W.}~\bibnamefont
  {Liu}}\ and\ \bibinfo {author} {\bibfnamefont {D.}~\bibnamefont {Peng}},\
  }\href {\doibase 10.1063/1.3159445} {\bibfield  {journal} {\bibinfo
  {journal} {J. Chem. Phys.}\ }\textbf {\bibinfo {volume} {131}},\ \bibinfo
  {pages} {031104} (\bibinfo {year} {2009})}\BibitemShut {NoStop}%
\bibitem [{\citenamefont {Liu}\ \emph {et~al.}(2018)\citenamefont {Liu},
  \citenamefont {Shen}, \citenamefont {Asthana},\ and\ \citenamefont
  {Cheng}}]{mfso:cc:liu2018}%
  \BibitemOpen
  \bibfield  {author} {\bibinfo {author} {\bibfnamefont {J.}~\bibnamefont
  {Liu}}, \bibinfo {author} {\bibfnamefont {Y.}~\bibnamefont {Shen}}, \bibinfo
  {author} {\bibfnamefont {A.}~\bibnamefont {Asthana}}, \ and\ \bibinfo
  {author} {\bibfnamefont {L.}~\bibnamefont {Cheng}},\ }\href {\doibase
  10.1063/1.5009177} {\bibfield  {journal} {\bibinfo  {journal} {J. Chem.
  Phys.}\ }\textbf {\bibinfo {volume} {148}},\ \bibinfo {pages} {034106}
  (\bibinfo {year} {2018})}\BibitemShut {NoStop}%
\bibitem [{\citenamefont {Liu}\ and\ \citenamefont
  {Cheng}(2018)}]{x2camf:cc:liu2018}%
  \BibitemOpen
  \bibfield  {author} {\bibinfo {author} {\bibfnamefont {J.}~\bibnamefont
  {Liu}}\ and\ \bibinfo {author} {\bibfnamefont {L.}~\bibnamefont {Cheng}},\
  }\href {\doibase 10.1063/1.5023750} {\bibfield  {journal} {\bibinfo
  {journal} {J. Chem. Phys.}\ }\textbf {\bibinfo {volume} {148}},\ \bibinfo
  {pages} {144108} (\bibinfo {year} {2018})}\BibitemShut {NoStop}%
\bibitem [{\citenamefont {Cheng}(2019)}]{mfso:eomcc:cheng2019}%
  \BibitemOpen
  \bibfield  {author} {\bibinfo {author} {\bibfnamefont {L.}~\bibnamefont
  {Cheng}},\ }\href {\doibase 10.1063/1.5113796} {\bibfield  {journal}
  {\bibinfo  {journal} {J. Chem. Phys.}\ }\textbf {\bibinfo {volume} {151}},\
  \bibinfo {pages} {104103} (\bibinfo {year} {2019})}\BibitemShut {NoStop}%
\bibitem [{\citenamefont {Asthana}, \citenamefont {Liu},\ and\ \citenamefont
  {Cheng}(2019)}]{x2camf:cc:asthana2019}%
  \BibitemOpen
  \bibfield  {author} {\bibinfo {author} {\bibfnamefont {A.}~\bibnamefont
  {Asthana}}, \bibinfo {author} {\bibfnamefont {J.}~\bibnamefont {Liu}}, \ and\
  \bibinfo {author} {\bibfnamefont {L.}~\bibnamefont {Cheng}},\ }\href
  {\doibase 10.1063/1.5081715} {\bibfield  {journal} {\bibinfo  {journal} {J.
  Chem. Phys.}\ }\textbf {\bibinfo {volume} {150}},\ \bibinfo {pages} {074102}
  (\bibinfo {year} {2019})}\BibitemShut {NoStop}%
\bibitem [{\citenamefont {Sikkema}\ \emph {et~al.}(2009)\citenamefont
  {Sikkema}, \citenamefont {Visscher}, \citenamefont {Saue},\ and\
  \citenamefont {Ilia{\v{s}}}}]{Sikkema2009}%
  \BibitemOpen
  \bibfield  {author} {\bibinfo {author} {\bibfnamefont {J.}~\bibnamefont
  {Sikkema}}, \bibinfo {author} {\bibfnamefont {L.}~\bibnamefont {Visscher}},
  \bibinfo {author} {\bibfnamefont {T.}~\bibnamefont {Saue}}, \ and\ \bibinfo
  {author} {\bibfnamefont {M.}~\bibnamefont {Ilia{\v{s}}}},\ }\href {\doibase
  10.1063/1.3239505} {\bibfield  {journal} {\bibinfo  {journal} {J. Chem.
  Phys.}\ }\textbf {\bibinfo {volume} {131}},\ \bibinfo {pages} {124116}
  (\bibinfo {year} {2009})}\BibitemShut {NoStop}%
\bibitem [{\citenamefont {F{\ae}gri}\ and\ \citenamefont
  {Saue}(2001)}]{Fgri2001}%
  \BibitemOpen
  \bibfield  {author} {\bibinfo {author} {\bibfnamefont {K.}~\bibnamefont
  {F{\ae}gri}}\ and\ \bibinfo {author} {\bibfnamefont {T.}~\bibnamefont
  {Saue}},\ }\href {\doibase 10.1063/1.1385366} {\bibfield  {journal} {\bibinfo
   {journal} {J. Chem. Phys.}\ }\textbf {\bibinfo {volume} {115}},\ \bibinfo
  {pages} {2456} (\bibinfo {year} {2001})}\BibitemShut {NoStop}%
\bibitem [{\citenamefont {Gomes}\ and\ \citenamefont
  {Visscher}(2004)}]{pereira_gomes_influence_2004}%
  \BibitemOpen
  \bibfield  {author} {\bibinfo {author} {\bibfnamefont {A.~S.~P.}\
  \bibnamefont {Gomes}}\ and\ \bibinfo {author} {\bibfnamefont
  {L.}~\bibnamefont {Visscher}},\ }\href {\doibase
  10.1016/j.cplett.2004.09.132} {\bibfield  {journal} {\bibinfo  {journal}
  {Chem. Phys. Lett.}\ }\textbf {\bibinfo {volume} {399}},\ \bibinfo {pages}
  {1} (\bibinfo {year} {2004})}\BibitemShut {NoStop}%
\bibitem [{\citenamefont {Thierfelder}\ \emph {et~al.}(2009)\citenamefont
  {Thierfelder}, \citenamefont {Schwerdtfeger}, \citenamefont {Koers},
  \citenamefont {Borschevsky},\ and\ \citenamefont
  {Fricke}}]{thierfelder_scalar_2009}%
  \BibitemOpen
  \bibfield  {author} {\bibinfo {author} {\bibfnamefont {C.}~\bibnamefont
  {Thierfelder}}, \bibinfo {author} {\bibfnamefont {P.}~\bibnamefont
  {Schwerdtfeger}}, \bibinfo {author} {\bibfnamefont {A.}~\bibnamefont
  {Koers}}, \bibinfo {author} {\bibfnamefont {A.}~\bibnamefont {Borschevsky}},
  \ and\ \bibinfo {author} {\bibfnamefont {B.}~\bibnamefont {Fricke}},\ }\href
  {\doibase 10.1103/PhysRevA.80.022501} {\bibfield  {journal} {\bibinfo
  {journal} {Phys. Rev. A}\ }\textbf {\bibinfo {volume} {80}},\ \bibinfo
  {pages} {022501} (\bibinfo {year} {2009})}\BibitemShut {NoStop}%
\bibitem [{\citenamefont {Pershina}(2010)}]{Pershina2010}%
  \BibitemOpen
  \bibfield  {author} {\bibinfo {author} {\bibfnamefont {V.}~\bibnamefont
  {Pershina}},\ }in\ \href {\doibase 10.1007/978-1-4020-9975-5_11} {\emph
  {\bibinfo {booktitle} {Challenges and Advances in Computational Chemistry and
  Physics}}}\ (\bibinfo  {publisher} {Springer Netherlands},\ \bibinfo {year}
  {2010})\ pp.\ \bibinfo {pages} {451--520}\BibitemShut {NoStop}%
\bibitem [{\citenamefont {Autschbach}\ \emph {et~al.}(2015)\citenamefont
  {Autschbach}, \citenamefont {Govind}, \citenamefont {Atta-Fynn},
  \citenamefont {Bylaska}, \citenamefont {Weare},\ and\ \citenamefont
  {de~Jong}}]{Autschbach2015}%
  \BibitemOpen
  \bibfield  {author} {\bibinfo {author} {\bibfnamefont {J.}~\bibnamefont
  {Autschbach}}, \bibinfo {author} {\bibfnamefont {N.}~\bibnamefont {Govind}},
  \bibinfo {author} {\bibfnamefont {R.}~\bibnamefont {Atta-Fynn}}, \bibinfo
  {author} {\bibfnamefont {E.~J.}\ \bibnamefont {Bylaska}}, \bibinfo {author}
  {\bibfnamefont {J.~W.}\ \bibnamefont {Weare}}, \ and\ \bibinfo {author}
  {\bibfnamefont {W.~A.}\ \bibnamefont {de~Jong}},\ }in\ \href {\doibase
  10.1002/9781118688304.ch12} {\emph {\bibinfo {booktitle} {Computational
  Methods in Lanthanide and Actinide Chemistry}}}\ (\bibinfo  {publisher} {John
  Wiley {\&} Sons Ltd},\ \bibinfo {year} {2015})\ pp.\ \bibinfo {pages}
  {299--342}\BibitemShut {NoStop}%
\bibitem [{\citenamefont {Halbert}\ \emph {et~al.}(2021)\citenamefont
  {Halbert}, \citenamefont {Vidal}, \citenamefont {Shee}, \citenamefont
  {Coriani},\ and\ \citenamefont {Gomes}}]{Halbert2021}%
  \BibitemOpen
  \bibfield  {author} {\bibinfo {author} {\bibfnamefont {L.}~\bibnamefont
  {Halbert}}, \bibinfo {author} {\bibfnamefont {M.~L.}\ \bibnamefont {Vidal}},
  \bibinfo {author} {\bibfnamefont {A.}~\bibnamefont {Shee}}, \bibinfo {author}
  {\bibfnamefont {S.}~\bibnamefont {Coriani}}, \ and\ \bibinfo {author}
  {\bibfnamefont {A.~S.~P.}\ \bibnamefont {Gomes}},\ }\href {\doibase
  10.1021/acs.jctc.0c01203} {\bibfield  {journal} {\bibinfo  {journal} {J.
  Chem. Theory Comput.}\ }\textbf {\bibinfo {volume} {17}},\ \bibinfo {pages}
  {3583} (\bibinfo {year} {2021})}\BibitemShut {NoStop}%
\bibitem [{\citenamefont {Sucher}(1980)}]{Sucher1980}%
  \BibitemOpen
  \bibfield  {author} {\bibinfo {author} {\bibfnamefont {J.}~\bibnamefont
  {Sucher}},\ }\href {\doibase 10.1103/physreva.22.348} {\bibfield  {journal}
  {\bibinfo  {journal} {Phys. Rev. A}\ }\textbf {\bibinfo {volume} {22}},\
  \bibinfo {pages} {348} (\bibinfo {year} {1980})}\BibitemShut {NoStop}%
\bibitem [{\citenamefont {Pototschnig}\ \emph {et~al.}(2021)\citenamefont
  {Pototschnig}, \citenamefont {Papadopoulos}, \citenamefont {Lyakh},
  \citenamefont {Repisky}, \citenamefont {Halbert}, \citenamefont {Gomes},
  \citenamefont {Jensen},\ and\ \citenamefont
  {Visscher}}]{pototschnig_implementation_2021}%
  \BibitemOpen
  \bibfield  {author} {\bibinfo {author} {\bibfnamefont {J.~V.}\ \bibnamefont
  {Pototschnig}}, \bibinfo {author} {\bibfnamefont {A.}~\bibnamefont
  {Papadopoulos}}, \bibinfo {author} {\bibfnamefont {D.~I.}\ \bibnamefont
  {Lyakh}}, \bibinfo {author} {\bibfnamefont {M.}~\bibnamefont {Repisky}},
  \bibinfo {author} {\bibfnamefont {L.}~\bibnamefont {Halbert}}, \bibinfo
  {author} {\bibfnamefont {A.~S.~P.}\ \bibnamefont {Gomes}}, \bibinfo {author}
  {\bibfnamefont {H.~J.~A.}\ \bibnamefont {Jensen}}, \ and\ \bibinfo {author}
  {\bibfnamefont {L.}~\bibnamefont {Visscher}},\ }\href {\doibase
  10.1021/acs.jctc.1c00260} {\bibfield  {journal} {\bibinfo  {journal} {J.
  Chem. Theory. Comput.}\ }\textbf {\bibinfo {volume} {17}},\ \bibinfo {pages}
  {5509} (\bibinfo {year} {2021})}\BibitemShut {NoStop}%
\bibitem [{\citenamefont {Lyakh}(2019)}]{lyakh_domainspecific_2019}%
  \BibitemOpen
  \bibfield  {author} {\bibinfo {author} {\bibfnamefont {D.~I.}\ \bibnamefont
  {Lyakh}},\ }\href {\doibase 10.1002/qua.25926} {\bibfield  {journal}
  {\bibinfo  {journal} {Int. J. Quantum. Chem}\ }\textbf {\bibinfo {volume}
  {119}},\ \bibinfo {pages} {e25926} (\bibinfo {year} {2019})}\BibitemShut
  {NoStop}%
\bibitem [{\citenamefont {Helmich-Paris}, \citenamefont {Repisky},\ and\
  \citenamefont {Visscher}(2016)}]{Helmich-Paris2016ev}%
  \BibitemOpen
  \bibfield  {author} {\bibinfo {author} {\bibfnamefont {B.}~\bibnamefont
  {Helmich-Paris}}, \bibinfo {author} {\bibfnamefont {M.}~\bibnamefont
  {Repisky}}, \ and\ \bibinfo {author} {\bibfnamefont {L.}~\bibnamefont
  {Visscher}},\ }\href {\doibase 10.1063/1.4955106} {\bibfield  {journal}
  {\bibinfo  {journal} {J. Chem. Phys.}\ }\textbf {\bibinfo {volume} {145}},\
  \bibinfo {pages} {014107} (\bibinfo {year} {2016})},\ \Eprint
  {http://arxiv.org/abs/1606.06498} {1606.06498} \BibitemShut {NoStop}%
\bibitem [{\citenamefont {Visscher}, \citenamefont {Aerts},\ and\ \citenamefont
  {Visser}(1991)}]{visscher1991contraction}%
  \BibitemOpen
  \bibfield  {author} {\bibinfo {author} {\bibfnamefont {L.}~\bibnamefont
  {Visscher}}, \bibinfo {author} {\bibfnamefont {P.~J.~C.}\ \bibnamefont
  {Aerts}}, \ and\ \bibinfo {author} {\bibfnamefont {O.}~\bibnamefont
  {Visser}},\ }in\ \href {\doibase 10.1007/978-1-4615-3702-1_13} {\emph
  {\bibinfo {booktitle} {The Effects of Relativity in Atoms, Molecules, and the
  Solid State}}}\ (\bibinfo  {publisher} {Springer {US}},\ \bibinfo {year}
  {1991})\ pp.\ \bibinfo {pages} {197--205}\BibitemShut {NoStop}%
\bibitem [{\citenamefont {Belpassi}\ \emph {et~al.}(2005)\citenamefont
  {Belpassi}, \citenamefont {Tarantelli}, \citenamefont {Sgamellotti},\ and\
  \citenamefont {Quiney}}]{Belpassi2005}%
  \BibitemOpen
  \bibfield  {author} {\bibinfo {author} {\bibfnamefont {L.}~\bibnamefont
  {Belpassi}}, \bibinfo {author} {\bibfnamefont {F.}~\bibnamefont
  {Tarantelli}}, \bibinfo {author} {\bibfnamefont {A.}~\bibnamefont
  {Sgamellotti}}, \ and\ \bibinfo {author} {\bibfnamefont {H.~M.}\ \bibnamefont
  {Quiney}},\ }\href {\doibase 10.1063/1.1897383} {\bibfield  {journal}
  {\bibinfo  {journal} {J. Chem. Phys.}\ }\textbf {\bibinfo {volume} {122}},\
  \bibinfo {pages} {184109} (\bibinfo {year} {2005})}\BibitemShut {NoStop}%
\bibitem [{\citenamefont {Löwdin}(1955)}]{lowdin_quantum_1955}%
  \BibitemOpen
  \bibfield  {author} {\bibinfo {author} {\bibfnamefont {P.-O.}\ \bibnamefont
  {Löwdin}},\ }\href {\doibase 10.1103/PhysRev.97.1474} {\bibfield  {journal}
  {\bibinfo  {journal} {Phys. Rev.}\ }\textbf {\bibinfo {volume} {97}},\
  \bibinfo {pages} {1474} (\bibinfo {year} {1955})}\BibitemShut {NoStop}%
\bibitem [{\citenamefont {Davidson}(1972)}]{davidson_properties_1972}%
  \BibitemOpen
  \bibfield  {author} {\bibinfo {author} {\bibfnamefont {E.~R.}\ \bibnamefont
  {Davidson}},\ }\href {\doibase 10.1103/RevModPhys.44.451} {\bibfield
  {journal} {\bibinfo  {journal} {Rev. Mod. Phys.}\ }\textbf {\bibinfo {volume}
  {44}},\ \bibinfo {pages} {451} (\bibinfo {year} {1972})}\BibitemShut
  {NoStop}%
\bibitem [{\citenamefont {Barr}\ and\ \citenamefont
  {Davidson}(1970)}]{barr_nature_1970}%
  \BibitemOpen
  \bibfield  {author} {\bibinfo {author} {\bibfnamefont {T.~L.}\ \bibnamefont
  {Barr}}\ and\ \bibinfo {author} {\bibfnamefont {E.~R.}\ \bibnamefont
  {Davidson}},\ }\href {\doibase 10.1103/PhysRevA.1.644} {\bibfield  {journal}
  {\bibinfo  {journal} {Phys. Rev. A}\ }\textbf {\bibinfo {volume} {1}},\
  \bibinfo {pages} {644} (\bibinfo {year} {1970})}\BibitemShut {NoStop}%
\bibitem [{\citenamefont {Sosa}\ \emph {et~al.}(1989)\citenamefont {Sosa},
  \citenamefont {Geertsen}, \citenamefont {Trucks}, \citenamefont {Bartlett},\
  and\ \citenamefont {Franz}}]{sosa_selection_1989}%
  \BibitemOpen
  \bibfield  {author} {\bibinfo {author} {\bibfnamefont {C.}~\bibnamefont
  {Sosa}}, \bibinfo {author} {\bibfnamefont {J.}~\bibnamefont {Geertsen}},
  \bibinfo {author} {\bibfnamefont {G.~W.}\ \bibnamefont {Trucks}}, \bibinfo
  {author} {\bibfnamefont {R.~J.}\ \bibnamefont {Bartlett}}, \ and\ \bibinfo
  {author} {\bibfnamefont {J.~A.}\ \bibnamefont {Franz}},\ }\href {\doibase
  10.1016/0009-2614(89)87399-3} {\bibfield  {journal} {\bibinfo  {journal}
  {Chem. Phys. Lett.}\ }\textbf {\bibinfo {volume} {159}},\ \bibinfo {pages}
  {148} (\bibinfo {year} {1989})}\BibitemShut {NoStop}%
\bibitem [{\citenamefont {Taube}\ and\ \citenamefont
  {Bartlett}(2005)}]{taube_frozen_2005}%
  \BibitemOpen
  \bibfield  {author} {\bibinfo {author} {\bibfnamefont {A.~G.}\ \bibnamefont
  {Taube}}\ and\ \bibinfo {author} {\bibfnamefont {R.~J.}\ \bibnamefont
  {Bartlett}},\ }\href {\doibase 10.1135/cccc20050837} {\bibfield  {journal}
  {\bibinfo  {journal} {Collect. Czech. Chem. Commun.}\ }\textbf {\bibinfo
  {volume} {70}},\ \bibinfo {pages} {837} (\bibinfo {year} {2005})}\BibitemShut
  {NoStop}%
\bibitem [{\citenamefont {Jensen}\ \emph
  {et~al.}(1988{\natexlab{a}})\citenamefont {Jensen}, \citenamefont
  {Jørgensen}, \citenamefont {Ågren},\ and\ \citenamefont
  {Olsen}}]{jensen_secondorder_1988}%
  \BibitemOpen
  \bibfield  {author} {\bibinfo {author} {\bibfnamefont {H.~J.~A.}\
  \bibnamefont {Jensen}}, \bibinfo {author} {\bibfnamefont {P.}~\bibnamefont
  {Jørgensen}}, \bibinfo {author} {\bibfnamefont {H.}~\bibnamefont {Ågren}},
  \ and\ \bibinfo {author} {\bibfnamefont {J.}~\bibnamefont {Olsen}},\ }\href
  {\doibase 10.1063/1.453884} {\bibfield  {journal} {\bibinfo  {journal} {J.
  Chem. Phys.}\ }\textbf {\bibinfo {volume} {88}},\ \bibinfo {pages} {3834}
  (\bibinfo {year} {1988}{\natexlab{a}})}\BibitemShut {NoStop}%
\bibitem [{\citenamefont {Jensen}\ \emph
  {et~al.}(1988{\natexlab{b}})\citenamefont {Jensen}, \citenamefont
  {Jørgensen}, \citenamefont {Ågren},\ and\ \citenamefont
  {Olsen}}]{jensen_erratum_1988}%
  \BibitemOpen
  \bibfield  {author} {\bibinfo {author} {\bibfnamefont {H.~J.~A.}\
  \bibnamefont {Jensen}}, \bibinfo {author} {\bibfnamefont {P.}~\bibnamefont
  {Jørgensen}}, \bibinfo {author} {\bibfnamefont {H.}~\bibnamefont {Ågren}},
  \ and\ \bibinfo {author} {\bibfnamefont {J.}~\bibnamefont {Olsen}},\ }\href
  {\doibase 10.1063/1.455749} {\bibfield  {journal} {\bibinfo  {journal} {J.
  Chem. Phys.}\ }\textbf {\bibinfo {volume} {89}},\ \bibinfo {pages} {5354}
  (\bibinfo {year} {1988}{\natexlab{b}})}\BibitemShut {NoStop}%
\bibitem [{\citenamefont {Taube}\ and\ \citenamefont
  {Bartlett}(2008)}]{taube_frozen_2008}%
  \BibitemOpen
  \bibfield  {author} {\bibinfo {author} {\bibfnamefont {A.~G.}\ \bibnamefont
  {Taube}}\ and\ \bibinfo {author} {\bibfnamefont {R.~J.}\ \bibnamefont
  {Bartlett}},\ }\href {\doibase 10.1063/1.2902285} {\bibfield  {journal}
  {\bibinfo  {journal} {J. Chem. Phys.}\ }\textbf {\bibinfo {volume} {128}},\
  \bibinfo {pages} {164101} (\bibinfo {year} {2008})}\BibitemShut {NoStop}%
\bibitem [{\citenamefont {Verma}\ \emph {et~al.}(2021)\citenamefont {Verma},
  \citenamefont {Huntington}, \citenamefont {Coons}, \citenamefont {Kawashima},
  \citenamefont {Yamazaki},\ and\ \citenamefont
  {Zaribafiyan}}]{verma_scaling_2021}%
  \BibitemOpen
  \bibfield  {author} {\bibinfo {author} {\bibfnamefont {P.}~\bibnamefont
  {Verma}}, \bibinfo {author} {\bibfnamefont {L.}~\bibnamefont {Huntington}},
  \bibinfo {author} {\bibfnamefont {M.~P.}\ \bibnamefont {Coons}}, \bibinfo
  {author} {\bibfnamefont {Y.}~\bibnamefont {Kawashima}}, \bibinfo {author}
  {\bibfnamefont {T.}~\bibnamefont {Yamazaki}}, \ and\ \bibinfo {author}
  {\bibfnamefont {A.}~\bibnamefont {Zaribafiyan}},\ }\href {\doibase
  10.1063/5.0054647} {\bibfield  {journal} {\bibinfo  {journal} {J. Chem.
  Phys.}\ }\textbf {\bibinfo {volume} {155}},\ \bibinfo {pages} {034110}
  (\bibinfo {year} {2021})}\BibitemShut {NoStop}%
\bibitem [{\citenamefont {Gordon}\ \emph {et~al.}(1999)\citenamefont {Gordon},
  \citenamefont {Schmidt}, \citenamefont {Chaban}, \citenamefont {Glaesemann},
  \citenamefont {Stevens},\ and\ \citenamefont
  {Gonzalez}}]{gordon_natural_1999}%
  \BibitemOpen
  \bibfield  {author} {\bibinfo {author} {\bibfnamefont {M.~S.}\ \bibnamefont
  {Gordon}}, \bibinfo {author} {\bibfnamefont {M.~W.}\ \bibnamefont {Schmidt}},
  \bibinfo {author} {\bibfnamefont {G.~M.}\ \bibnamefont {Chaban}}, \bibinfo
  {author} {\bibfnamefont {K.~R.}\ \bibnamefont {Glaesemann}}, \bibinfo
  {author} {\bibfnamefont {W.~J.}\ \bibnamefont {Stevens}}, \ and\ \bibinfo
  {author} {\bibfnamefont {C.}~\bibnamefont {Gonzalez}},\ }\href {\doibase
  10.1063/1.478301} {\bibfield  {journal} {\bibinfo  {journal} {J. Chem.
  Phys.}\ }\textbf {\bibinfo {volume} {110}},\ \bibinfo {pages} {4199}
  (\bibinfo {year} {1999})}\BibitemShut {NoStop}%
\bibitem [{\citenamefont {Chamoli}\ \emph {et~al.}(2022)\citenamefont
  {Chamoli}, \citenamefont {Surjuse}, \citenamefont {Nayak},\ and\
  \citenamefont {Dutta}}]{chamoli2022lower}%
  \BibitemOpen
  \bibfield  {author} {\bibinfo {author} {\bibfnamefont {S.}~\bibnamefont
  {Chamoli}}, \bibinfo {author} {\bibfnamefont {K.}~\bibnamefont {Surjuse}},
  \bibinfo {author} {\bibfnamefont {M.~K.}\ \bibnamefont {Nayak}}, \ and\
  \bibinfo {author} {\bibfnamefont {A.~K.}\ \bibnamefont {Dutta}},\ }\href@noop
  {} {\bibfield  {journal} {\bibinfo  {journal} {arXiv preprint
  arXiv:2201.07752}\ } (\bibinfo {year} {2022})}\BibitemShut {NoStop}%
\bibitem [{\citenamefont {van Stralen}\ \emph {et~al.}(2005)\citenamefont {van
  Stralen}, \citenamefont {Visscher}, \citenamefont {Larsen},\ and\
  \citenamefont {Jensen}}]{vanStralen2005}%
  \BibitemOpen
  \bibfield  {author} {\bibinfo {author} {\bibfnamefont {J.~N.}\ \bibnamefont
  {van Stralen}}, \bibinfo {author} {\bibfnamefont {L.}~\bibnamefont
  {Visscher}}, \bibinfo {author} {\bibfnamefont {C.~V.}\ \bibnamefont
  {Larsen}}, \ and\ \bibinfo {author} {\bibfnamefont {H.~J.~A.}\ \bibnamefont
  {Jensen}},\ }\href {\doibase 10.1016/j.chemphys.2004.10.018} {\bibfield
  {journal} {\bibinfo  {journal} {Chem. Phys}\ }\textbf {\bibinfo {volume}
  {311}},\ \bibinfo {pages} {81} (\bibinfo {year} {2005})}\BibitemShut
  {NoStop}%
\bibitem [{\citenamefont {Kramers}(1930)}]{Kramers1930}%
  \BibitemOpen
  \bibfield  {author} {\bibinfo {author} {\bibfnamefont {H.~A.}\ \bibnamefont
  {Kramers}},\ }\href@noop {} {\bibfield  {journal} {\bibinfo  {journal} {Proc.
  Acad. Amsterdam}\ }\textbf {\bibinfo {volume} {33}} (\bibinfo {year}
  {1930})}\BibitemShut {NoStop}%
\bibitem [{\citenamefont {Visscher}(1996)}]{Visscher1996}%
  \BibitemOpen
  \bibfield  {author} {\bibinfo {author} {\bibfnamefont {L.}~\bibnamefont
  {Visscher}},\ }\href {<Go to ISI>://A1996UK63500004} {\bibfield  {journal}
  {\bibinfo  {journal} {Chem. Phys. Lett.}\ }\textbf {\bibinfo {volume}
  {253}},\ \bibinfo {pages} {20 } (\bibinfo {year} {1996})}\BibitemShut
  {NoStop}%
\bibitem [{\citenamefont {Saue}\ and\ \citenamefont
  {Jensen}(1999)}]{saue_quaternion_1999}%
  \BibitemOpen
  \bibfield  {author} {\bibinfo {author} {\bibfnamefont {T.}~\bibnamefont
  {Saue}}\ and\ \bibinfo {author} {\bibfnamefont {H.~J.~A.}\ \bibnamefont
  {Jensen}},\ }\href {\doibase 10.1063/1.479958} {\bibfield  {journal}
  {\bibinfo  {journal} {J. Chem. Phys.}\ }\textbf {\bibinfo {volume} {111}},\
  \bibinfo {pages} {6211} (\bibinfo {year} {1999})}\BibitemShut {NoStop}%
\bibitem [{\citenamefont {Repisky}\ \emph
  {et~al.}(2020{\natexlab{b}})\citenamefont {Repisky}, \citenamefont
  {Komorovsky}, \citenamefont {Kadek}, \citenamefont {Konecny}, \citenamefont
  {Ekström}, \citenamefont {Malkin}, \citenamefont {Kaupp}, \citenamefont
  {Ruud}, \citenamefont {Malkina},\ and\ \citenamefont
  {Malkin}}]{repisky_respect_2020}%
  \BibitemOpen
  \bibfield  {author} {\bibinfo {author} {\bibfnamefont {M.}~\bibnamefont
  {Repisky}}, \bibinfo {author} {\bibfnamefont {S.}~\bibnamefont {Komorovsky}},
  \bibinfo {author} {\bibfnamefont {M.}~\bibnamefont {Kadek}}, \bibinfo
  {author} {\bibfnamefont {L.}~\bibnamefont {Konecny}}, \bibinfo {author}
  {\bibfnamefont {U.}~\bibnamefont {Ekström}}, \bibinfo {author}
  {\bibfnamefont {E.}~\bibnamefont {Malkin}}, \bibinfo {author} {\bibfnamefont
  {M.}~\bibnamefont {Kaupp}}, \bibinfo {author} {\bibfnamefont
  {K.}~\bibnamefont {Ruud}}, \bibinfo {author} {\bibfnamefont {O.~L.}\
  \bibnamefont {Malkina}}, \ and\ \bibinfo {author} {\bibfnamefont {V.~G.}\
  \bibnamefont {Malkin}},\ }\href {\doibase 10.1063/5.0005094} {\bibfield
  {journal} {\bibinfo  {journal} {J. Chem. Phys.}\ }\textbf {\bibinfo {volume}
  {152}},\ \bibinfo {pages} {184101} (\bibinfo {year}
  {2020}{\natexlab{b}})}\BibitemShut {NoStop}%
\bibitem [{\citenamefont {Yoshimine}(1973)}]{yoshimine_construction_1973}%
  \BibitemOpen
  \bibfield  {author} {\bibinfo {author} {\bibfnamefont {M.}~\bibnamefont
  {Yoshimine}},\ }\href {\doibase 10.1016/0021-9991(73)90085-5} {\bibfield
  {journal} {\bibinfo  {journal} {J. Comput. Phys}\ }\textbf {\bibinfo {volume}
  {11}},\ \bibinfo {pages} {449} (\bibinfo {year} {1973})}\BibitemShut
  {NoStop}%
\bibitem [{\citenamefont {Shee}, \citenamefont {Visscher},\ and\ \citenamefont
  {Saue}(2016)}]{shee_analytic_2016}%
  \BibitemOpen
  \bibfield  {author} {\bibinfo {author} {\bibfnamefont {A.}~\bibnamefont
  {Shee}}, \bibinfo {author} {\bibfnamefont {L.}~\bibnamefont {Visscher}}, \
  and\ \bibinfo {author} {\bibfnamefont {T.}~\bibnamefont {Saue}},\ }\href
  {\doibase 10.1063/1.4966643} {\bibfield  {journal} {\bibinfo  {journal} {J.
  Chem. Phys.}\ }\textbf {\bibinfo {volume} {145}},\ \bibinfo {pages} {184107}
  (\bibinfo {year} {2016})}\BibitemShut {NoStop}%
\bibitem [{\citenamefont {Jensen}\ \emph {et~al.}(2022)\citenamefont {Jensen},
  \citenamefont {Bast}, \citenamefont {Gomes}, \citenamefont {Saue},
  \citenamefont {Visscher}, \citenamefont {Aucar}, \citenamefont {Bakken},
  \citenamefont {Chibueze}, \citenamefont {Creutzberg}, \citenamefont {Dyall},
  \citenamefont {Dubillard}, \citenamefont {Ekstr\"{o}m}, \citenamefont
  {Eliav}, \citenamefont {Enevoldsen}, \citenamefont {Faßhauer}, \citenamefont
  {Fleig}, \citenamefont {Fossgaard}, \citenamefont {Halbert}, \citenamefont
  {Hedegård}, \citenamefont {Helgaker}, \citenamefont {Helmich-Paris},
  \citenamefont {Henriksson}, \citenamefont {van Horn}, \citenamefont {Iliaš},
  \citenamefont {Jacob}, \citenamefont {Knecht}, \citenamefont {Komorovský},
  \citenamefont {Kullie}, \citenamefont {Lærdahl}, \citenamefont {Larsen},
  \citenamefont {Lee}, \citenamefont {List}, \citenamefont {Nataraj},
  \citenamefont {Nayak}, \citenamefont {Norman}, \citenamefont {Olejniczak},
  \citenamefont {Olsen}, \citenamefont {Olsen}, \citenamefont {Papadopoulos},
  \citenamefont {Park}, \citenamefont {Pedersen}, \citenamefont {Pernpointner},
  \citenamefont {Pototschnig}, \citenamefont {Di~Remigio}, \citenamefont
  {Repiský}, \citenamefont {Ruud}, \citenamefont {Sałek}, \citenamefont
  {Schimmelpfennig}, \citenamefont {Senjean}, \citenamefont {Shee},
  \citenamefont {Sikkema}, \citenamefont {Sunaga}, \citenamefont {Thorvaldsen},
  \citenamefont {Thyssen}, \citenamefont {van Stralen}, \citenamefont {Vidal},
  \citenamefont {Villaume}, \citenamefont {Visser}, \citenamefont {Winther},
  \citenamefont {Yamamoto},\ and\ \citenamefont {Yuan}}]{Dirac_22}%
  \BibitemOpen
  \bibfield  {author} {\bibinfo {author} {\bibfnamefont {H.~J.~A.}\
  \bibnamefont {Jensen}}, \bibinfo {author} {\bibfnamefont {R.}~\bibnamefont
  {Bast}}, \bibinfo {author} {\bibfnamefont {A.~S.~P.}\ \bibnamefont {Gomes}},
  \bibinfo {author} {\bibfnamefont {T.}~\bibnamefont {Saue}}, \bibinfo {author}
  {\bibfnamefont {L.}~\bibnamefont {Visscher}}, \bibinfo {author}
  {\bibfnamefont {I.~A.}\ \bibnamefont {Aucar}}, \bibinfo {author}
  {\bibfnamefont {V.}~\bibnamefont {Bakken}}, \bibinfo {author} {\bibfnamefont
  {C.}~\bibnamefont {Chibueze}}, \bibinfo {author} {\bibfnamefont
  {J.}~\bibnamefont {Creutzberg}}, \bibinfo {author} {\bibfnamefont {K.~G.}\
  \bibnamefont {Dyall}}, \bibinfo {author} {\bibfnamefont {S.}~\bibnamefont
  {Dubillard}}, \bibinfo {author} {\bibfnamefont {U.}~\bibnamefont
  {Ekstr\"{o}m}}, \bibinfo {author} {\bibfnamefont {E.}~\bibnamefont {Eliav}},
  \bibinfo {author} {\bibfnamefont {T.}~\bibnamefont {Enevoldsen}}, \bibinfo
  {author} {\bibfnamefont {E.}~\bibnamefont {Faßhauer}}, \bibinfo {author}
  {\bibfnamefont {T.}~\bibnamefont {Fleig}}, \bibinfo {author} {\bibfnamefont
  {O.}~\bibnamefont {Fossgaard}}, \bibinfo {author} {\bibfnamefont
  {L.}~\bibnamefont {Halbert}}, \bibinfo {author} {\bibfnamefont {E.~D.}\
  \bibnamefont {Hedegård}}, \bibinfo {author} {\bibfnamefont {T.}~\bibnamefont
  {Helgaker}}, \bibinfo {author} {\bibfnamefont {B.}~\bibnamefont
  {Helmich-Paris}}, \bibinfo {author} {\bibfnamefont {J.}~\bibnamefont
  {Henriksson}}, \bibinfo {author} {\bibfnamefont {M.}~\bibnamefont {van
  Horn}}, \bibinfo {author} {\bibfnamefont {M.}~\bibnamefont {Iliaš}},
  \bibinfo {author} {\bibfnamefont {C.~R.}\ \bibnamefont {Jacob}}, \bibinfo
  {author} {\bibfnamefont {S.}~\bibnamefont {Knecht}}, \bibinfo {author}
  {\bibfnamefont {S.}~\bibnamefont {Komorovský}}, \bibinfo {author}
  {\bibfnamefont {O.}~\bibnamefont {Kullie}}, \bibinfo {author} {\bibfnamefont
  {J.~K.}\ \bibnamefont {Lærdahl}}, \bibinfo {author} {\bibfnamefont {C.~V.}\
  \bibnamefont {Larsen}}, \bibinfo {author} {\bibfnamefont {Y.~S.}\
  \bibnamefont {Lee}}, \bibinfo {author} {\bibfnamefont {N.~H.}\ \bibnamefont
  {List}}, \bibinfo {author} {\bibfnamefont {H.~S.}\ \bibnamefont {Nataraj}},
  \bibinfo {author} {\bibfnamefont {M.~K.}\ \bibnamefont {Nayak}}, \bibinfo
  {author} {\bibfnamefont {P.}~\bibnamefont {Norman}}, \bibinfo {author}
  {\bibfnamefont {G.}~\bibnamefont {Olejniczak}}, \bibinfo {author}
  {\bibfnamefont {J.}~\bibnamefont {Olsen}}, \bibinfo {author} {\bibfnamefont
  {J.~M.~H.}\ \bibnamefont {Olsen}}, \bibinfo {author} {\bibfnamefont
  {A.}~\bibnamefont {Papadopoulos}}, \bibinfo {author} {\bibfnamefont {Y.~C.}\
  \bibnamefont {Park}}, \bibinfo {author} {\bibfnamefont {J.~K.}\ \bibnamefont
  {Pedersen}}, \bibinfo {author} {\bibfnamefont {M.}~\bibnamefont
  {Pernpointner}}, \bibinfo {author} {\bibfnamefont {J.~V.}\ \bibnamefont
  {Pototschnig}}, \bibinfo {author} {\bibfnamefont {R.}~\bibnamefont
  {Di~Remigio}}, \bibinfo {author} {\bibfnamefont {M.}~\bibnamefont
  {Repiský}}, \bibinfo {author} {\bibfnamefont {K.}~\bibnamefont {Ruud}},
  \bibinfo {author} {\bibfnamefont {P.}~\bibnamefont {Sałek}}, \bibinfo
  {author} {\bibfnamefont {B.}~\bibnamefont {Schimmelpfennig}}, \bibinfo
  {author} {\bibfnamefont {B.}~\bibnamefont {Senjean}}, \bibinfo {author}
  {\bibfnamefont {A.}~\bibnamefont {Shee}}, \bibinfo {author} {\bibfnamefont
  {J.}~\bibnamefont {Sikkema}}, \bibinfo {author} {\bibfnamefont
  {A.}~\bibnamefont {Sunaga}}, \bibinfo {author} {\bibfnamefont {A.~J.}\
  \bibnamefont {Thorvaldsen}}, \bibinfo {author} {\bibfnamefont
  {J.}~\bibnamefont {Thyssen}}, \bibinfo {author} {\bibfnamefont
  {J.}~\bibnamefont {van Stralen}}, \bibinfo {author} {\bibfnamefont {M.~L.}\
  \bibnamefont {Vidal}}, \bibinfo {author} {\bibfnamefont {S.}~\bibnamefont
  {Villaume}}, \bibinfo {author} {\bibfnamefont {O.}~\bibnamefont {Visser}},
  \bibinfo {author} {\bibfnamefont {T.}~\bibnamefont {Winther}}, \bibinfo
  {author} {\bibfnamefont {S.}~\bibnamefont {Yamamoto}}, \ and\ \bibinfo
  {author} {\bibfnamefont {X.}~\bibnamefont {Yuan}},\ }\href {\doibase
  10.5281/ZENODO.6010450} {\enquote {\bibinfo {title} {Dirac22},}\ } (\bibinfo
  {year} {2022})\BibitemShut {NoStop}%
\bibitem [{Dir(2021)}]{Dirac_21}%
  \BibitemOpen
  \href@noop {} {\emph {\bibinfo {title} {DIRAC, a relativistic ab initio
  electronic structure program, Release DIRAC21, Written by Bast, R.; Gomes, A.
  S. P.; Saue, T.; Visscher, L.; Jensen, H. J. Aa.; Aucar, I. A.; Bakken, V.;
  Dyall, K. G.; Dubillard, S.; Ekström, U.; Eliav, E.; Enevoldsen, T.;
  Faßhauer, E.; Fleig, T.; Fossgaard, O.; Halbert, L.; Hedegård, E. D.;
  Helgaker, T.; Helmich-Paris, B.; Henriksson, J.; Iliaš, M.; Jacob, Ch. R.;
  Knecht, S.; Komorovský, S.; Kullie, O.; Lærdahl, J. K.; Larsen, C. V.; Lee,
  Y. S.; List, N. H.; Nataraj, H. S.; Nayak, M. K.; Norman, P.; Olejniczak, G.;
  Olsen, J.; Olsen, J. M. H.; Papadopoulos, A.; Park, Y. C.; Pedersen, J. K.;
  Pernpointner, M.; Pototschnig, J. V.; Di Remigio, R.; Repiský, M.; Ruud, K.;
  Sałek, P.; Schimmelpfennig, B.; Senjean, B.; Shee, A.; Sikkema, J.; Sunaga,
  A.; Thorvaldsen, A. J.; Thyssen, J.; van Stralen, J.; Vidal, M. L.; Villaume,
  S.; Visser, O.; Winther, T.; Yamamoto, S. see http://www.diracprogram.org}}}
  (\bibinfo {year} {2021})\BibitemShut {NoStop}%
\bibitem [{\citenamefont {Lyakh}()}]{exatensor-url}%
  \BibitemOpen
  \bibfield  {author} {\bibinfo {author} {\bibfnamefont {D.~I.}\ \bibnamefont
  {Lyakh}},\ }\href@noop {} {\enquote {\bibinfo {title} {Exatensor, a basic
  numerical tensor algebra library for distributed heterogeneous hpc
  platforms},}\ }\bibinfo {howpublished}
  {\url{https://github.com/ORNL-QCI/ExaTENSOR}}\BibitemShut {NoStop}%
\bibitem [{\citenamefont {Dyall}(2002)}]{dyall_relativistic_2002}%
  \BibitemOpen
  \bibfield  {author} {\bibinfo {author} {\bibfnamefont {K.~G.}\ \bibnamefont
  {Dyall}},\ }\href {\doibase 10.1007/s00214-002-0388-0} {\bibfield  {journal}
  {\bibinfo  {journal} {Theor Chem Acc}\ }\textbf {\bibinfo {volume} {108}},\
  \bibinfo {pages} {335} (\bibinfo {year} {2002})}\BibitemShut {NoStop}%
\bibitem [{\citenamefont {Dyall}(2006)}]{dyall_relativistic_2006}%
  \BibitemOpen
  \bibfield  {author} {\bibinfo {author} {\bibfnamefont {K.~G.}\ \bibnamefont
  {Dyall}},\ }\href {\doibase 10.1007/s00214-006-0126-0} {\bibfield  {journal}
  {\bibinfo  {journal} {Theor Chem Acc}\ }\textbf {\bibinfo {volume} {115}},\
  \bibinfo {pages} {441} (\bibinfo {year} {2006})}\BibitemShut {NoStop}%
\bibitem [{\citenamefont {Dyall}(2012)}]{dyall_relativistic_2012}%
  \BibitemOpen
  \bibfield  {author} {\bibinfo {author} {\bibfnamefont {K.~G.}\ \bibnamefont
  {Dyall}},\ }\href {\doibase 10.1007/s00214-012-1172-4} {\bibfield  {journal}
  {\bibinfo  {journal} {Theor Chem Acc}\ }\textbf {\bibinfo {volume} {131}},\
  \bibinfo {pages} {1172} (\bibinfo {year} {2012})}\BibitemShut {NoStop}%
\bibitem [{\citenamefont {Dunning}(1989)}]{Dunning1989}%
  \BibitemOpen
  \bibfield  {author} {\bibinfo {author} {\bibfnamefont {T.~H.}\ \bibnamefont
  {Dunning}},\ }\href {\doibase 10.1063/1.456153} {\bibfield  {journal}
  {\bibinfo  {journal} {J. Chem. Phys.}\ }\textbf {\bibinfo {volume} {90}},\
  \bibinfo {pages} {1007} (\bibinfo {year} {1989})}\BibitemShut {NoStop}%
\bibitem [{\citenamefont {Kendall}, \citenamefont {Dunning},\ and\
  \citenamefont {Harrison}(1992)}]{kendall_electron_1992}%
  \BibitemOpen
  \bibfield  {author} {\bibinfo {author} {\bibfnamefont {R.~A.}\ \bibnamefont
  {Kendall}}, \bibinfo {author} {\bibfnamefont {T.~H.}\ \bibnamefont
  {Dunning}}, \ and\ \bibinfo {author} {\bibfnamefont {R.~J.}\ \bibnamefont
  {Harrison}},\ }\href {\doibase 10.1063/1.462569} {\bibfield  {journal}
  {\bibinfo  {journal} {J. Chem. Phys.}\ }\textbf {\bibinfo {volume} {96}},\
  \bibinfo {pages} {6796} (\bibinfo {year} {1992})}\BibitemShut {NoStop}%
\bibitem [{\citenamefont {Woon}\ and\ \citenamefont
  {Dunning}(1993)}]{woon_gaussian_1993}%
  \BibitemOpen
  \bibfield  {author} {\bibinfo {author} {\bibfnamefont {D.~E.}\ \bibnamefont
  {Woon}}\ and\ \bibinfo {author} {\bibfnamefont {T.~H.}\ \bibnamefont
  {Dunning}},\ }\href {\doibase 10.1063/1.464303} {\bibfield  {journal}
  {\bibinfo  {journal} {J. Chem. Phys.}\ }\textbf {\bibinfo {volume} {98}},\
  \bibinfo {pages} {1358} (\bibinfo {year} {1993})}\BibitemShut {NoStop}%
\bibitem [{\citenamefont {Woon}\ and\ \citenamefont
  {Dunning}(1995)}]{Woon1995}%
  \BibitemOpen
  \bibfield  {author} {\bibinfo {author} {\bibfnamefont {D.~E.}\ \bibnamefont
  {Woon}}\ and\ \bibinfo {author} {\bibfnamefont {T.~H.}\ \bibnamefont
  {Dunning}},\ }\href {\doibase 10.1063/1.470645} {\bibfield  {journal}
  {\bibinfo  {journal} {J. Chem. Phys.}\ }\textbf {\bibinfo {volume} {103}},\
  \bibinfo {pages} {4572} (\bibinfo {year} {1995})}\BibitemShut {NoStop}%
\bibitem [{\citenamefont {Peterson}\ and\ \citenamefont
  {Dunning}(2002)}]{Peterson2002}%
  \BibitemOpen
  \bibfield  {author} {\bibinfo {author} {\bibfnamefont {K.~A.}\ \bibnamefont
  {Peterson}}\ and\ \bibinfo {author} {\bibfnamefont {T.~H.}\ \bibnamefont
  {Dunning}},\ }\href {\doibase 10.1063/1.1520138} {\bibfield  {journal}
  {\bibinfo  {journal} {J. Chem. Phys.}\ }\textbf {\bibinfo {volume} {117}},\
  \bibinfo {pages} {10548} (\bibinfo {year} {2002})}\BibitemShut {NoStop}%
\bibitem [{\citenamefont {Hess}\ \emph {et~al.}(1996)\citenamefont {Hess},
  \citenamefont {Marian}, \citenamefont {Wahlgren},\ and\ \citenamefont
  {Gropen}}]{HE1996365}%
  \BibitemOpen
  \bibfield  {author} {\bibinfo {author} {\bibfnamefont {B.~A.}\ \bibnamefont
  {Hess}}, \bibinfo {author} {\bibfnamefont {C.~M.}\ \bibnamefont {Marian}},
  \bibinfo {author} {\bibfnamefont {U.}~\bibnamefont {Wahlgren}}, \ and\
  \bibinfo {author} {\bibfnamefont {O.}~\bibnamefont {Gropen}},\ }\href
  {\doibase https://doi.org/10.1016/0009-2614(96)00119-4} {\bibfield  {journal}
  {\bibinfo  {journal} {Chem. Phys. Lett.}\ }\textbf {\bibinfo {volume}
  {251}},\ \bibinfo {pages} {365} (\bibinfo {year} {1996})}\BibitemShut
  {NoStop}%
\bibitem [{\citenamefont {Marian}(2001)}]{marian2001}%
  \BibitemOpen
  \bibfield  {author} {\bibinfo {author} {\bibfnamefont {C.~M.}\ \bibnamefont
  {Marian}},\ }in\ \href {\doibase 10.1002/0471224413.ch3} {\emph {\bibinfo
  {booktitle} {Reviews in Computational Chemistry}}}\ (\bibinfo  {publisher}
  {John Wiley {\&} Sons, Inc.},\ \bibinfo {year} {2001})\ pp.\ \bibinfo {pages}
  {99--204}\BibitemShut {NoStop}%
\bibitem [{\citenamefont {Schimmelpfennig}()}]{prog:amfi}%
  \BibitemOpen
  \bibfield  {author} {\bibinfo {author} {\bibfnamefont {B.}~\bibnamefont
  {Schimmelpfennig}},\ }\href@noop {} {\enquote {\bibinfo {title}
  {{\textit{AMFI, an atomic mean-field spin-orbit integral program}}},}\
  }\bibinfo {note} {{University of Stockholm, Stockholm, Sweden;
  1999}}\BibitemShut {NoStop}%
\bibitem [{\citenamefont {Huber}(1979)}]{huber1979constants}%
  \BibitemOpen
  \bibfield  {author} {\bibinfo {author} {\bibfnamefont {K.-P.}\ \bibnamefont
  {Huber}},\ }\href@noop {} {\bibfield  {journal} {\bibinfo  {journal}
  {Molecular spectra and molecular structure}\ } (\bibinfo {year}
  {1979})}\BibitemShut {NoStop}%
\bibitem [{\citenamefont {Laerdahl}\ and\ \citenamefont
  {Schwerdtfeger}(1999)}]{laerdahl_fully_1999}%
  \BibitemOpen
  \bibfield  {author} {\bibinfo {author} {\bibfnamefont {J.~K.}\ \bibnamefont
  {Laerdahl}}\ and\ \bibinfo {author} {\bibfnamefont {P.}~\bibnamefont
  {Schwerdtfeger}},\ }\href {\doibase 10.1103/PhysRevA.60.4439} {\bibfield
  {journal} {\bibinfo  {journal} {Phys. Rev. A}\ }\textbf {\bibinfo {volume}
  {60}},\ \bibinfo {pages} {4439} (\bibinfo {year} {1999})}\BibitemShut
  {NoStop}%
\bibitem [{\citenamefont {Yuan}, \citenamefont {Visscher},\ and\ \citenamefont
  {Gomes}()}]{dataset-mp2fnos}%
  \BibitemOpen
  \bibfield  {author} {\bibinfo {author} {\bibfnamefont {X.}~\bibnamefont
  {Yuan}}, \bibinfo {author} {\bibfnamefont {L.}~\bibnamefont {Visscher}}, \
  and\ \bibinfo {author} {\bibfnamefont {A.~S.~P.}\ \bibnamefont {Gomes}},\
  }\href {\doibase 10.5281/zenodo.5939520} {\enquote {\bibinfo {title}
  {Dataset: Assessing mp2 frozen natural orbitals in relativistic correlated
  electronic structure calculations},}\ }\bibinfo {howpublished} {\url
  {https://doi.org/10.5281/zenodo.5939520}}\BibitemShut {NoStop}%
\bibitem [{\citenamefont {DePrince}\ and\ \citenamefont
  {Sherrill}(2013)}]{DePrince2013}%
  \BibitemOpen
  \bibfield  {author} {\bibinfo {author} {\bibfnamefont {A.~E.}\ \bibnamefont
  {DePrince}}\ and\ \bibinfo {author} {\bibfnamefont {C.~D.}\ \bibnamefont
  {Sherrill}},\ }\href {\doibase 10.1021/ct400250u} {\bibfield  {journal}
  {\bibinfo  {journal} {J. Chem. Theory Comput.}\ }\textbf {\bibinfo {volume}
  {9}},\ \bibinfo {pages} {2687} (\bibinfo {year} {2013})}\BibitemShut
  {NoStop}%
\bibitem [{\citenamefont {Le~Roy}(2017)}]{le_roy_level_2017}%
  \BibitemOpen
  \bibfield  {author} {\bibinfo {author} {\bibfnamefont {R.~J.}\ \bibnamefont
  {Le~Roy}},\ }\href {\doibase 10.1016/j.jqsrt.2016.05.028} {\bibfield
  {journal} {\bibinfo  {journal} {J. Quant. Spectrosc. Ra}\ }\textbf {\bibinfo
  {volume} {186}},\ \bibinfo {pages} {167} (\bibinfo {year}
  {2017})}\BibitemShut {NoStop}%
\bibitem [{\citenamefont {Kello}(1996)}]{kello_determination_nodate}%
  \BibitemOpen
  \bibfield  {author} {\bibinfo {author} {\bibfnamefont {V.}~\bibnamefont
  {Kello}},\ }\href@noop {} {\bibfield  {journal} {\bibinfo  {journal} {Mol.
  Phys}\ }\textbf {\bibinfo {volume} {89}},\ \bibinfo {pages} {127} (\bibinfo
  {year} {1996})}\BibitemShut {NoStop}%
\bibitem [{\citenamefont {Autschbach}, \citenamefont {Zheng},\ and\
  \citenamefont {Schurko}(2010)}]{Autschbach2010}%
  \BibitemOpen
  \bibfield  {author} {\bibinfo {author} {\bibfnamefont {J.}~\bibnamefont
  {Autschbach}}, \bibinfo {author} {\bibfnamefont {S.}~\bibnamefont {Zheng}}, \
  and\ \bibinfo {author} {\bibfnamefont {R.~W.}\ \bibnamefont {Schurko}},\
  }\href {\doibase https://doi.org/10.1002/cmr.a.20155} {\bibfield  {journal}
  {\bibinfo  {journal} {Concepts Magn. Reson.}\ }\textbf {\bibinfo {volume}
  {36A}},\ \bibinfo {pages} {84} (\bibinfo {year} {2010})}\BibitemShut
  {NoStop}%
\bibitem [{\citenamefont {Larsson}\ and\ \citenamefont
  {Pyykk{\"o}}(1986)}]{larsson1986relativistically}%
  \BibitemOpen
  \bibfield  {author} {\bibinfo {author} {\bibfnamefont {S.}~\bibnamefont
  {Larsson}}\ and\ \bibinfo {author} {\bibfnamefont {P.}~\bibnamefont
  {Pyykk{\"o}}},\ }\href@noop {} {\bibfield  {journal} {\bibinfo  {journal}
  {Chem. Phys.}\ }\textbf {\bibinfo {volume} {101}},\ \bibinfo {pages} {355}
  (\bibinfo {year} {1986})}\BibitemShut {NoStop}%
\bibitem [{\citenamefont {Visscher}\ \emph {et~al.}(1998)\citenamefont
  {Visscher}, \citenamefont {Enevoldsen}, \citenamefont {Saue},\ and\
  \citenamefont {Oddershede}}]{visscher_molecular_1998}%
  \BibitemOpen
  \bibfield  {author} {\bibinfo {author} {\bibfnamefont {L.}~\bibnamefont
  {Visscher}}, \bibinfo {author} {\bibfnamefont {T.}~\bibnamefont
  {Enevoldsen}}, \bibinfo {author} {\bibfnamefont {T.}~\bibnamefont {Saue}}, \
  and\ \bibinfo {author} {\bibfnamefont {J.}~\bibnamefont {Oddershede}},\
  }\href {\doibase 10.1063/1.477637} {\bibfield  {journal} {\bibinfo  {journal}
  {J. Chem. Phys.}\ }\textbf {\bibinfo {volume} {109}},\ \bibinfo {pages}
  {9677} (\bibinfo {year} {1998})}\BibitemShut {NoStop}%
\bibitem [{\citenamefont {Bast}\ \emph {et~al.}(2011)\citenamefont {Bast},
  \citenamefont {Koers}, \citenamefont {Gomes}, \citenamefont {Iliaš},
  \citenamefont {Visscher}, \citenamefont {Schwerdtfeger},\ and\ \citenamefont
  {Saue}}]{bast_analysis_2011}%
  \BibitemOpen
  \bibfield  {author} {\bibinfo {author} {\bibfnamefont {R.}~\bibnamefont
  {Bast}}, \bibinfo {author} {\bibfnamefont {A.}~\bibnamefont {Koers}},
  \bibinfo {author} {\bibfnamefont {A.~S.~P.}\ \bibnamefont {Gomes}}, \bibinfo
  {author} {\bibfnamefont {M.}~\bibnamefont {Iliaš}}, \bibinfo {author}
  {\bibfnamefont {L.}~\bibnamefont {Visscher}}, \bibinfo {author}
  {\bibfnamefont {P.}~\bibnamefont {Schwerdtfeger}}, \ and\ \bibinfo {author}
  {\bibfnamefont {T.}~\bibnamefont {Saue}},\ }\href {\doibase
  10.1039/C0CP01483D} {\bibfield  {journal} {\bibinfo  {journal} {Phys. Chem.
  Chem. Phys.}\ }\textbf {\bibinfo {volume} {13}},\ \bibinfo {pages} {864}
  (\bibinfo {year} {2011})}\BibitemShut {NoStop}%
\bibitem [{\citenamefont {Rauhut}\ and\ \citenamefont
  {Schwerdtfeger}(2021)}]{rauhut_parity-violation_2021}%
  \BibitemOpen
  \bibfield  {author} {\bibinfo {author} {\bibfnamefont {G.}~\bibnamefont
  {Rauhut}}\ and\ \bibinfo {author} {\bibfnamefont {P.}~\bibnamefont
  {Schwerdtfeger}},\ }\href {\doibase 10.1103/PhysRevA.103.042819} {\bibfield
  {journal} {\bibinfo  {journal} {Phys. Rev. A}\ }\textbf {\bibinfo {volume}
  {103}},\ \bibinfo {pages} {042819} (\bibinfo {year} {2021})}\BibitemShut
  {NoStop}%
\bibitem [{\citenamefont {Berger}\ and\ \citenamefont
  {Quack}(2000)}]{BergerPV2000}%
  \BibitemOpen
  \bibfield  {author} {\bibinfo {author} {\bibfnamefont {R.}~\bibnamefont
  {Berger}}\ and\ \bibinfo {author} {\bibfnamefont {M.}~\bibnamefont {Quack}},\
  }\href {\doibase 10.1063/1.480900} {\bibfield  {journal} {\bibinfo  {journal}
  {J. Chem. Phys.}\ }\textbf {\bibinfo {volume} {112}},\ \bibinfo {pages}
  {3148} (\bibinfo {year} {2000})}\BibitemShut {NoStop}%
\end{thebibliography}%

\end{document}



\title{Supplementary information: Assessing MP2 frozen natural orbitals in relativistic correlated electronic structure calculations}

\author{Xiang Yuan}
    \email{xiang.yuan@univ-lille.fr}
    \affiliation{
        Université de Lille, CNRS, UMR 8523 - PhLAM - Physique des Lasers, Atomes et Molécules, F-59000 Lille,  France.
    }

    \affiliation{%
        Department of Chemistry and Pharmaceutical Sciences, Faculty of Science, Vrije Universiteit Amsterdam, de Boelelaan 1083, 1081 HV Amsterdam, The Netherlands.
    }%

\author{Lucas Visscher}
    \email{l.visscher@vu.nl}
    \affiliation{%
        Department of Chemistry and Pharmaceutical Sciences, Faculty of Science, Vrije Universiteit Amsterdam, de Boelelaan 1083, 1081 HV Amsterdam, The Netherlands.
    }%

\author{Andr\'{e} Severo Pereira Gomes}%
    \email{andre.gomes@univ-lille.fr}
    \affiliation{
        Université de Lille, CNRS, UMR 8523 - PhLAM - Physique des Lasers, Atomes et Molécules, F-59000 Lille, France.
    }%
    
\let\cleardoublepage\clearpage
\begin{abstract}
\end{abstract}
\maketitle

        \begin{figure*}[htb]
            \centering
            \includegraphics[width=1.0\linewidth]{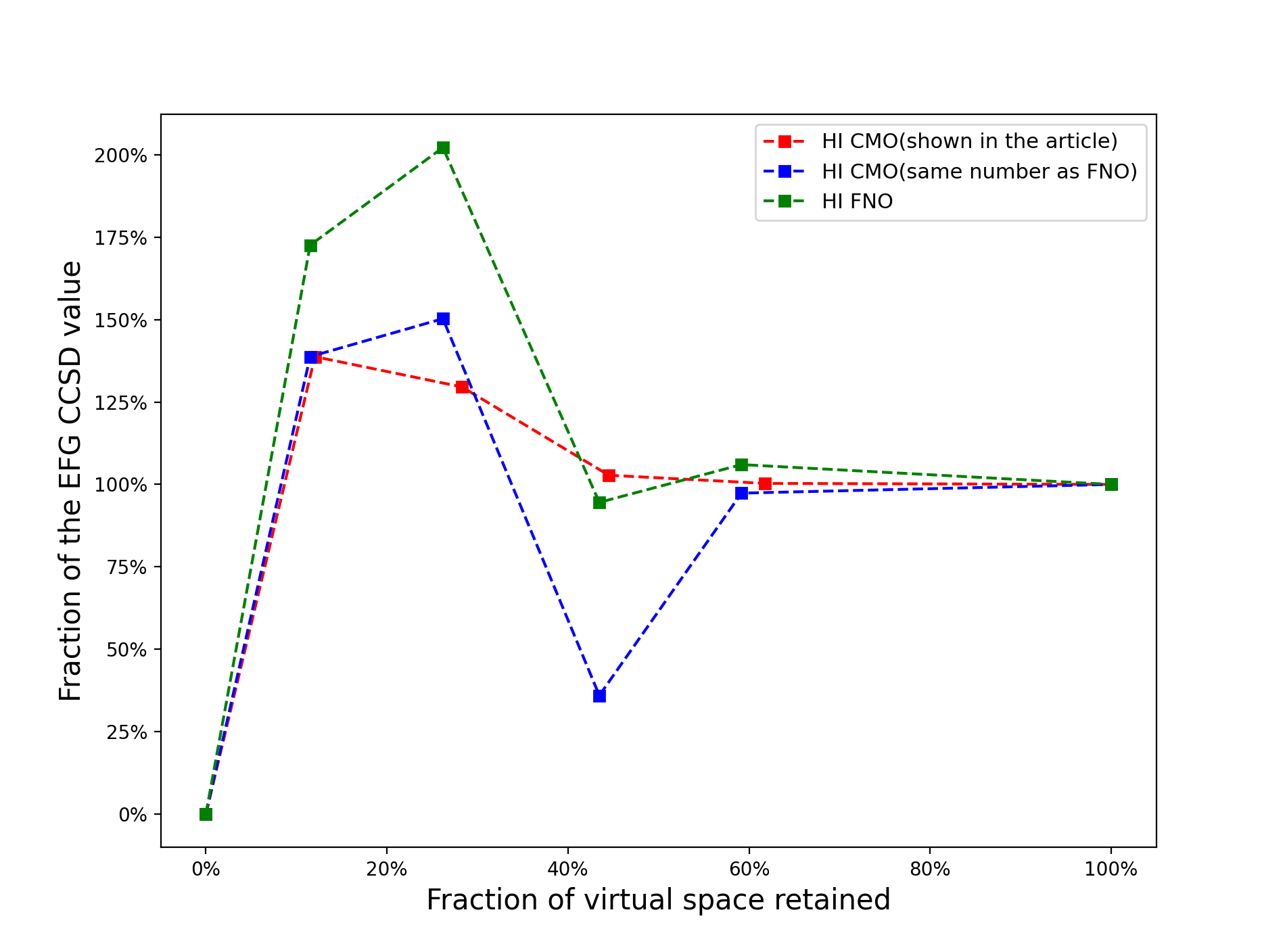}
            \captionsetup{font=normal}
            \caption{Convergence of the CCSD EFG at the I atom with respect to the size of the virtual orbital space, for the X2C Hamiltonian, comparing FNO and CMO, the latter being truncated at exactly the same number of FNO and approximately respecting atomic shell boundaries as shown in the article body. The X axis indicates the fraction of the virtual space retained, while the Y axis gives the fraction of the correlation energy recovered with respect to the value obtained with the untruncated virtual space.}\label{fig:Fig-s1}
        \end{figure*}

        \begin{table*}[htb]
        \large
        \centering
        \captionsetup{font=Large}
        \caption{Correlation energy and expectation value using uncontracted aug-cc-pVTZ and aug-cc-pCVTZ basis sets for HCl with full orbital space}\label{tab:table-s1}
        \begin{ruledtabular}
        \begin{tabular}{ccc}
            \hline
            MP2 Correlation energy (a.u.)       & aug-cc-pVTZ     & aug-cc-pCVTZ    \\
            All Electrons                             & -0.33206 & -0.51702 \\
            Valence Electrons only                   & -0.20986 & -0.21094 \\
            \hline
            CCSD Correlation energy (a.u.)      & aug-cc-pVTZ     & aug-cc-pCVTZ    \\
            All Electrons                             & -0.34904 & -0.53415   \\
            Valence Electrons only                    & -0.22911 & -0.23077 \\
            \hline
            CCSD Dipole Moment (a.u.)           & aug-cc-pVTZ     & aug-cc-pCVTZ \\
            All Electrons                             & 0.73 & 0.73  \\
            Valence Electrons only                    & 0.76  & 0.77  \\
            \hline
            CCSD Quadupole Moment (a.u.)        & aug-cc-pVTZ     & aug-cc-pCVTZ \\
            All Electrons                             & 2.74  & 2.74  \\
            Valence Electrons only                    & 2.87   & 2.87  \\
            \hline
            CCSD Electric field gradient (a.u.) & aug-cc-pVTZ     & aug-cc-pCVTZ \\
            All Electrons                             & 3.59     & 3.56     \\
            Valence Electrons only                    & 3.41     & 3.38    \\
            \hline
        \end{tabular}
        \end{ruledtabular}
        \end{table*}

        \begin{table*}[htb]
        \begin{threeparttable}
        \large
        \centering
        \captionsetup{font=large}
        \caption{Energy and molecular properties of HTs using different Hamiltonians with Hartree-Fock wave functions}\label{tab:table-s2}
        \begin{ruledtabular}
        \begin{tabular}{ccccccc}
            \hline
            Property(a.u.)&DC\tnote{a} \  &DCSSSS\tnote{b} &X2C    &X2C\tnote{c} &X2C spinfree & NONREL\\
            \hline
            Total energy         &      -53750.9    & -53718.7& -53671.3 &-53448.8&-53241.6&-45401.0\\
            HOMO-LUMO gap        &0.2669  &0.2668 &0.2670                  &0.2694&0.3428&0.3633\\
            Electric Dipole moment &    0.992              & 0.988 & 0.986 &0.955&0.358&-0.074\\
            Electric Quadrupole moment\tnote{d} & 1566.65    & 1566.66     & 1566.68   &1566.83&1568.00&1570.81\\
            Electric field gradient\tnote{e}  &   60.1              & 59.9& 59.8   &56.5&78.4&21.7\\
            \hline
        \end{tabular}
        \begin{tablenotes}
        \item [a] Dirac–Coulomb Hamiltonian in which (SS|SS) integrals are replaced by an interatomic SS correction.
        \item [b] Dirac-Coulomb Hamiltonian with explicitly including (SS|SS) type Coulomb integrals.
        \item [c] Using Gaussian charge nuclear model.
        \item [d] ZZ component.
        \item [e] ZZ component on Ts nucleus. 
        \end{tablenotes}
        \end{ruledtabular}
        \end{threeparttable}
        \end{table*}

\begin{table*}[htb]
\large
    \centering
    \captionsetup{font=Large}
    \caption{The fraction of the MP2 and CCSD correlation energy of hydrogen halides molecules using frozen natural orbitals}\label{tab:table-s2}
\begin{ruledtabular}
\begin{tabular}{ccc}
Retained Virtual Orbital Space & MP2 Correlation Energy  & CCSD Correlation Energy  \\
\hline
&HF&\\
\hline
0.00\%                 & 0.00\%                       & 0.00\%                         \\
11.54\%                & 64.93\%                      & 64.33\%                        \\
32.05\%                & 85.02\%                      & 85.09\%                        \\
64.10\%                & 98.36\%                      & 98.46\%                        \\
82.05\%                & 99.88\%                      & 99.89\%                        \\
100.00\%               & 100.00\%                     & 100.00\%                       \\
\hline
&HCl&\\
\hline
0.00\%                 & 0.00\%                       & 0.00\%                         \\
10.99\%                & 43.08\%                      & 44.59\%                        \\
35.16\%                & 84.80\%                      & 85.63\%                        \\
62.64\%                & 94.91\%                      & 95.13\%                        \\
81.32\%                & 99.39\%                      & 99.43\%                        \\
100.00\%               & 100.00\%                     & 100.00\%                       \\
\hline
&HBr&\\
\hline
0.00\%                 & 0.00\%                       & 0.00\%                         \\
16.13\%                & 57.53\%                      & 56.12\%                        \\
29.68\%                & 77.27\%                      & 76.52\%                        \\
48.39\%                & 92.01\%                      & 91.71\%                        \\
67.74\%                & 98.03\%                      & 97.97\%                        \\
100.00\%               & 100.00\%                     & 100.00\%                       \\
\hline
&HI&\\
\hline
0.00\%                 & 0.00\%                       & 0.00\%                         \\
11.52\%                & 30.97\%                      & 31.32\%                        \\
26.18\%                & 67.66\%                      & 66.99\%                        \\
43.46\%                & 89.49\%                      & 89.23\%                        \\
59.16\%                & 96.74\%                      & 96.66\%                        \\
100.00\%               & 100.00\%                     & 100.00\%                       \\
\hline
&HAt&\\
\hline
0.00\%                 & 0.00\%                       & 0.00\%                         \\
11.76\%                & 35.97\%                      & 33.95\%                        \\
23.90\%                & 63.53\%                      & 61.22\%                        \\
44.85\%                & 87.62\%                      & 86.66\%                        \\
62.87\%                & 96.86\%                      & 96.62\%                        \\
100.00\%               & 100.00\%                     & 100.00\%                       \\
\hline
&HTs&\\
\hline
0.00\%                 & 0.00\%                       & 0.00\%                         \\
10.54\%                & 25.86\%                      & 24.22\%                        \\
23.47\%                & 55.31\%                      & 52.89\%                        \\
42.86\%                & 80.29\%                      & 79.00\%                        \\
59.52\%                & 93.50\%                      & 93.08\%                        \\
100.00\%               & 100.00\%                     & 100.00\%                      
\end{tabular}
\end{ruledtabular}
\end{table*}
